\documentclass[11pt, oneside]{article}  

\usepackage{bm}
\usepackage{enumitem}
\usepackage{amssymb}
\usepackage{amsmath}
\usepackage{amsfonts}
\usepackage{bbold}
\usepackage{bm}
\usepackage{slashed}
\usepackage{graphicx}
\usepackage{subcaption}
\usepackage{cite}
\usepackage{bbold}
\usepackage{color}
\usepackage{hyperref} 
\hypersetup{
    colorlinks=true,       
    linkcolor=blue,        
    citecolor=blue,        
    filecolor=magenta,     
    urlcolor=blue          
}

\newcommand{\V}{V^\prime}
\newcommand{\nn}{\nonumber}
\newcommand{\dNeff}{\Delta N_{\rm eff}}

\newcommand{\kaB}{\kappa_{B^\prime}}
\newcommand{\kal}{\kappa_{\ell^\prime}}

\newcommand{\TeV}{{\rm \, TeV}}
\newcommand{\GeV}{{\rm \, GeV}}
\newcommand{\MeV}{{\rm \, MeV}}
\newcommand{\keV}{{\rm \, keV}}
\newcommand{\eV}{{\rm \, eV}}
\newcommand{\s}{{\rm \, s}}

\newcommand{\eps}{\epsilon}
\newcommand{\eq}[1]{\begin{align}#1\end{align}}

\newcommand{\vpq}{f_{\rm PQ}}
\newcommand{\one}{{\cal O}(1)}

\newcommand{\F}{\Lambda}
\newcommand{\FA}{\Lambda^{\rm A}}
\newcommand{\FV}{\Lambda^{\rm V}}
\newcommand{\FVA}{\Lambda^{\rm V,A}}
\newcommand{\FL}{\Lambda^{\rm L}}
\newcommand{\FR}{\Lambda^{\rm R}}
\newcommand{\CA}{C^{\rm A}}
\newcommand{\CV}{C^{\rm V}}

\newcommand{\CVA}{C^{\rm V,A}}
\newcommand{\CP}{C^{\rm P}}
\newcommand{\CS}{C^{\rm S}}
\newcommand{\CAs}{C^{{\rm A}*}}

\newcommand{\CtV}{\tilde{C}^{\rm V}}
\newcommand{\CtA}{\tilde{C}^{\rm A}}
\newcommand{\CtVA}{\tilde{C}^{\rm V,A}}

\newcommand{\CbV}{{\bf C}^{\rm V}}
\newcommand{\CbA}{{\bf C}^{\rm A}}
\newcommand{\CbVA}{{\bf C}^{\rm V,A}}

\newcommand{\Cga}{C_\gamma}

\begin{document}

\title{
\begin{flushright}
{\small TTP26-019,  P3H-26-043}\\
\end{flushright}
\begin{flushright}
\begin{minipage}{0.2\linewidth}
\normalsize 
\end{minipage}
\end{flushright}
 {\huge\bf Flavor phenomenology  of \\ light dark particles}\\[0.2cm]}
\date{}

\author{
Robert Ziegler \\*[20pt]
\centerline{
\begin{minipage}{\linewidth}
\begin{center}
{\small
Institute for Theoretical Particle Physics, \\ Karlsruhe Institute of Technology,  76131 Karlsruhe, Germany} \\
\end{center}
\end{minipage}}
\\[8mm]}
\maketitle
\thispagestyle{empty}

\centerline{\large\bf Abstract}
\begin{quote}
\indent
We review  the flavor phenomenology of light dark particles, focusing on axion-like particles with sub-GeV masses and generic flavor-violating couplings. Such states can  naturally emerge from the spontaneous breaking of generic flavored symmetries, and are motivated by dark matter or the Strong CP Problem, with the QCD axion serving as a paradigmatic example. Light dark particles can be produced in two-body decays of Standard Model particles,  giving rise to missing energy signals that can not only be observed  in high-precision flavor experiments, but  also be probed in core-collapse supernovae and the cosmic microwave background. These decays are controlled by dimension-five operators,  which makes dedicated laboratory searches  sensitive to very large UV scales up to $10^{12} {\rm GeV}$ and thus highly complementary to astrophysical and cosmological probes. We provide a comprehensive survey of the resulting  limits and  prospects across all relevant channels, highlighting the central role of flavor physics  in exploring the landscape of light dark matter.
\end{quote}
\newpage

\tableofcontents

\numberwithin{equation}{section}
\setlength\parindent{1.5em}

\newpage
\section{Introduction}
 
Close to the end of Run 3 of the Large Hadron Collider (LHC) there is no crack in the Standard Model (SM) of particle physics, establishing its reign supreme of fundamental phenomena down to scales of $10^{-18}$ m. Hopes to find new particles  with masses just beyond the electroweak (EW) scale have been  disappointed so far, raising doubts on the traditional paradigm of low-scale supersymmetry and similar frameworks addressing the hierarchy problem. At the same time, no direct hints for dark matter (DM) particles have been seen in   low-background underground detectors, challenging the  WIMP paradigm of DM as a thermal relic with weak scale mass. 

While it may well be that great discoveries lie just ahead, the  present situation may  hint to  a shift of paradigms. In fact new particles beyond the SM are not required to have masses above the weak scale, and can be significantly lighter if they are more weakly coupled than neutrinos. Although difficult to find  at the LHC and high-threshold direct detection experiments, light dark  particles have an extremely rich phenomenology  crossing the boundaries between particle physics, astrophysics and cosmology. 

In this review, we focus on the flavor phenomenology of light dark particles, where ``light" refers to masses below 100 MeV, and ``dark" means stability on the scale of meters or above. Although we will mainly talk about axions, the discussion holds to large extent also for light CP-even scalars and light vector bosons. Building upon several surveys of axion flavor physics~\cite{Feng:1997tn, Kamenik:2011vy, Bjorkeroth:2018dzu, Ziegler:2019gjr, MartinCamalich:2020dfe, Calibbi:2020jvd,  Bauer:2021mvw, Ziegler:2023aoe, MartinCamalich:2025srw},  we aim to give an  overview of the field in its full breadth that covers limits and  prospects from laboratory searches as well as 
probes from astrophysics and cosmology. One purpose of this discussion is to highlight the importance of  searches for new light states in flavor-violating decays, in particular to the experimental communities.  To this extent we will argue in the following that searches for two-body decays with missing energy are especially promising: they are firmly motivated by DM and the Strong CP Problem, sensitive to very large UV scales up to $10^{12} \GeV$, and highly complementary to other experimental probes, for example observations of the Cosmic Microwave Background (CMB) sensitive to dark radiation.

Particularly motivated examples of light dark particles are  pseudoscalars, which can  arise as (pseudo-) Nambu-Goldstone bosons (pNGBs) from the spontaneous breaking of   global symmetries. This symmetry simultaneously  protects the pNGB mass, which can naturally be well \emph{below}  the weak scale, and ensures that   its interactions  with the SM are  suppressed by the Goldstone decay constant, which can be well \emph{above} the weak scale. Given this  regime of its mass and decay constant,  a pNGB can be easily stable  on cosmological scales. Light pseudoscalars  are thus vanilla DM candidates that in contrast to heavy WIMPs do not need to be stabilized by additional symmetries. Besides the DM relic abundance, pNGBs may be connected to other  open issues in particle physics, such as  the hierarchical structure of the SM Yukawa sector,  the origin of neutrino masses or the hierarchy problem. Their archetype is  the QCD axion~\cite{Weinberg:1977ma,Wilczek:1977pj}, originally motivated to address the Strong CP Problem~\cite{Peccei:1977hh,Peccei:1977ur}, and distinguished by a direct relation between  mass  and  decay constant. It has been customary to call generic pNGBs (where mass and decay constant are treated as independent parameters)  ``axion-like particles" (ALPs) and we will  follow this terminology, using the terms pNGB, Goldstone, ALP and axion interchangeably in this review.

The importance of flavor physics in probing new light particles was  emphasized already in the 1980s~\cite{Reiss:1982sq, Gelmini:1982zz, Wilczek:1982rv}. This was reflected in the  efforts of several experimental groups to search for missing energy (e.g. invisible axions) in flavor-violating two-body decays~\cite{Jodidio:1986mz, SINDRUM:1986klz, Bryman:1988ut, Bolton:1988af}, and it is remarkable that searches conducted at TRIUMF in 1986~\cite{Jodidio:1986mz} still provide the strongest limits on lepton-flavor-violating (LFV) axion couplings to muons. In many contemporary  flavor experiments the focus seems to have shifted to probes of heavy new physics above the electroweak scale motivated by e.g. the hierarchy problem. For example the flagship search of the MEG experiment for $\mu \to e \gamma$ decays probes dimension-six operators and is sensitive to effective UV scales of the order of $10^8 \GeV$, giving very stringent constraints on the flavor-violating pattern of low-energy supersymmetry. It is remarkable that searches for light new physics in  
processes such as $\mu \to e a$ probe even larger effective scales up to $10^{10} \GeV$, since they can arise from dimension-five operators,  and because of the different scaling with the UV scale, future experiments will  further broaden this gap. Similarly in the quark sector searches for $K \to \pi a$ carried out by the NA62 collaboration probe effective UV scales up to $10^{12} \GeV$, which is many orders of magnitude larger than the limits on dimension-six operators tested by searches for $K \to \pi \nu \overline \nu$. Notice that often the sensitivity of two-body decays appears to be much weaker, since the searches are interpreted not in terms of model-independent flavor-violating operators, but within explicit models that mimic the flavor suppression of the SM, e.g.~\cite{Dolan:2014ska, Izaguirre:2016dfi}. We emphasize that there is no need for such an unnecessary theory bias, since there is no reason why the particular pattern of flavor suppression in the SM should  also hold for new physics. In contrast to heavy physics addressing the hierarchy problem, there is no requirement for the axion decay constant to be close to the electroweak scale. 

Similar to flavor-violating decays  mediated by heavy particles, precision flavor experiments looking for  light new physics are also highly supplementary to other experiments. Just as searches for rare SM decays such as $B_s \to \mu^+ \mu^-$ were important for constraining supersymmetric particles  hunted  at the LHC~\cite{Choudhury:1998ze}, searches for light axions  in $B \to K a$ complement cosmological probes sensitive to the very same decay as a source of dark radiation in the early universe~\cite{Baumann:2016wac}. More generally, decays of SM particles can provide a unique opportunity to produce light  axion dark matter in the laboratory~\cite{Bird:2004ts}, which otherwise is matched only in extreme environments such as core-collapse supernovae and the very early universe at $10^{-4} {\rm\,  s}$. This is particularly relevant as  light DM particles in our local halo are very hard to see with conventional  DM searches, because of their non-relativistic energies and the large energy thresholds in direct or indirect detection experiments. Instead flavor physics is perfectly sensitive even to very light axions, complementing dedicated searches e.g.  with axion haloscopes. Note that in contrast to those searches, which heavily rely on resonant enhancement and thus are limited to a small range of  axion masses, flavor experiments are basically independent of the axion mass up to the kinematical threshold.

This review is organized as follows. In the next section we will set up the general effective ALP Lagrangian~(Section~\ref{sec:EFTsetup}), and discuss at length its theoretical motivation~(Section~\ref{sec:motivation}), with an emphasis on the origin of flavor-violating couplings~(Section~\ref{sec:origin}). Their anatomy will be investigated in Section 3, beginning with a simple estimate of  the limits that can be inferred from lifetimes of SM particles~(Section~\ref{sec:lifetimes}). Stronger  constraints can be obtained from dedicated searches for two-body decays of SM particles with missing energy~(Section~\ref{sec:decayrates}), which have to be distinguished from the corresponding SM decay with a pair of neutrinos~ (Section~\ref{sec:background}). This requires the ALP to be sufficiently long-lived in collider experiments, and we will argue that this situation is quite generic for ALPs below 100 MeV, taking into account existing experimental limits on couplings to electrons and photons~(Section~\ref{sec:decaylengths}).  In Section 4 we will discuss the model-independent limits on flavor-violating couplings that can be obtained from all systems  that feature a sizable number of heavy SM flavors:  particle colliders~(Section~\ref{sec:lab}), the extreme environments of core collapse supernovae~(Section~\ref{sec:astro}) and the early universe~(Section~\ref{sec:cosmolimits}). The  limits are compared  in Section~\ref{sec:summarylimits}, in particular with respect to their complementarity regarding ALP mass and decay length. As arguably the best motivation for light dark axions  is their ability to fully account for the observed DM abundance, we study this possibility in Section 5, discussing axion stability~(Section~\ref{sec:stability}) and production in the early universe~(Section~\ref{sec:production}). Section 6 is dedicated to complete models, where explicit targets for flavor-violating couplings are obtained from a direct connection to SM flavor hierarchies~(Section~\ref{sec:axiflavon}), the pattern of neutrino masses within type-I seesaw models~(Section~\ref{sec:majoron}) or the observed DM relic abundance~(Section~\ref{sec:lfvfreezein}). We present additional material in various appendices. The most general axion Lagrangian and the field redefinitions that can be used to switch between physically equivalent bases are discussed in Appendix~\ref{app:axionbasis}, while the relation to  a spontaneously broken Peccei-Quinn symmetry is reviewed in Appendix~\ref{app:PQ}. We also list the complete two-body decay rates~(Appendix~\ref{app:2bodydecays}) with  hadronic matrix elements defined in Appendix~\ref{app:FFs}. Appendix~\ref{app:diphoton} contains  complete expressions for  the axion decay rate into two photons for the general effective Lagrangian, while Appendix~\ref{app:cosmo} gives a brief summary of various results of thermal axion production in the early universe.


\section{Light dark particles}

In this section we define the basic setup in terms of an effective Lagrangian below the EW scale. We  restrict to local operators of lowest dimension,  only including light pseudoscalars, and consider the most general interactions with SM particles (Section~\ref{sec:EFTsetup}), focusing in particular on flavor-violating couplings. The phenomenology of these interactions can be easily mapped to other light bosons, such as CP-even scalars or light vector bosons. Nevertheless light pseudoscalars enjoy the best theoretical motivation, as they can arise as (pseudo-) Nambu-Goldstone bosons of  spontaneously broken symmetries, the main representative being the QCD axion (Section~\ref{sec:motivation}). We finally discuss the UV origin of flavor-violating couplings in Section~\ref{sec:origin}.

\subsection{Effective Lagrangian}
\label{sec:EFTsetup}

The interactions of  light dark particles relevant for flavor physics can be conveniently described  by an effective  field theory (EFT) below the electroweak  scale, defined by a local Lagrangian that contains the most general set of  flavor-violating interactions operators that respect the unbroken part of the SM gauge group, $SU(3)_c\times U(1)_{\rm em}$. The operators themselves are constructed from the dark fields (taken as SM singlets) and  SM fermions. The latter can be taken  as mass eigenstates, since without loss of generality one can work in the basis where SM Yukawa interactions have been diagonalized. The EFT is then defined by an expansion in the UV scale $\Lambda$, which we take much larger than the weak scale, and we keep only the leading operators with least inverse powers of $\Lambda$. 

Following this logic, we  restrict our analysis to light dark bosons, since operators involving dark fermions coupled to SM fermion currents are at least dimension-six (the flavor phenomenology of such operators has been discussed in e.g. Ref.~\cite{Badin:2010uh}). Moreover, as we will motivate below, we specifically consider the effective Lagrangian of a light pseudoscalar $a$, to which we will refer to interchangeably as axion, pNGB or ALP.  The relevant flavor-violating dimension-five effective Lagrangian then reads
\begin{align}
\label{axionFV}
{\cal L}_{{\rm axion, FV}}  =  
 \sum_{i \ne j} \frac{\partial_\mu a}{2 \Lambda} \, \overline{f}_i \gamma^\mu \left( \CV_{ij} + \CA_{ij} \gamma_5 \right)f_j \, ,
\end{align}
where $\Lambda$ is some UV scale and $\CVA_{ij}$ are hermitian matrices in flavor space. In fact this Lagrangian is the most general for \emph{any} scalar when considering flavor-off diagonal couplings, since operators involving scalar and pseudoscalar fermion currents can always be brought into this form via field redefinitions (see Appendix~\ref{app:axionbasis}). Note in particular that the naive Yukawa-like operators of the form $a \overline{f}_i f_j$ violate $SU(2)_L$, and thus must involve (at least) an additional power of the electroweak scale, making these operator  dimension-five. 

At the same order in the EFT expansion generic scalar fields  have also flavor-diagonal couplings that cannot be brought into derivative form, as well as couplings to gauge fields strength like $FF \equiv F^{\mu \nu}F_{\mu \nu}$ and   $F \tilde{F} \equiv \tilde{F}^{\mu \nu}F_{\mu \nu}$. The former couplings are subject to very stringent  limits due to long-range forces that modify the Newtonian potential at short distances~\cite{OHare:2020wah, Fiorillo:2025zzx}, while light pseudoscalars instead only give rise to spin-dependent $1/r^3$ potentials~\cite{Reiss:1982sq, Gelmini:1982zz}, which give only extremely weak constraints of the order of the EW scale~\cite{Mostepanenko:2020lqe}. Taking $a$ explicitly as a pseudoscalar and restricting to CP-conserving interactions  selects a subset of these flavor-conserving dimension-five operators, which are given by
\begin{align}
\label{axionFC}
{\cal L}_{{\rm axion, FC}}  =  C_{GG} \frac{a}{\Lambda} \frac{\alpha_s}{4 \pi} G \tilde{G}+ C_{\gamma \gamma} \frac{a}{\Lambda} \frac{\alpha_{\rm em}}{4 \pi} F  \tilde{F}  +  \sum_{i } C_{i} \frac{\partial_\mu a}{2 \Lambda} \, \overline{f}_i \gamma^\mu  \gamma_5 f_j \, ,
\end{align}
with real parameters $C_{GG}, C_{\gamma \gamma}, C_{i}$ and we have used the fact that derivative couplings to flavor-diagonal vector currents can be completely absorbed via field redefinitions, see Appendix~\ref{app:axionbasis}. Moreover, imposing CP on the interactions in Eq.~\eqref{axionFV} implies that the hermitian couplings $\CV_{ij}$ and $\CA_{ij}$ are also symmetric. 

In this review we will be interested mainly in the flavor-violating axion couplings in Eq.~\eqref{axionFV}, for which we will often use the shorthand notation 
\begin{align}
\label{eq:shorthandF}
\FVA_{ij} & \equiv \frac{2 \Lambda}{\CVA_{ij}} \, , &
\F_{ij} \equiv \frac{2 \Lambda}{\sqrt{|\CV_{ij}|^2 + {|\CA_{ij}|^2}}} \, ,
\end{align}
with $i \ne j$. The flavor phenomenology controlled by these couplings can be easily matched to analogous couplings of generic scalar particles, which couple to off-diagonal fermion currents as
\begin{equation}
\label{scalar}
{\cal L}_{\rm scalar, FV}  =  -   \frac{a}{\Lambda} \sum_{i \ne j} \overline{f}_i \left( \CS_{ij} +i \CP_{ij} \gamma_5 \right) f_j   \, ,
\end{equation}
upon using the identifications (see Appendix~\ref{app:axionbasis})
\eq{
\CS_{ij} & = i (m_i - m_j) \CV_{ij} \, , &  \CP_{ij} & = (m_i + m_j) \CA_{ij}  \, .
}
Moreover, our analysis also applies to light vector particles $V^\prime_\mu$ associated with some extra group $U(1)^\prime$, which  couple to flavor-violating vector and axial currents according to (cf.~Ref.~\cite{Eguren:2024oov})
\begin{equation}
\label{vector}
{\cal L}_{\rm vector, FV}  =  \frac{m_{\V}}{2 \Lambda} \V_\mu \sum_{i \ne j} \overline{f}_i \gamma^\mu \left( \CtV_{ij} + \CtA_{ij} \gamma_5 \right)f_j \, .
\end{equation}
Note that flavor-violating currents are not conserved and thus must be proportional to at least one power of the $U(1)^\prime$-breaking order parameter, which
we take as the same vacuum expectation value that controls the vector boson mass $m_{\V}$. This normalization  is convenient, since it  ensures that the growth of amplitudes with longitudinally polarized vectors in initial or final states proportional to $E/m_{\V}$ as $m_{\V}\to 0$ is cancelled by the $m_{\V}$ dependence in the interaction vertex. This leads to finite amplitudes in the $m_{\V} \to 0$ limit, which are just the amplitudes with the associated Goldstone bosons as initial or final states according to the Goldstone boson equivalence theorem. In the limit of small vector masses, one  recovers the axion Lagrangian in Eq.~\eqref{axionFV} case upon the replacement $V_\mu^\prime \to \partial_\mu a /m_V^\prime$ and identifying $\CtVA_{ij} = \CVA_{ij}$. Thus in the limit of small vectors boson masses the amplitudes for axion and vector processes are identical. For more details and the flavor phenomenology of light dark vectors beyond $m_{\V} \to 0$  see Ref.~\cite{Eguren:2024oov}, which  considers both flavor-violating vector~\cite{Heeck:2016xkh, Ibarra:2021xyk} and dipole interactions~\cite{Gabrielli:2016cut, Fabbrichesi:2017vma, Su:2019ipw, Su:2020xwt}.

To summarize our setup, the flavor phenomenology of  light axion is determined by the effective Lagrangians in Eq.~\eqref{axionFV} with the shorthand in notation in Eq.~\eqref{eq:shorthandF}. We will frequently refer also to the flavor-conserving ALP couplings in Eq.~\eqref{axionFC}, for which we use the shorthand  notation $\F_{\alpha} \equiv 2 \Lambda/C_{\alpha}$ for $\alpha =GG, \gamma \gamma, i$. For later reference, the ALP Lagrangian that we are going to discuss reads
\begin{align}
\label{eq:axionfull}
{\cal L}_{{\rm axion}}  & =  
\frac{a}{\F_{GG}} \frac{\alpha_s}{2 \pi} G \tilde{G}+ \frac{a}{\F_{\gamma \gamma}}\frac{\alpha_{\rm em}}{2 \pi} F  \tilde{F}  +  \sum_{i }  \frac{\partial_\mu a}{\F_{i}} \, \overline{f}_i \gamma^\mu  \gamma_5 f_j \nn \\
& +  \sum_{i \ne j} \partial_\mu a  \overline{f}_i \gamma^\mu \left( \frac{1}{\FV_{ij}} +  \frac{\gamma_5}{\FA_{ij}}  \right)f_j \, .
\end{align}
Note that the mere fact that ALPs couple to SM fermions at dimension five signalizes an enormous sensitivity to the associated UV scale. This is because in the SM  flavor-violating decay rates and neutrino interactions are controlled by dimension-six operators below the weak scale. Since axions are light, they can open new channels for flavor-violating decays of SM particles, and since they are dark, they can enable new energy-loss processes in stellar plasmas  beyond neutrino emission. Very roughly, this means that measurements of particle lifetimes and  stellar cooling rates can probe UV scales of order $1/(E G_F)$, where $G_F$ is the Fermi constant and $E$ is the relevant energy scale, given by the particle mass or star temperature. For typical  core temperatures of 10 keV this gives sensitivity of order $10^9 \GeV$, while the  muon lifetime can be used to probe scales of order $10^6 \GeV$.    Existing searches for SM decays with missing energy further improve the reach of flavor physics, reaching scales comparable or even larger than those probed by astrophysics. Besides flavor experiments and astrophysics, also cosmology plays a plays a key  role in probing feebly interacting light particles. 
The Planck mission has demonstrated that there is little space for additional particles besides SM neutrinos that are relativistic at recombination, which can be translated to limits on UV scales of about  $10^8 \GeV$. In  Section 4  we will make these estimates more precise. 

\subsection{Theoretical motivation}
\label{sec:motivation}

The main theoretical motivation for light dark particles is a (pseudo-) Nambu-Goldstone boson that arises from the spontaneous breaking of a global symmetry. In the following we will refer to this broken symmetry as the Peccei-Quinn (PQ) symmetry, borrowing the terminology from the QCD axion~\cite{Peccei:1977hh,Peccei:1977ur}. The PQ symmetry is non-linearly realized as a shift symmetry on the Goldstone boson, which in the symmetry limit forbids any mass terms. A potential is  generated from symmetry breaking effects, either explicitly by e.g. $M_{\rm Pl}$-suppressed quantum gravity effects, or non-perturbatively from chiral anomalies with confining gauge groups. As long as these effects are small compared to the PQ breaking scale, the pNGB acquires a mass that can  naturally be much  smaller than the masses of other UV degrees of freedom (such as the radial mode), since it is protected from quantum corrections  by the shift symmetry. The classical SM example is the pion as the pNGB of the global $SU(2)_A$ symmetry of massless two-flavor QCD, which is spontaneously broken by the quark condensate $\langle \overline{u} u \rangle = \langle   \overline{d} d \rangle = -f_\pi^2 B_0$ and explicitly broken  by the light quark mass terms $m_{u,d}$. These scales  collectively control the pion mass according to the Gell-Mann-Oakes-Renner relation 
\eq{
\label{GMOR}
m_\pi^2 = B_0 (m_u + m_d) \, , } 
which gives a pion mass parametrically  smaller than the other QCD hadrons. The interactions of pions at low energies are  well described by chiral perturbation theory ($\chi$PT), which only uses the relevant symmetries and their breaking patterns. An important component are the Wess-Zumino-Witten terms~\cite{Wess:1971yu, Witten:1983tw}, which describe the coupling of the Goldstone to gauge fields, and arise from the explicit breaking of the PQ symmetry by chiral anomalies.

 In the same spirit we have written down the most general  effective Lagrangian of a light pNGB in Eq.~\eqref{eq:axionfull} (which necessarily includes flavor-violating couplings), 
that arises as the low-energy theory of a more complete theory with a linearly realized PQ symmetry. The couplings in the effective theory are related to the PQ breaking scales and charges as described in Appendix~\ref{app:PQ}, but from a phenomenological point of view they are just free parameters. In a similar way, the effective theory includes a mass term $m_a$ for the pNGB, which is taken as a free parameter, although it may be related to the PQ breaking scale (and thus the pNGB couplings) in some UV-complete theory. This is the case for the QCD axion~\cite{Weinberg:1977ma,Wilczek:1977pj}, where the dominant contribution to explicit PQ breaking is the anomalous coupling to SM gluons, which generates a pNGB potential via non-perturbative QCD effects. Thus by definition, the QCD axion requires $C_{GG} \ne 0$ in Eq.~\eqref{axionFC}, and it is customary to absorb this parameter in the definition of the axion decay constant $f_a \equiv \Lambda/2 C_{GG}$, so that the minimal QCD axion Lagrangian is 
\begin{align}
\label{min}
{\cal L}_{ a} =   \frac{a}{f_a} \frac{\alpha_s}{8 \pi} G \tilde{G} \, , \end{align}
which in particular determines the QCD axion mass. For $f_a \gg \Lambda_{\rm QCD}$, one can accurately calculate the axion potential in chiral perturbation theory~\cite{GrillidiCortona:2015jxo}.  A simple estimate can be obtained from two-flavor $\chi$PT in the isospin limit where $m_u = m_d \equiv m$,  using a chiral field redefinition that eliminates the gluon coupling in Eq.~\eqref{min} and generates derivative couplings and axion-dependent quark masses (see Appendix~\ref{app:axionbasis})
\begin{equation}
{\cal L}_{ a} =   \frac{\partial_\mu a}{4 f_a}  \overline{q} \gamma^\mu \gamma_5 q - \left( m e^{- i a/2f_a} \overline{q}_L q_R + {\rm h.c.} \right)  \, . 
\end{equation}
Plugging in the chiral quark condensate gives the axion potential 
\begin{align}
\label{eq:Va}
V_a & =  - m f_\pi^2 B_0  \left( e^{- i a/2f_a}  + {\rm h.c.} \right)  = -m_\pi^2 f_\pi^2   \cos  \frac{a (x)}{2 f_a}    \, ,
\end{align}
where we used Eq.~\eqref{GMOR}, so that the axion mass is given by $m_a = m_\pi f_\pi/2 f_a $ in the isospin limit. The numerical value of this estimate  differs only by about 10\% from  the result of the state-of-the-art calculation, which reads~\cite{Gorghetto:2018ocs}
\eq{\label{eq:QCDaxionmass} m_a & = 0.5691 (51) \eV \left( \frac{10^7 \GeV}{f_a} \right) \, . } The QCD axion mass is tied  to the solution of the strong CP Problem, since the QCD $\theta$-term can be absorbed into the axion field. Its vacuum expectation value is obtained by  minimizing Eq.~\eqref{eq:Va}, which by construction is the only contribution to the axion potential. This mechanism thus   dynamically enforces $\theta_{\rm eff} = 0$, elegantly explaining the smallness of the neutron EDM, and more generally the absence of CP violation in strong interactions. Moreover, the same potential also provides a natural way of producing QCD axions in the early universe through the misalignment mechanism~\cite{Preskill:1982cy, Abbott:1982af, Dine:1982ah}. At temperatures around the QCD phase transition the  potential in Eq.~\eqref{eq:Va} is active with an effective axion mass close to the $ T =0$ mass in Eq.~\eqref{eq:QCDaxionmass}. At these temperatures the Hubble friction is small compared to the mass, so that the axion field oscillates coherently, and the energy density in these oscillations just behaves as cold DM. In a pre-inflationary scenario this is the only contribution to the axion energy density, so 
apart from an initial condition the observed relic abundance  depends only on $f_a$, and can be reproduced for $f_a \gtrsim 10^{11} \GeV$. Note that for these values the QCD axion is stable on cosmological scales since the lifetime scales as $\tau_a \propto \Lambda_{\rm QCD}^4/m_a^5$, which is a direct consequence of the particular relation of mass and couplings in Eq.~\eqref{eq:QCDaxionmass}.  

Being independently motivated by the Strong CP Problem and DM, the QCD axion is arguably the most popular example for a light dark particle, and a vast amount of theoretical and experimental activity has been devoted to study its phenomenology, see e.g.~Ref.~\cite{DiLuzio:2020wdo} for a review. Many aspects of the QCD axion can be generalized to the (simpler) case where the axion mass is unrelated to $f_a$ and simply treated as a constant free parameter, defining the ``axion-like-particle"~\cite{Jaeckel:2010ni}. This is useful, since many experimental searches for the QCD axion are also sensitive to the ALP, by probing  the  entire $m_a-f_a$ parameter plane~\cite{Irastorza:2018dyq} in which the QCD axion is  just a straight line.  Note that the ALP in general does \emph{not} solve the Strong CP Problem without further model-building,  e.g. introducing new gauge sectors confining at large energies with aligned vacuum angles~\cite{DiLuzio:2020wdo}. Nevertheless, the ALP can still be a perfectly viable DM candidate, which is stable on cosmological scales because of its small mass $m_a$ (typically much below an MeV) and a large UV scale   $\Lambda$ (see Section~\ref{sec:stability}). It can then be produced in the early universe via misalignment similar to the QCD axion, or  from its couplings to SM particles in the thermal bath, provided it is sufficiently heavy to avoid limits on warm dark matter (see Section~\ref{sec:production}).  Theoretical motivations for light ALPs are models with spontaneously broken global continuous symmetries, with small sources of explicit breaking provided by e.g. quantum gravity effects. The existence of such symmetries may in turn be well motivated by other problems or shortcomings of the SM, for example lepton number connected to the origin of neutrino masses (``majoron")~\cite{Chikashige:1980ui, Schechter:1981cv} (see Section~\ref{sec:majoron}), or flavor symmetries that might explain the hierarchical structure of SM Yukawa couplings (``familon")~\cite{Reiss:1982sq, Gelmini:1982zz, Anselm:1977jf, Wilczek:1982rv}. Note that these notions are not mutually exclusive with the possibility of solving the Strong CP problem  (see Section~\ref{sec:axiflavon}), since any high-quality global symmetry that is anomalous under QCD solves the Strong CP Problem.  Light ALPs can also be motivated as a possible explanation of the smallness of the EW scale due to cosmological relaxation (``relaxion")~\cite{Graham:2015cka}
or by string theory compactifications, in which light pseudoscalars with axion-like properties generically arise as Kaluza-Klein zero modes of antisymmetric tensor fields~\cite{Conlon:2006tq, Svrcek:2006yi, Arvanitaki:2009fg, Cicoli:2013ana}. 
 
 \subsection{Flavor-violating couplings}
\label{sec:origin}

We finally elaborate on the motivation of  flavor-violating ALP couplings  in the last line of the reference Lagrangian in Eq.~\eqref{eq:axionfull}. In fact, such couplings are often ignored in the literature, presumably because of a  bias against flavor-changing neutral currents (FCNCs) at tree-level, which are absent in the SM but a common feature of its extensions. Indeed flavor-violating ALP couplings are part of the most general ALP effective Lagrangian, and arise in complete PQ models whenever the PQ charges of SM fermions constitute a new source of flavor violation, i.e. are not diagonal in the same basis as SM Yukawa matrices. According to Eq.~\eqref{eq:CVAbasis} the flavor-violating couplings $\CbVA_{i \ne j}$ are determined by rotating the flavor-diagonal PQ charge matrices ${\bf X}_f$ to the fermion mass basis
\begin{align}
\label{eq:CVAbasis}
\CbVA_f & = - \frac{1}{2N} \left( {\bf U}_{f_R}^\dagger {\bf X}_{f_R} {\bf U}_{f_R}  \pm {\bf U}_{f_L}^\dagger {\bf X}_{f_L} {\bf U}_{f_L} \right) \, , 
\end{align}
where $ {\bf U}_{f}$ are unitary matrices that diagonalize the Yukawa matrices ${\bf Y}_f$ according to $ {\bf U}_{f_L}^\dagger  {\bf Y}_{f}  {\bf U}_{f_R} =  {\bf Y}_{f}^{\rm diag}$. Therefore flavor-violating couplings arise whenever SM fermions carry PQ charges that are not aligned to Yukawas, i.e. when
\begin{align}
[ {\bf X}_{f_L} ,   {\bf Y}_{f}  {\bf Y}_{f}^\dagger]  \ne 0  \qquad   {\rm or} \qquad [ {\bf X}_{f_R} ,  {\bf Y}_{f}^\dagger  {\bf Y}_{f} ] \ne 0 \, .
\end{align}
Usual QCD axion benchmark models are flavor aligned,  either because PQ charges of the SM fermions  vanish, ${\bf X}_{f} = 0$, as in KSVZ models~\cite{Kim:1979if,Shifman:1979if}, or PQ charges are taken to as flavor-universal, ${\bf X}_{f} \propto \mathbb{1}_{3\times3}$, as in  standard DFSZ models~\cite{Dine:1981rt,Zhitnitsky:1980tq}. However, in a generic situation PQ charges are not aligned to SM Yukawas, for example in the broad class of variant DFSZ models with flavor non-universal charges~\cite{Davidson:1984ik, Bardeen:1986yb,Peccei:1986pn, Krauss:1986wx, Geng:1988nc,Hindmarsh:1997ac, Celis:2014iua}. In this case the size of flavor-violating axion couplings  depends on the associated misalignment, i.e. on the unitary matrices ${\bf U}_f$ that diagonalize SM Yukawas. Since only the CKM and PMNS combinations are physical in the SM,  in the absence of a theory of flavor (which would provide ${\bf Y}_{f}$ in a preferred basis) the bulk of these rotations is just described by free parameters. These can be chosen suitably to realize an arbitrary pattern of flavor structures $\CbVA_{f}$, so that one can directly treat these couplings as free parameters on the same ground as the flavor-diagonal effective axion couplings in Eq.~\eqref{eq:axionfull}. Particular flavor patterns in variant DFSZ models  may be motivated in specific scenarios, for example to suppress  axion couplings to nucleons~\cite{DiLuzio:2017ogq,Bjorkeroth:2019jtx, Badziak:2023fsc, Badziak:2024szg} or address stellar cooling anomalies~\cite{Saikawa:2019lng, Badziak:2021apn}. 

Particularly motivated and predictive scenarios emerge when the PQ symmetry is a subgroup of a flavor symmetry that explains SM flavor hierarchies~\cite{Davidson:1981zd, Wilczek:1982rv, Berezhiani:1989fp, Ema:2016ops, Calibbi:2016hwq}. In the simplest realization, the PQ symmetry is identified with a $U(1)_F$ Froggatt-Nielsen symmetry~\cite{Froggatt:1978nt}, which necessarily possesses a QCD anomaly if the effective Yukawas satisfy a holomorphy condition~\cite{Ibanez:1994ig,Binetruy:1994ru, Calibbi:2016hwq}. In this  setup the parametric size of flavor-violating axion couplings is predicted, since both the PQ charges ${\bf X}_f$ as well as the unitary rotations ${\bf U}_f$ are fixed  up to model-dependent ${\cal O}(1)$ coefficients~\cite{Ema:2016ops,Calibbi:2016hwq}. In particular, it determines the most phenomenologically relevant coupling, $\CV_{sd}$ (cf.~Section~\ref{sec:lab}), to be on the order of the Cabibbo angle, $\CV_{sd} \sim V_{us} \sim 0.2$.  The PQ symmetry could also be a proper subgroup of a larger flavor symmetry, see e.g. Refs.~\cite{Berezhiani:1990wn, Berezhiani:1990jj, Berezhiani:1989fp, Ahn:2014gva, Nomura:2016nfi, Bjorkeroth:2017tsz, Ahn:2018nfb, Ahn:2018cau, Linster:2018avp, Carone:2019lfc}, which can result in a significant CKM suppression of light quark transitions, for example $\CV_{sd} \sim V_{td} V_{ts} \sim 10^{-4}$ in the case of $U(2)_F$ models~\cite{Linster:2018avp}. Alternatively, the PQ symmetry could be related to SM flavor hierarchies indirectly, by enforcing texture zeros in the Yukawa matrices~\cite{Davidson:1983fy, Davidson:1983tp, Bjorkeroth:2018ipq}. Flavored PQ symmetries can arise also in the context of Minimal Flavor Violation~\cite{Albrecht:2010xh, Arias-Aragon:2017eww} or as (potentially high-quality) accidental symmetries in models with gauged flavor symmetries~\cite{Babu:1992cu, Cheung:2010hk, Suematsu:2018hbu, Bonnefoy:2019lsn, Babu:2026yqp}. Effective flavor-violating axion couplings are also present in warped 5d models, and are controlled by non-trivial wave-function profiles of bulk fermions chosen to reproduce fermion mass hierarchies~\cite{Bonnefoy:2020llz}.

Finally we note that even in scenarios with flavor alignment, renormalization group (RG) running induces flavor violation proportional to small CKM matrix elements~\cite{Choi:2017gpf,MartinCamalich:2020dfe,Chala:2020wvs,Bauer:2021mvw}. This leads to such a strong suppression  of flavor-violating couplings, that for light ALPs the resulting limits on the UV couplings are typically irrelevant compared to astrophysical limits on  flavor-diagonal couplings, which are generated by the same RG effects without CKM suppression (see also Ref.~\cite{Goudzovski:2022vbt}). For example top-Yukawa enhanced RG effects generate the coupling~\cite{MartinCamalich:2020dfe}
\eq{\Delta \CV_{sd} \sim \frac{y_t^2}{16 \pi^2} V_{td} V_{ts} C_{t} \log \frac{\Lambda_{\rm UV}}{M_Z} \, , } 
but necessarily induce also  flavor-diagonal couplings to electrons and light quarks without CKM suppression~\cite{MartinCamalich:2020dfe}
\eq{\Delta C_{e} \sim \Delta C_{u} \sim \Delta C_{d} \sim \frac{6 y_t^2}{16 \pi^2}  C_{t} \log \frac{\Lambda_{\rm UV}}{M_Z}  \sim 10^{4} \Delta \CV_{sd} \, . }
For axion masses $m_a \lesssim \keV$ astrophysical constraints on $C_{e}$ from Red Giants are weaker than limits on $\CV_{sd}$ only by about two orders of magnitude (cf.~Section~\ref{sec:astro}), so that they  dominate by far the phenomenological constraints on the UV coupling $C_{t}$. Heavier axions with masses $\keV \lesssim m_a \lesssim 100 \MeV$ are  constrained by SN1987A through their couplings to  nucleons of order $\Delta C_N \sim \Delta C_{q}$ at a  level similar to electron couplings (factor 3 weaker),  which renders RG-induced flavor constraints largely irrelevant also for heavy ALPs.

\section{Flavor phenomenology}

The couplings in Eq.~\eqref{eq:axionfull} give rise to two-body decays of SM particles, such as $K \to \pi a $ or $\mu \to e  a$. The associated rates are controlled by dimension-five operators, in contrast to the  dominant decays of kaons and muons in the SM, which are induced by  effective dimension-six operators below the weak scale. This implies that  lifetime measurements alone yield stringent constraints on the axion UV scale, as we discuss in Section~\ref{sec:lifetimes}. Even stronger limits can be obtained from dedicated searches, which in our case involves an invisible ALP (as motivated by dark matter), which would appear as missing energy in a laboratory experiment\footnote{For the flavor phenomenology of ALPs that decay visibly see e.g. Refs.~\cite{Cornella:2019uxs, Escribano:2020wua, Bauer:2021mvw}}. Thus, the experimental signature resembles three-body SM decays with a final state neutrino pair, which can be distinguished using the two-body decay kinematics (Section~\ref{sec:background}). After presenting the expressions of the associated decay rates in Section~\ref{sec:decayrates}, we quantitatively discuss  possible ALP decays into SM particles in Section~\ref{sec:decaylengths}, showing that in typical flavor experiments the ALP is indeed invisible if lighter than about 40 MeV.

\subsection{Lifetimes}
\label{sec:lifetimes}
Before we discuss kinematics and  SM background  in more detail, we make a simple argument that the lifetime alone is sufficient to derive stringent constraints on the inverse couplings $\Lambda_{ij}$ that control  two-body decays. This is because  flavor-violating SM decays are  mediated by the  $W$-boson, and thus the corresponding rates  are strongly suppressed by four powers of the EW scale, scaling as $M^5 G_F^2$, where $M$ is the mass of the decaying particle and $G_F$ is the Fermi constant. No such suppression is present for the two-body decay rate, which scales as $M^3/\Lambda_{ij}^2$, so that merely requiring the two-body decay rate below the total measured rate results in a lower limit on the inverse coupling roughly given by 
\begin{align}
\label{Lambdatau}
\Lambda_{ij} \gtrsim \frac{1}{M G_F} \approx 90 \TeV \left( \frac{\GeV}{M} \right) \, .
\end{align}
In essence, this impressive reach is simply due to the fact that two-body decays are induced by  dimension-five operators, while flavor-violating SM decays arise from dimension-six operators. For the same reason the sensitivity to UV scales is much better than probing corrections to three-body SM decays with neutrinos induced by heavy new particles. These  are also controlled by  dimension-six operators, giving sensitivity to the associated UV scales merely at the order of $1/\sqrt G_F \approx 300 \GeV$ using only lifetime measurements. 
Realistic SM amplitudes are further reduced due to chirality suppression, phase space, loop factors and/or small CKM matrix elements, leading to a significant suppression relative to the two-body decay rate. Consequently, two-body decays  have an enormous sensitivity to the flavor-violating couplings in Eq.~\eqref{eq:axionfull}, probing effective scales of at least $10^6 \GeV$ just using measured lifetimes of SM particles. Note that these limits also apply to unstable light particles as long as they can be produced on-shell. 

As an example one can consider muon decays, with a branching ratio of  the two-body decay given by
\begin{equation}
{\rm BR}_{\mu \to e  a}
  \approx \frac{m_\mu^3/(16 \pi \Lambda_{\mu e}^2)}{m_\mu^5G_F^2/(192 \pi^3)} = \left(\frac{9 \times 10^6 \GeV}{\F_{\mu e}}  \right)^2 \, , 
\end{equation} 
where we have neglected  electron and ALP mass for simplicity. This means that the muon lifetime alone already sets a bound on the inverse coupling of order $10^7$ GeV. Note that the  estimate for the  bound on the UV scale from dimensional analysis  in Eq.~\eqref{Lambdatau} is strengthened by a phase space factor.

Similarly for charged kaons decays, with the main decay  $K^+ \to \mu^+ \nu$ with a branching fraction of  64\%,  one has up to $\one$ factors
\begin{equation}
\label{eq:Klifetime}
{\rm BR}_{K \to \pi  a}
\approx \frac{m_K^3  /(16 \pi (\Lambda^V_{sd})^{2})}{m_{K} m_\mu^2  G_F^2 V_{us}^2 f_{K}^2/(4 \pi)} = \left(\frac{6 \times 10^6 \GeV}{\Lambda^{\rm V}_{s d}}  \right)^2 \, , 
\end{equation} 
where  $f_K \approx 156 \MeV$ is the kaon decay constant, $V_{us} \approx 0.23$ the Cabibbo angle and  we have neglected pion and ALP mass and dropped the hadronic form factor for $K \to \pi$ transitions  $f_+(m_a^2) \approx 1$. Thus the measured charged kaon lifetime sets a bound on the UV scale of similar order as for $\mu \to e$ transitions. Note that here  estimate for the  bound from dimensional analysis  in Eq.~\eqref{Lambdatau} is enhanced by the product of helicity and CKM suppression factors, $m_K/m_\mu \times m_K/f_K \times  1/V_{us} \sim 60$. 

Since the branching ratio of two-body decays with missing energy has to be much smaller than unity, one gains sensitivity to even larger UV scales than from  lifetimes alone, with an enhancement factor given by  the square root of the experimental limit on the two-body branching fraction. This   bound  is mainly  affected by the SM background from three-body decays with neutrinos, which is very different in the quark and lepton sectors. 

Before we discuss the flavor phenomenology of two-body decays in more detail, we briefly compare the above estimates to the limits that can be obtained from neutral meson mixing, which gives the most stringent constraints on dimension-six SMEFT operators induced by heavy new physics~\cite{Bona:2024bue}. The ALP contributes to meson mixing  already at tree-level, giving a contribution to e.g. the neutral kaon mass difference that  can be estimated as (cf.~Refs.~\cite{Gelmini:1982zz, Feng:1997tn})
\begin{align}
\Delta m_K \sim \frac{f_K^2 m_K}{\Lambda_{sd}^2}  = 3 \times 10^{-15} \GeV \left( \frac{2 \times 10^6 \GeV}{\Lambda_{sd}}\right)^2\, .
\end{align}
Requiring this contribution to saturate the observed neutral kaon mass difference $\Delta m_K^{\rm exp} \equiv m_{K_L} - m_{K_S} \approx  3.5 \times 10^{-15} \GeV$~\cite{ParticleDataGroup:2024cfk} thus gives a limit on the inverse coupling $\Lambda_{sd}$ that is weaker than the limit from the kaon lifetime in Eq.~\eqref{eq:Klifetime}. This is welcome, because the above estimate is unreliable because of non-perturbative uncertainties, which nevertheless can be expected to not change the result by orders of magnitudes. Moreover, in contrast to two-body meson decays neutral meson mixing is also sensitive to contributions of \emph{heavy} degrees of freedom associated with the UV scale $\Lambda_{ij}$, which are parametrically of the same order as the ALP contribution~\cite{MartinCamalich:2020dfe}. For example, the radial mode in PQ models has a mass of order $\Lambda$ and flavor-violating coupling set by $C_{ij}$, giving the same parametric suppression of $\Lambda_{ij}^2$ for  mixing amplitudes as the tree-level ALP exchange. Nevertheless it is important to calculate the ALP contribution to neutral meson mixing as accurately as possible, which has been addressed in Ref.~\cite{MartinCamalich:2020dfe} employing $\chi$PT for kaon mixing, and an operator product expansion in the heavy quark mass for $B$ and $D$ mesons. The resulting limits are much weaker than those that can be obtained from two-body decays, with the only exception for charm mixing. Since also this sector is subject to large non-perturbative uncertainties, two-body decays are much more promising as an experimental probe of light flavor-violating axions than neutral meson mixing.

\subsection{Two-body decay rates}
\label{sec:decayrates}
We now discuss in detail the flavor-changing two-body decay rates, which result from the ALP couplings in the effective Lagrangian in Eq.~\eqref{eq:axionfull}. We  distinguish  decays of  hadrons and leptons, as the former  involve hadronic matrix elements that describe the non-perturbative physics of  quark-hadron duality. These matrix elements are parametrized in terms of  a few form factors that are functions of the momentum transfer $q^2$, which here is just the squared mass of the axion.

The hadrons which are  most accessible experimentally are pseudoscalar mesons like $K^+, D^+, B^0$, which can decay to lighter mesons and a dark boson. Since strong interactions conserve parity, it is clear that hadronic matrix elements vanish identically between parity eigenstates and currents that are $P$-odd, for example pseudoscalar currents between two pseudoscalar mesons. Therefore, such decays will only be sensitive to ALP couplings $\CV_{ij}$, but not $\CA_{ij}$. In order to probe these latter couplings, one either has to consider pseudoscalar decays to vector mesons, or  decays of baryons which are not parity eigenstates~\cite{MartinCamalich:2020dfe}. In Appendix~\ref{app:2bodydecays} we provide the complete two-body rates for all relevant processes, including also polarized baryon decays, as a function of the couplings in Eq.~\eqref{axionFV} and the relevant form factors that parametrize the  hadronic matrix elements as defined in Appendix~\ref{app:FFs}. Below we   focus below for simplicity on the limit  $m_a = 0$,  referring to  Appendix~\ref{app:2bodydecays} for the complete expressions. Here we also neglect the mass of the visible final state particle, in order to make the expressions concise. 

The  rates for decays  of pseudoscalars $P$ to pseudoscalar  $P^\prime$ or vector meson $\mathcal{V} $, and baryonic decays $B \to B^\prime$ for an underlying flavor-changing $q \to q^\prime$ transition with final state ALP $a$ read
\begin{align}
\label{eq:rates}
\Gamma(P \to P^\prime a) &\approx
\frac{ m_P^3}{64\pi \Lambda^2} 
 |f_+ |^2| \CV_{ q^\prime q } |^2 \,, \nn \\
\Gamma (P \to \mathcal{V}  a) &\approx \frac{m_P^3}{64\pi \Lambda^2}
|A_0|^2 |\CA_{ q^\prime q }|^2   \, ,  \\
\Gamma (B \to B^\prime  a) &\approx
\frac{ m_B^3}{64\pi \Lambda^2}
\left[|f_1|^2|\CV_{ q^\prime q}|^2+|g_1|^2|\CA_{q^\prime q}|^2\right]\, , \nn
\end{align}
where the form factors $f_+^{P P^\prime} (q^2)$, $A_0^{P  \mathcal{V} }(q^2)$, $f_1^{B B^\prime}(q^2)$, $g_1^{B B^\prime}(q^2)$ depend on the  hadron transition and are evaluated at nearly vanishing momentum transfer $q^2 \approx 0$. For convenience of the reader we collect their numerical values (which are ${\cal O}(1)$ numbers) for the relevant hadronic transitions in Table~\ref{tab:FFtable}  of Appendix~\ref{app:FFs}. 

Turning to leptons, the total rate for $\ell \to \ell^\prime$ decays in the limit of small axion  masses is given by
\begin{equation}
\label{leptonunpol}
\Gamma (\ell \to \ell^\prime  a)=
\frac{ m_\ell^3}{64\pi \Lambda^2}
\left(\left|\CV_{\ell^\prime\ell} \right|^2  + \left|\CA_{\ell^\prime\ell}\right|^2 \right) \, ,
\end{equation}
where again we have neglected $m_\ell^\prime/m_\ell $ corrections. As discussed above, a useful experimental handle  to suppress SM background is  provided by the angular distribution of final lepton momenta  for  decays of polarized leptons $\ell$. While the full expressions are given in Appendix~\ref{app:2bodydecays},  the  differential two-body rates for massless axions are given by
\begin{equation}
\label{leptonpol}
\frac{d\Gamma_{\ell \to \ell^\prime  a} }{d\cos\theta} =
\frac{ m_\ell^3}{128\pi \Lambda^2}
\left(\left|\CV_{\ell^\prime\ell} \right|^2  + \left|\CA_{\ell^\prime\ell}\right|^2 +2 P_{\ell^-} \cos\theta \, {\rm Re}( \CV_{\ell^\prime\ell} \CAs_{\ell^\prime\ell})\right)\, .
\end{equation}
Here $\theta$ denotes the angle between the spin direction of the decaying lepton $\vec{s}_\ell$ with polarization fraction $P_{\ell^-}$ and the direction of the final state lepton $\vec{p}_{\ell^\prime}$. For polarized anti-leptons the same expressions hold with $P_{\ell^-} \to -P_{\ell^+}$. For unpolarized leptons $P_{\ell^-} = 0$, or either $\CV_{\ell^\prime\ell}$ or $\CA_{\ell^\prime\ell}$  vanishing, the decay is isotropic, and we recover the total rate in Eq.~\eqref{leptonunpol} after integrating over the polarization angle $\theta$. For non-zero polarization instead the angular distribution depends on the relative size of chiral couplings
\begin{align}
\frac{d\Gamma (\ell \to \ell^\prime  X) }{d\cos\theta} \propto 1 +  P_{\ell^-} \cos\theta\cdot\frac{ \CV_{\ell^\prime\ell} \CAs_{\ell^\prime\ell} + C^{{\rm V*}}_{\ell^\prime\ell} \CA_{\ell^\prime\ell} }{\left| \CV_{\ell^\prime\ell} \right|^2  + \left|\CA_{\ell^\prime\ell}\right|^2 } \, , 
\end{align}
to be compared with the  SM decay rate close to the endpoint, i.e. with a final state lepton energy close to the maximal value $E_{\ell^\prime} = m_\ell/2$, which reads (cf.~Eq.~\eqref{eq:Michel})
\begin{align}
\frac{d \Gamma(\ell \to \ell^\prime  \overline \nu_{\ell^\prime}  {\nu}_\ell)}{d \cos\theta} \propto 1- P_{\ell^-} \cos\theta  \, .
\end{align}
In general the experimental limits will strongly depend on whether the chiral structure of the dark boson couplings is aligned to the SM  (i.e. left-handed couplings with ${\CV_{ij} = - \CA_{ij}} $) or not.

\subsection{Kinematics and SM background}
\label{sec:background}
In the two-body decay the visible particle is monochromatic, with an energy  in the rest frame of the decaying particle given by
\begin{equation}
E_{\rm vis}  = \frac{M}{2} \left( 1 + \frac{m_{\rm vis}^2- m_X^2}{ M^2} \right) \, , 
\end{equation}
where $M$ ($m_X$) is the mass of the decaying (invisible) particle and $m_{\rm vis}$ the mass of the visible particle in the final state. If the invisible particle can be treated as massless, i.e., has  a mass below the experimental resolution, the energy of the visible particle only depends on the decay channel and is approximately given by half of the mother particle. The finite energy resolution in a realistic experiment will result in a smearing and produce a narrow peak around $E_{\rm vis}$. This constitutes a clear experimental signal, that can be well distinguished from the continuous energy spectrum of the visible particle in the  3-body SM decay with neutrinos, which  peaks precisely at the 2-body decay energy for $m_X = 0$, since the electron momentum is maximized when recoiling against collinear neutrinos.  

Without additional handles, one may expect that experimental limits on two-body decays with missing energy can be set at most of the order of the di-neutrino decays, and we summarize in Table~\ref{di-neutrino} the SM prediction and current experimental limits for various relevant meson and baryon decays. Note that in $K \to \pi$ and $B \to K$ transitions the di-neutrino decay has been measured, with a value slightly above the SM prediction, although not statistically significant. In other channels only upper limits exist, which are of the order of $10^{-5}$ in $B$-decays and $10^{-4}$ in $D$-decays, while there are no experimental limits on di-neutrino baryon decays (apart from  total lifetime).  Because of the highly suppressed SM background, $ b \to d$ and $c \to u$ flavor transitions are therefore very promising channels to search for light dark particles. 
\begin{table}[h]
\renewcommand{\arraystretch}{1.6}
  \setlength{\arrayrulewidth}{.35mm}
\centering
\setlength{\tabcolsep}{1.3 mm}
\centering
\begin{tabular}{|c||c|c|}
\hline
Decay channel & SM prediction & Experiment \\
\hline
${\rm BR} (K^+ \to \pi^+ \nu \overline{\nu})$ & $(8.60 \pm 0.42) \times 10^{-11}$~\cite{Buras:2022wpw}  & $(1.14^{+0.40}_{-0.33}) \times 10^{-10}$~\cite{ParticleDataGroup:2024cfk}  \\
${\rm BR} (B^+ \to K^+ \nu \overline{\nu})$ &  $(5.58 \pm 0.37 ) \times 10^{-6}$~\cite{Parrott:2022zte}  &  $(2.3 \pm 0.7) \times 10^{-5}$~\cite{Belle-II:2023esi}   \\
${\rm BR} (B^+ \to K^{*+} \nu \overline{\nu})$ &  $ (8.93 \pm 1.07) \times 10^{-6}$~\cite{Bause:2021cna}& $< 4.0 \times 10^{-5}$~\cite{Belle:2013tnz} 
 \\ 
${\rm BR} (B^+ \to \pi^+ \nu \overline{\nu})$ & $ (1.2 \pm 0.1) \times 10^{-7}$~\cite{Bause:2021cna} & $< 1.4 \times 10^{-5}$~\cite{Belle:2017oht} 
 \\
${\rm BR} (B^+ \to \rho^+ \nu \overline{\nu})$ & $ (4.8 \pm 1.8) \times 10^{-7}$~\cite{Bause:2021cna} & $< 3.0 \times 10^{-5}$~\cite{Belle:2017oht}  \\
${\rm BR} (D^0 \to \pi^0 \nu \overline{\nu})$ & $\simeq 5 \times 10^{-16}$~\cite{Burdman:2001tf} & $< 2.1 \times 10^{-4}$~\cite{BESIII:2021slf}  \\
${\rm BR} (\Lambda \to n \nu \overline{\nu})$ & $ 7.1 \times 10^{-13}$~\cite{Tandean:2019tkm} &   \\
${\rm BR} (\Sigma^+ \to p \nu \overline{\nu})$ & $ 4.3 \times 10^{-13}$~\cite{Tandean:2019tkm} &  \\
\hline
\end{tabular}
\caption{Selected SM predictions and experimental limits for di-neutrino decay channels. The last column is either a measurement or denotes the upper 90\% CL limit. \label{di-neutrino}}
\end{table}

Going beyond the irreducible three-body SM background requires an extremely good understanding of the di-neutrino event shape, or other experimental techniques in order to enhance the signal over background ratio. One such possibility is to use  the angular information of the visible particle in decays of polarized decaying particles, either leptons or baryons. This is particular relevant for muon decays, as the SM background is huge, and available  beams of (anti-)muons at TRIUMF or PSI are usually polarized to high degree\footnote{Since  produced from pion decays,  momentum conservation dictates that muons (anti-muons) are polarized along the positive (negative) direction of their momenta.}.  In the following we briefly sketch how this feature can be used to distinguish signal from background, referring to more details to Ref.~\cite{Calibbi:2020jvd}.

 In the SM  anti-muons decay as $\mu^+\to e^+ \nu_e \overline{\nu}_\mu$, with a differential rate described by the Michel spectrum~\cite{Scheck:1977yg,Kuno:1999jp}  
\begin{equation}
\label{eq:Michel}
\frac{d^2\Gamma_{\mu^+\to e^+ \nu_e \overline{\nu}_\mu}}{dx_e\, d \cos\theta_e} \approx \Gamma_{\mu}\big[\left(3-2x_e\right) + P_{\mu^+}(2x_e-1)\cos\theta_e\big]x_e^2 \, , 
\end{equation}
where $ \Gamma_{\mu} \approx {m_\mu^5G_F^2}/{(192\pi^3)}=3\times 10^{-10}\text{ eV}$ is the total muon decay width and we have neglected the electron mass. Here $P_{\mu^+}$ is the $\mu^+$  polarization along a given axis (usually the beam direction) and $\theta_e$ denotes the angle between this  axis  and the positron momentum in the muon rest frame. In the limit of vanishing electron mass, the positron energy fraction $x_e=2E_e/m_\mu$ takes the values $0\leq x_e\leq 1$. 

For an unpolarized  beam, $ P_{\mu^+} =0$,  the SM background in Eq.~\eqref{eq:Michel} peaks at $E_{e}^{\text{max}}=m_\mu/2$ corresponding to $x_e=1$. Therefore most  positrons from SM decays of unpolarized muons have a momentum very close to the monochromatic positrons produced in the two-body $\mu^+\to e^+  X$ decay for a massless dark boson. However, for fully polarized anti-muons ($ P_{\mu^+} = - 1$) the decay rate for positrons with maximal energy ($x_e = 1$)  emitted along the  beam direction ($\cos \theta_e=  1$) vanishes. In a realistic experimental setting one has to take into account finite resolution in the positron energy and angle, as well as non-perfect muon polarization, which  makes the SM background nonzero. Still, for suitable cuts one can optimize the signal to background ratio, using  as discriminants the positron momentum and its angle. This strategy has been employed in the late 80's to constrain two-body branching ratios at the level of $10^{-6}$~\cite{Jodidio:1986mz}, which remains the best  limit to date. Improving this bound may be feasible by e.g. adding 
a dedicated calorimeter in the MEG forward region~\cite{Calibbi:2020jvd}, and in general requires an extremely accurate control of theoretical uncertainties due to the irreducible SM background, which has been addressed in Ref.~\cite{Banerjee:2022nbr}.

\subsection{Visible decay lengths}
\label{sec:decaylengths}
While total lifetime limits apply for any light particle produced on-shell,  flavor constraints on  missing energy  require that the ALP is  either long-lived  or decays to invisible particles (e.g. dark fermions). This implies that possible decay channels to visible particles like photons or electrons need to be sufficiently suppressed, in order to have decay lengths much larger than the typical scale of collider experiments, very roughly of the order  of meters. In this section we therefore consider  the relevant flavor-conserving  ALP couplings in Eq.~\eqref{axionFC}, and quantify  how small they need to be in order to have missing energy signals. As we are going to see, existing experimental constraints on these couplings imply that that ALPs lighter than about 40 MeV are  always invisible in collider experiments.

\subsubsection*{Decays to electrons}
We first consider ALP couplings to electrons defined as
\begin{equation}
\label{eaxion}
{\cal L}_{a}  \supset  C_e  \frac{\partial_\mu a}{2 \Lambda} \, \overline{e} \gamma^\mu \gamma_5 e  \, ,
\end{equation}
with $C_e$ real, which sets the  decay rate to electrons as 
\eq{
\label{ratescalartoee}
\Gamma (a \to e^+ e^-) & = m_a \frac{C_e^2}{8\pi} \frac{m_e^2}{\Lambda^2} \sqrt{1- \frac{4 m_e^2}{m_a^2}} \, ,
}
For ALPs significantly heavier than the decay threshold, one can neglect the phase space suppression and the electron decay length reads
\begin{align}
c \tau_{a \to e^+e^-}\approx 20 \, {\rm cm} \left( \frac{1 \MeV}{m_a} \right)  \left( \frac{\Lambda/C_e}{100 \GeV} \right)^2 \, .
\end{align}

\subsubsection*{Decays to photons}

For the most general ALP couplings at the EW scale, the di-photon decay rate of a light spinless particle with mass $m_a \ll \GeV$ has been calculated at one-loop in Ref.~\cite{Feruglio:2025xvc}, and in Appendix~\ref{app:diphoton}  we collect the expressions for a CP-conserving ALP. Focussing on a light axion with $m_a \ll m_\mu$, the decay rate reads
\begin{align}
\Gamma(a \to \gamma \gamma) & = \frac{\alpha^2}{64\pi^3}\,\frac{m_a^3}{\Lambda^2} |C^{\rm eff}_\gamma |^2 \,, 
\end{align}
with, neglecting terms of order $m_a^2/m_{\mu}^2$
\begin{align}
\Cga^{\rm eff} & \approx
C_{\gamma \gamma} - 1.96 \, C_{GG}  + C_e J_e (m_a)  \,.
\end{align}
where $J_e (m_a)$ is a  (complex) loop function defined in Eq.~\eqref{eq:Jedef} and $C_{\gamma \gamma}$ and $C_{GG}$ are the ALP couplings to photons and gluons at the electroweak scale in Eq.~\eqref{axionFC}. The photon decay length is given by 
\begin{align}
c \tau_{a \to \gamma \gamma}\approx 7 \times 10^4 \, {\rm m} \left( \frac{1 \MeV}{m_a} \right)^3  \left( \frac{\Lambda/|C^{\rm eff}_\gamma|}{100 \GeV} \right)^2 \, .
\end{align}

\subsubsection*{ALP decay lengths}

For a given experiment with typical  fiducial detector volume of size $L_{\rm exp}$, the fraction of ALPs  that decay within the detector if produced at the center is given by
\begin{align}
P_{\rm in} = 1 - \exp \left( -\frac{m_a \Gamma_a^{\rm vis} L_{\rm exp} }{p_a} \right) \, ,
\end{align}
where $ \Gamma_a^{\rm vis}$ is total ALP decay width into visible particles. Apart from $ L_{\rm exp}$, this probability depends on its momentum $p_a$, which for 2-body decays from a mother particle at rest in the lab frame is  given by
\begin{align}
p_a & = \frac{M}{2} \sqrt{1-2 \frac{m_a^2+m^2}{M^2}+\frac{(m_a^2 - m^2)^2}{M^4}}  \, , 
\end{align}
where $M$ is the mass of the mother particle and $m$ the mass of the other daughter particle. The rest frame is a good approximation for low-energy  experiments where particles produced with small momenta,  while in the ultra-relativistic regime the ALP essentially inherits the energy of the mother particle. 

If the fraction of visible decays in the detector is small, $P_{\rm in}  \ll 1$,  it can be approximated by
\begin{align}
P_{\rm in} \approx \frac{m_a \Gamma_a^{\rm vis} L_{\rm exp} }{p_a} \, ,
\end{align}
which gives the following condition on the visible decay length $\ell^{\rm vis}_a$ in the laboratory frame
\begin{align}
 \ell^{\rm vis}_a  \equiv \frac{p_a}{m_a}  \tau_a^{\rm vis} = \beta_a \gamma_a c \tau_a^{\rm vis} \gg L_{\rm exp} \, .
\end{align}
This expression is the quantitative notion of a ``dark" particle, which  depends on the experimental setup.

\begin{figure}
\centering
  \includegraphics[width=0.49 \textwidth]{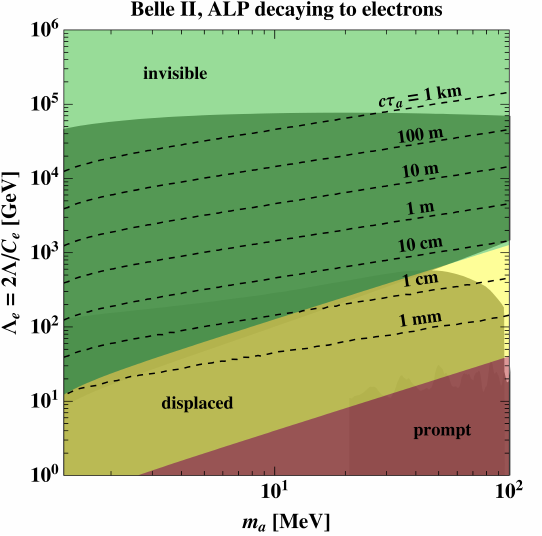}
   \includegraphics[width=0.49 \textwidth]{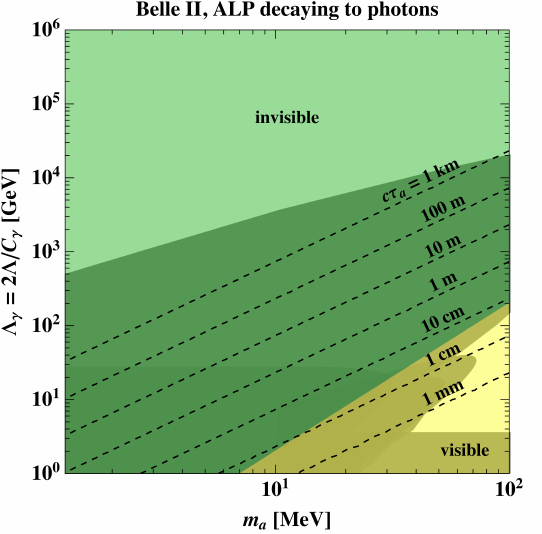} 
   
   \vspace{0.2cm}
     \includegraphics[width=0.49 \textwidth]{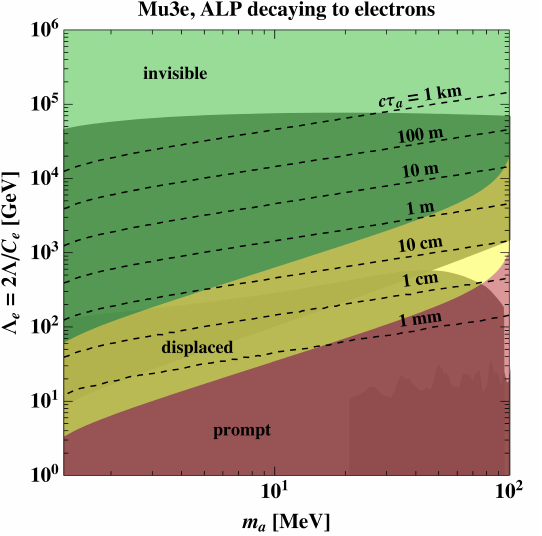}
   \includegraphics[width=0.49 \textwidth]{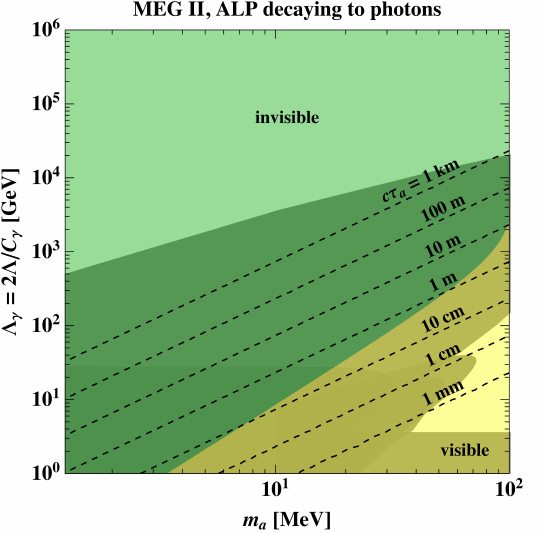}
   \caption{Parameter space for effective electron (left panel) and photon couplings (right panel) with dashed contour lines denoting proper decay lengths. Dark shaded regions indicate excluded regions by laboratory experiments, see text for details.  In the left panel we  show in color the regions where the  ALP decay to $e^+ e^-$ is prompt (red), displaced (yellow) or invisible (green) at Belle~II  and Mu3e, while in the red panel we only distinguish visible (yellow) or invisible (green) di-photon decays at Belle~II  and MEG~II. }
  \label{fig:decaylength}
\end{figure}

In the following we discuss Belle~II, Mu3e and MEG~II as representative examples, which give relevant constraints on $B$ meson and muon decays with light axions, respectively. At Belle~II $B$ mesons are produced almost at rest, so that ALPs from $B$-decays have momentum $p_a^{\rm Belle II} \approx m_B/2 = 2.6 \GeV$. Similar at Mu3e and MEG-II ALPs are produced from muons at rest, so $p_a^{\rm Mu3e} = p_a^{\rm MEG} \approx m_\mu/2 = 53 \MeV$, where we have neglected the ALP mass. For Mu3e an ALP produced from muon decay appears ``invisible" if it decays to charged particles outside the instrumented region, which we approximate very roughly as a sphere with radius $L_{\rm invis}^{\rm Mu3e} = 1 {\rm \, m}$~\cite{Mu3e:2020gyw}. MEG~II  is also sensitive to ALPs decaying into photons, provided  the decay occurs within the fiducial detector volume,  again approximated as a sphere  with radius of $L_{\rm invis}^{\rm MEG} = 0.7 {\rm \, m}$~\cite{MEG:2020zxk}. Belle~II  is sensitive to both decay modes unless they happen outside of the detector, with a size of  roughly $L_{\rm invis}^{\rm Belle II} = 2 {\rm \, m}$~\cite{Belle-II:2018jsg}. For ALPs decaying to electrons one can further distinguish ``prompt" and ``displaced" decays, although this does not mean that the displaced vertex can always be reconstructed (see Ref.~\cite{Heeck:2017xmg} for a study of displaced LFV decays). Electrons from ALP decays with decay lenghts below $ L_{\rm invis}$ leave charged tracks that can be used to reconstruct displaced vertices if the decay is not classified as prompt, which corresponds to  decay lengths roughly above  $L_{\rm prompt} = 3 {\rm \, mm}$ in Mu3e ~\cite{Knapen:2023iwg} and $L_{\rm prompt} \approx  2 {\rm \, mm}$ at Belle~II ~\cite{Belle-II:2023ueh}. Photons from ALP decays can typically not be used to reconstruct displaced ALP decay vertices, so we refer to ALPs decaying into photons with  decay lenghts below $ L_{\rm invis}$ decay simply as ``visible".

In Fig.~\ref{fig:decaylength} we show the parameter regions for which the decay into an electron-positron pair (left panel) is prompt (red), displaced (yellow) or invisible (green) at Belle~II and Mu3e, while for photons (right panel) we only distinguish between visible (yellow) or invisible (green) decays at Belle~II and MEG~II,. The dashed lines are the contour lines for the indicated proper decay lengths and we also show as dark shaded regions the parameter space excluded by laboratory experiments\footnote{There are also astrophysical limits extending to smaller couplings, which however are not relevant for the present discussion.}.  For limits on photon couplings we use the collection of limits provided in the GIT repository~\cite{AxionLimits}, which in the relevant parameter space are set by LEP-II~\cite{Knapen:2016moh}, FASER~\cite{FASER:2024bbl} and BaBar and various beam dump experiments taken from Ref.~\cite{Dolan:2017osp}. For electron couplings instead we use the limits collected in Ref.~\cite{Eberhart:2025lyu}, which for the relevant couplings and masses are limits set on dark photons by BaBar~\cite{BaBar:2014zli}, which have been re-interpreted for ALPs in Ref.~\cite{Eberhart:2025lyu},  limits on pion decays such as $\pi^+ \to e^+ \nu a$ by SINDRUM~\cite{SINDRUM:1986klz} (as pointed out in Ref.~\cite{Altmannshofer:2022ckw}) and the re-calculated E137 beam-dump limits~\cite{Bjorken:1988as}.

As can be seen from Fig.~\ref{fig:decaylength}, ALPs lighter than 40 MeV are always invisible, with decay lengths above km. Heavier ALPs between 40 and 100 MeV can in principle decay visibly, which however requires very low effective scales, $\F_e \equiv 2\Lambda/C_e< \TeV$ for electrons and $\F_\gamma  \equiv 2\Lambda/|C_\gamma^{\rm eff}| < 100 \GeV$ for photons. While  direct limits on these ALP couplings from perturbative unitarity are rather weak (see e.g. Ref.~\cite{Cornella:2019uxs}), realistic models with e.g. universal couplings to fermions are more severely constrained, as  perturbative unitarity on the top coupling give limits $\F_e = \F_t \gtrsim 112 \GeV$ and $\F_\gamma \gtrsim 100\GeV$ for $5 \MeV \le m_a \le 100 \MeV$, see Appendix~\ref{app:diphoton}. Therefore ALPs below 100 MeV can be expected to be invisible in collider experiments, unless the associated UV scales are extremely low, which requires elaborate model-building since these are potentially in conflict with perturbative unitarity. This justifies the focus on light dark  ALPs for flavor phenomenology.

\section{Model-independent limits}
\label{limits}

In this section we discuss general limits on flavor-violating  couplings of a light invisible ALP that can be obtained from laboratory searches (Section~\ref{sec:lab}), astrophysics (Section~\ref{sec:astro}) and cosmology  (Section~\ref{sec:cosmolimits}). The resulting constraints on the effective UV scales are summarized in Section~\ref{sec:summarylimits}, see Fig.~\ref{fig:limits}.

\subsection{Limits from  laboratory searches}
\label{sec:lab}

As discussed in the previous section, stringent limits on  effective flavor-changing couplings of light invisible axions can be derived  from  meson and lepton lifetimes, which can be tighten further using dedicated searches for two-body decays with missing energy. While the theoretical calculations of the tree-level decay rates in Section~\ref{sec:decayrates} are very straightforward, the  challenge remains on the experimental side. At present, many relevant searches for two-body decays with missing energy have not been carried out by the experimental collaborations, and limited information often prevents  the re-interpretation of  available data on SM di-neutrino decays  for monochromatic visible particles.

In the following, we discuss the experimental status in more detail, considering separately the couplings to quarks and to leptons. This separation is appropriate, since experimental searches face vastly different SM backgrounds and probe distinct chiral structures of ALP couplings. In the quark sector the most important limits come from pseudoscalar meson decays, which probe either parity even or odd quark currents, while in the lepton sector total decay rates do not depend on chirality. However, at least in the case of muons, one can rely on polarized decays, which gives sensitivity to different chiral couplings via the angular distribution of  final state leptons. As discussed in Section~\ref{sec:background}, this can give an important handle in order to suppressed SM background, which in contrast to the quark sector is huge. 

We summarize the current limits on two-body  meson, baryon and lepton decays with  a massless, invisible final-state particle $X$ in Table~\ref{tab:exp}, focusing on the process that gives the best constraint on the respective flavor-violating coupling presented in  Table~\ref{tab:lablimits}. Below we discuss these numbers and the resulting limits on  flavor-violating couplings in detail. Note that these constraints also apply when the ALP  but promptly decays into stable invisible particles, for example dark fermions or neutrinos.

\begin{table}[h!]
\renewcommand{\arraystretch}{1.6}
  \setlength{\arrayrulewidth}{.35mm}
\centering
\setlength{\tabcolsep}{5.3 mm}
\begin{center}
\begin{tabular}{@{}|c||c r|c|@{}}
\hline
Decay & 90\% CL Limit && Experiment  \\
\hline
${\rm BR} (K \to\pi X)$ &  $2.8 \times 10^{-11}$ &&NA62~\cite{NA62:2025upx}   \\
${\rm BR} (\Sigma  \to p X)$ & $3.2 \times 10^{-5}$ & &BESIII~\cite{BESIII:2023utd}  \\
${\rm BR} (D  \to \pi X)$ & $8.0 \times 10^{-6}$ & \cite{MartinCamalich:2020dfe}$_{\rm r}$ & CLEO~\cite{CLEO:2008ffk}  \\
${\rm BR} (D \to \omega X)$ & $1.1 \times 10^{-5}$& &BESIII~\cite{BESIII:2024rkp}   \\
${\rm BR} (\Lambda_c  \to p X)$ & $8.0 \times 10^{-5}$& &BESIII~\cite{BESIII:2022vrr}   \\
${\rm BR} (B  \to K X)$ &$3.4 \times 10^{-6}$ & &  Belle~\cite{Belle:2026kum}  \\
${\rm BR} (B  \to K^* X)$ &$4.2 \times 10^{-5}$ & \cite{Altmannshofer:2023hkn}$_{\rm r}$  & BaBar~\cite{BaBar:2013npw}    \\
${\rm BR} (B  \to \pi X)$ & $2.3 \times 10^{-5}$ &\cite{MartinCamalich:2020dfe}$_{\rm r}$ & BaBar~\cite{BaBar:2004xlo}  \\
${\rm BR} (B  \to \rho X)$ & $3.9 \times 10^{-4}$ & \cite{Alonso-Alvarez:2023mgc}$_{\rm r}$ & LEP~\cite{ALEPH:2000vvi}   \\
\hline
${\rm BR} (\mu \to e X)_{\rm iso}$ &  $2.6 \times 10^{-6}$ & &TRIUMF~\cite{Jodidio:1986mz} \\
${\rm BR} (\mu \to e X)_{\rm R}$ &  $2.5 \times 10^{-6}$ &\cite{Calibbi:2020jvd}$_{\rm r}$ & TRIUMF~\cite{Jodidio:1986mz}  \\
${\rm BR} (\mu \to e X)_{\rm L}$ &  $5.8 \times 10^{-5}$ &&TWIST~\cite{TWIST:2014ymv}    \\
${\rm BR} (\tau  \to \mu X)$ & $2.8 \times 10^{-4}$ & &  Belle II~\cite{Belle:2025bpu}  \\
${\rm BR} (\tau  \to e X)$ & $4.5 \times 10^{-4}$ & &  Belle II~\cite{Belle:2025bpu}  \\
\hline
\end{tabular}
\caption{Laboratory limits on  branching ratios of two-body meson, baryon and lepton decays with  a massless, invisible final-state particle $X$. When the limit was obtained by recasting public data (and no bound was published by the experimental collaboration), we indicate the  reference where the recast has been performed by   []$_{\rm r}$. For  polarized muon decays the limit depends on the  angular distribution of  electron momenta, denoted as ``${\rm iso}$" for isotropic decays, ``R" for $\propto 1 +  \cos \theta$ and ``L" for $\propto 1 -  \cos \theta$, see text for details. These bounds are valid for any dark boson  with mass below the experimental resolution and sufficiently long-lived to appear as missing energy.}
\label{tab:exp}
\end{center}
\end{table}

\subsubsection{Quark flavor violation}
\label{CL:quarks}
In the quark sector the most stringent limits typically arise from  searches for two-body decays of pseudoscalar mesons and baryons,  with associated decay rates scale in terms of  dark ALP couplings given in Appendix~\ref{app:2bodydecays}. In the following we focus on limits  for a massless invisible particle $X$ with rates   summarized in Section~\ref{sec:decayrates}, using the shorthand notation for effective ALP couplings in Eq.~\eqref{eq:shorthandF}. 

For $s \to d$ transitions, a limit on two-body kaon decays of ${\rm BR} (K^+ \to \pi^+ X) < 2.9 \times 10^{-11}$   has been set  by the NA62 collaboration in Ref.~\cite{NA62:2025upx}, superseding the old BNL result  ${\rm BR} (K^+ \to \pi^+  X) < 7.3 \times 10^{-11}$~\cite{E949:2007xyy} (for massless $X$). This bound can be used to set a  lower limit\footnote{In an updated version (v2) of Ref.~\cite{Guadagnoli:2025xnt}, the authors included publicly available NA62 data recorded between 2023 and 2024 to set a combined limit on ${\rm BR} (K^+ \to \pi^+  X)$  that is stronger than the official NA62 result (based on the dataset 2016 - 2022) by factor 2, which would improve the lower limit on $\FV_{sd}$ by $\sqrt 2$.}   on the inverse effective coupling $\FV_{sd}$  of $1.1 \times 10^{12} \GeV$, while the best limit on the inverse axial coupling, $\FA_{sd}  > 2.8 \times 10^{7} \GeV$, arises from hyperon decays $\Sigma^+ \to p  X$ limited by ${\rm BR} (\Sigma^+ \to p + {\rm invis.}) < 3.2 \times 10^{-5}$ as reported by  the BESIII collaboration~\cite{BESIII:2023utd}. Note that BESIII also performed  searches for $\Xi \to \Lambda  X$, finding a limit of ${\rm BR} (\Xi \to \Lambda  X) < 2.3 \times 10^{-4}$~\cite{BESIII:2026ztz}, which however gives a weaker bound on axion couplings than the searches discussed above. 

We note that three-body decays can be used to obtain slightly stronger limits on the axial coupling $\FA_{sd}$, while in general they are irrelevant compared to  two-body decays. For example, LHCb constraints on $B_{(s)} \to \mu^+ \mu^-  X$ cannot compete with limits on two-body decays $B \to K/K^*/\pi/\rho + X$ as discussed in Ref.~\cite{Albrecht:2019zul}. The only  relevant channel are kaon decays $K\to\pi   \pi  X$ controlled by $\FA_{sd}$, since two-body decays to light vector mesons are kinematically forbidden. Experimentally,  both $K^+\to\pi^0\pi^+ X$~\cite{Adler:2000ic,Sadovsky:2023cxu} and $K_L\to\pi^0\pi^0 X$~\cite{E391a:2011aa} have been searched for. Since the latter search  did not cover the range $m_X\leq50$ MeV, only the former decays can be used to obtain a limit on $\FA_{sd}$ for light $X$. Using the three-body decay rate calculated in Ref.~\cite{MartinCamalich:2020dfe}, the OKA upper limit on  ${\rm BR} (K^+\to\pi^0  \pi^+  X) < 2.5 \times 10^{-6}$~\cite{Sadovsky:2023cxu} translates into the limit $\FA_{sd}  > 6.4 \times 10^{7} \GeV$, which is slightly stronger than the above limit derived from baryon decays $\Sigma^+ \to p X$.  Nevertheless, for simplicity we restrict to two-body decays in Table~\ref{tab:lablimits}. 

In $c \to u$ transitions the  experimental situation is more involved (see also the discussions in Refs.~\cite{MartinCamalich:2020dfe, MartinCamalich:2025srw, Hiller:2025zgr}). Nominally the strongest limits can be derived from two-body meson decays such as $D\to \pi  X $  and $D\to \omega  X $. However, for the first decay no  limit  has been provided by the experimental collaborations, and only a recast of CLEO data for $D^+\to\tau^+(\to\pi^+\bar\nu)\nu$~\cite{CLEO:2008ffk} has been performed in Ref.~\cite{MartinCamalich:2020dfe}, giving 
${\rm BR} (D^+ \to \pi^+  X) < 8.0 \times 10^{-6}$.  It would be interesting to see if this value can be corroborated by the BESIII collaboration using their  data set. Taking this limit at face value, one obtains $\FV_{cu} > 6.1 \times 10^{7} \GeV$. On the other hand BESIII  published a limit on $D^0 \to \omega  X$ decays of $1.1 \times 10^{-5}$ for massless $X$~\cite{BESIII:2024rkp} (as suggested in Ref.~\cite{Su:2020yze}), however no lattice results are available for the relevant hadronic matrix elements. Approximating them by the same form factors as for the $D^0 \to \rho$ decay, Ref.~\cite{Hiller:2025zgr} obtains a  limit on the inverse effective coupling of $\FA_{cu}  > 2.9 \times 10^{7} \GeV$. Giving these uncertainties, we  choose to rely on the more robust, but slightly weaker limits that can be obtained from  BESIII searches for the baryonic two-body $\Lambda_c^+\to p  X$ decays with their complete data set~\cite{BESIII:2022vrr}, giving the limit ${\rm BR} (\Lambda_c  \to p X) < 8.0 \times 10^{-5}$ for massless dark bosons, which translates to a limit of $F^{V}_{cu}  > 1.5 \times 10^{7} \GeV$ and $F^{A}_{cu}  > 1.4 \times 10^{7} \GeV$. 

Turning to $b \to s$ transitions, searches for  invisible two-body decays of $B$ mesons have been carried out by the CLEO and Belle collaborations. CLEO obtained limits of ${\rm BR} (B\to K/K^* + X) < 4.9 \times 10^{-5}$ for both $K$ and $K^*$ decays~\cite{CLEO:2001acz}, which have been tightened to  ${\rm BR} (B\to K  X) < 3.4 \times 10^{-6}$ for decay to kaons using the Belle data set with 711 fb$^{-1}$~\cite{Belle:2026kum}. Similar limits have been obtained from recasts of searches of the three-body decay into neutrinos. Ref.~\cite{Altmannshofer:2023hkn} provides upper limits on ${\rm BR} (B\to K  X) < 7.0 \times 10^{-6}$  and ${\rm BR} (B\to K^* X) < 4.2 \times 10^{-5}$ combining data from Belle~II~\cite{Belle-II:2023esi} and BaBar~\cite{BaBar:2013npw}, while Ref.~\cite{Abumusabh:2025zsr} found ${\rm BR} (B\to K  X) < 1.4 \times 10^{-6}$ reinterpreting the same Belle~II data in Ref.~\cite{Belle-II:2023esi}. In absence of a limit by the Belle~II collaboration, we use the Belle limit for decays to kaons~\cite{Belle:2026kum} and the recast in Ref.~\cite{Altmannshofer:2023hkn} for decays to $K^*$, which translates to  limits on inverse effective axion couplings  of $\FV_{bs}  > 4.8 \times 10^{8} \GeV$ and $\FA_{bs}  > 1.4 \times 10^{8} \GeV$, respectively.

For $b \to d$ transitions  there are no analyses to date of  invisible two-body decays of $B$ mesons or bottom baryons. Therefore, all existing limits are obtained from recasts of searches of  neutrino three-body decays.  Ref.~\cite{MartinCamalich:2020dfe} used the BaBar analysis in Ref.~\cite{BaBar:2004xlo} to obtain ${\rm BR} (B\to \pi  X) < 2.3 \times 10^{-5}$, while an upper limit  ${\rm BR} (B\to \rho  X) < 3.9 \times 10^{-4}$ can be derived~\cite{Alonso-Alvarez:2023mgc} with data from the ALEPH experiment at LEP~\cite{ALEPH:2000vvi}. The resulting limits on  the inverse effective couplings are $\FV_{bd} > 1.2 \times 10^{8} \GeV$ and $\FA_{bd}  > 4.7 \times 10^{7} \GeV$,  respectively.

Finally we briefly comment on limits on flavor-violating couplings of top quarks. Direct limits from e.g. mono-top searches at the LHC give very weak constraints of the order of the EW scale, while  much more stringent (though not completely model-independent) bounds can be obtained from virtual corrections to the flavor-violating couplings $\CV_{sd}$~\cite{MartinCamalich:2020dfe}. Restricting for simplicity to real flavor-violating top couplings, the most relevant ($y_t^2$-enhanced) correction in the leading log approximation reads~\cite{MartinCamalich:2020dfe}\begin{align}
\Delta \CV_{sd} (\mu) & =   - \frac{y_t^2}{64 \pi^2} \log\frac{\Lambda}{\mu}   \sum_{q = u,c}  \left( V^*_{qs} V_{td}   + V^*_{ts} V_{qd} \right) (\CV_{t q} - \CA_{t q} )    \, , 
\label{eq:DeltacVsd}
\end{align}
where $\mu \gtrsim m_t$ is the RG scale, and on the right-hand side  couplings are evaluated at the UV scale $\Lambda$.  Assuming that this correction is much larger than the tree-level coupling $\CV_{sd}(\Lambda)$, the present constraint on the  $sd$ coupling, $\FV_{sd} \ge 1.1 \times 10^{12} \GeV$ translates to a bound on inverse flavor-violating top couplings $\FVA_{tc} \gtrsim 5.2 \times 10^{8} \GeV$ and $\FVA_{tu}\gtrsim 1.2 \times 10^{9} \GeV$. Here we have used $y_t \approx y_t ^{\rm SM}(\mu = M_Z) = 0.967$~\cite{Antusch:2025fpm} and  the values of the CKM matrix elements of the UTfit collaboration (Summer 2025)~\cite{UTfitweb}, and took  $\Lambda = 10^{10} \GeV$.

\subsubsection{Lepton flavor violation}
In the lepton sector the strongest limits  arise from  searches for two-body LFV decays,  with associated decay rates scale in terms of  ALP couplings given in Appendix~\ref{app:2bodydecays} and summarized in Section~\ref{sec:decayrates}. These decays look very similar to the corresponding SM di-neutrino decays, resulting in a charged lepton plus missing energy. As a consequence, the $\ell\to \ell'  X$ decays are not covered by standard LFV searches and require dedicated experimental strategies to suppress the large SM background. 

One such possibility is to study decays of polarized leptons, which gives an angular distribution of final state leptons that allows  distinguishing between different  LFV couplings, cf.~Eq.~\eqref{leptonpol}
\begin{align}
\frac{d\Gamma (\ell \to \ell^\prime X) }{d\cos\theta} \propto 1 +  P_{\ell^-} \cos\theta\cdot\frac{ \CV_{\ell^\prime\ell} \CAs_{\ell^\prime\ell} + C^{{\rm V}*}_{\ell^\prime\ell} \CA_{\ell^\prime\ell} }{\left| \CV_{\ell^\prime\ell} \right|^2  + \left|\CA_{\ell^\prime\ell}\right|^2 } \, .
\end{align}
 The experimental collaborations that search for $\mu \to e X$  constrain only a given benchmark scenario, but using the complete data sets one can obtain limits on any combination of chiral ALP couplings. Below we focus  on the limits on either isotropic decays  ($\CV_{\ell^\prime\ell} = 0$ or $\CA_{\ell^\prime\ell} = 0$), purely left-handed ($\CA_{\ell^\prime\ell} = - \CV_{\ell^\prime\ell}$) or purely right-handed ($\CA_{\ell^\prime\ell} =  \CV_{\ell^\prime\ell}$) couplings.

Isotropic muon decays with massless invisible particles have been severely constrained in experiments carried out at TRIUMF~\cite{Jodidio:1986mz}, limiting the branching ratio to ${\rm BR} (\mu \to e X)_{\rm iso} < 2.6 \times 10^{-6}$. Using the  information provided in this reference,  limits for right-handed (RH) and left-handed (LH) decays have been derived in Ref.~\cite{Calibbi:2020jvd}, giving ${\rm BR} (\mu \to e X)_R < 2.5 \times 10^{-6}$ and ${\rm BR} (\mu \to e X)_L < 2 \times 10^{-4}$. Here the latter number is expected to be strongly affected by  systematic uncertainties, since the signal is heavily polluted by SM background. More suitable to constrain light dark bosons with couplings aligned with the SM is the TWIST experiment, which relies  on the  electron spectrum rather than polarization to enhance signal-to-background ratio. This permits to derive the  limit ${\rm BR} (\mu \to e X)_L < 5.8 \times 10^{-5}$~\cite{TWIST:2014ymv} for an angular spectrum aligned to the Michel decay, while limits on isotropic or RH decays are weaker that the bounds obtained at TRIUMF. These constraints translate to limits on effective inverse axion couplings that read $\F^{\rm iso}_{\mu e} > 5.5 \times 10^{9} \GeV$ for isotropic decays with $\F^{\rm iso}_{\mu e} = \FV_{\mu e}$ or $\F^{\rm iso}_{\mu e}= \FA_{\mu e}$, $\FR_{\mu e}  > 5.6 \times 10^{9} \GeV$ for RH couplings with $\FR_{\mu e} \equiv \F_{\mu e} ({\CV_{\mu e} = \CA_{\mu e} })$, and  $\FL_{\mu e}  > 1.2 \times 10^{9} \GeV$ for LH couplings with $\FL_{\mu e} \equiv  \F_{\mu e} ({\CA_{\mu e} = - \CV_{\mu e} })$ in terms of the effective couplings defined in Eq.~\eqref{eq:shorthandF}. 

Another  strategy to probe LFV couplings of dark bosons to muons and electrons is to require an extra photon, $\mu \to e  \gamma  X$, which was pursued by the Crystal Box collaboration~\cite{Bolton:1988af}. The resulting constraint for ${\rm BR} (\mu \to e  \gamma X) < 1.1 \times 10^{-9}$ can be converted to limits on LFV ALP couplings, which are however  weaker than those from two-body decays as discussed in Ref.~\cite{Calibbi:2020jvd}. Nonetheless there are  interesting prospects to increase the sensitivity of this search using MEG-II data~\cite{Jho:2022snj, Grandoni:2026avn}.

Finally,  LFV transitions of $\tau$-leptons have been searched for by the Belle~II collaboration~\cite{Belle:2025bpu}, giving the limits ${\rm BR} (\tau \to \mu X) < 2.8 \times 10^{-4}$ and ${\rm BR} (\tau \to e X) < 4.5 \times 10^{-4}$. These constrain  the effective couplings in Eq.~\eqref{eq:shorthandF} as $\F_{\tau \mu} > 1.3 \times 10^{7} \GeV$ and $\F_{\tau e} > 3.5 \times 10^{7} \GeV$.


\subsubsection{Future prospects}
\label{sec:futurelab}

In general the sensitivity of searches for light particles in flavor-violating decays with missing energy can be improved by: i) performing existing searches with larger data sets; ii) applying dedicated search strategies to existing data sets; and, iii) performing entirely new searches.  The resulting prospects have been discussed in detail in Ref.~\cite{MartinCamalich:2025srw} (cf.~also Ref.~\cite{Knapen:2023zgi} for a summary of proposals to improve limits on $\mu \to e$ transitions), and  we  collect the expected limits provided in this reference in Table~\ref{tab:expfut}.

\begin{table}[h!]
\renewcommand{\arraystretch}{1.6}
  \setlength{\arrayrulewidth}{.35mm}
\centering
\setlength{\tabcolsep}{5.7 mm}
\begin{tabular}{@{}|c||c|c|}
\hline
Decay  & Prosp. Limit & Experiment   \\
\hline
${\rm BR}_{\rm proj} (K \to\pi X)$ &  $1 \times 10^{-11}$~\cite{MartinCamalich:2020dfe}  &NA62    \\
${\rm BR}_{\rm proj} (K \to\pi  \pi X)$ &  $7\times10^{-7}$~\cite{MartinCamalich:2020dfe} &E391a   \\
${\rm BR}_{\rm proj} (\Sigma  \to p X)$ &  $2 \times 10^{-6}$~\cite{MartinCamalich:2025srw}  & STCF\\
${\rm BR}_{\rm proj} (D  \to \pi X)$ & $1 \times  10^{-5}$~\cite{MartinCamalich:2025srw}  & STCF   \\
${\rm BR}_{\rm proj} (\Lambda_c  \to p X)$ &$1 \times  10^{-6}$~\cite{MartinCamalich:2025srw} & STCF   \\
${\rm BR}_{\rm proj} (B  \to K X)$ & $8\times10^{-7}$~\cite{MartinCamalich:2025srw}& Belle II \\
${\rm BR}_{\rm proj} (B  \to K^* X)$ & $4\times10^{-6}$~\cite{MartinCamalich:2025srw} &Belle II  \\
${\rm BR}_{\rm proj} (B  \to \pi X)$ &  $2\times10^{-6}$~\cite{MartinCamalich:2025srw} &Belle II  \\
${\rm BR}_{\rm proj} (B  \to \rho X)$ & $3 \times10^{-6}$~\cite{MartinCamalich:2025srw} & Belle II     \\
\hline
${\rm BR}_{\rm proj}  (\mu  \to e X)_{\rm iso}$ &$7 \times 10^{-8}$~\cite{Calibbi:2020jvd}  & MEGII-fwd   \\
${\rm BR}_{\rm proj}  (\mu  \to e X)_{{\rm iso}, {\rm L}, {\rm R}}$ &$7 \times 10^{-8}$~\cite{Perrevoort:2018ttp} & Mu3e   \\
${\rm BR}_{\rm proj}  (\mu  \to e X)_{\rm iso}$ &$1 \times  10^{-7}$~\cite{Hill:2023dym} & Mu2e/COMET-X    \\
${\rm BR}_{\rm proj}  (\tau  \to e X)$ & $4 \times 10^{-5}$~\cite{Badziak:2024szg} & Belle II   \\
${\rm BR}_{\rm proj}  (\tau  \to \mu X)$ & $3 \times 10^{-5}$~\cite{Badziak:2024szg} & Belle II    \\
\hline
\end{tabular}
\caption{Prospective  limits from Ref.~\cite{MartinCamalich:2025srw} on branching ratios of two-body meson, baryon and lepton decays with $X$ denoting a massless invisible particle. The last column denotes  the reference in which the forecast was performed.
\label{tab:expfut}
}
\end{table}

\subsubsection{Summary}

We summarize the best present and prospective laboratory limits on flavor-violating axion couplings in Table~\ref{tab:lablimits}. These are inferred from calculating the decay rates in Appendix~\ref{app:2bodydecays} for $m_a = 0$ with a single coupling switched on at a time, and  comparing to the experimental limits  in Table~\ref{tab:exp} and~\ref{tab:expfut}.  For hadron decays we neglect the systematic errors on the associated form factors for simplicity, which are small for the decays of interest and have no major impact in the bounds. These limits are valid also for ALPs with masses below the experimental resolution, which is typically of the order of a few MeV.

\begin{table}[h!]
\small
\renewcommand{\arraystretch}{1.6}
  \setlength{\arrayrulewidth}{.35mm}
\centering
\setlength{\tabcolsep}{5.3 mm}
\begin{tabular}{@{}|c||cc|cc|@{}}
\hline
& $\F_{\rm eff}$  & Experiment & $\F_{\rm eff}^{\rm proj}$  & Experiment \\
\hline
$C^{\rm V}_{sd}$   & $1.1 \times 10^{12}$ & NA62    & $1.8 \times 10^{12}$ & NA62     \\
$C^{\rm A}_{sd}$   & $2.8 \times 10^{7}$  & BESIII    & $1.1 \times 10^{8}$  & STCF     \\
$C^{\rm V}_{cu}$   & $1.5 \times 10^{7}$  & BESIII    & $1.4 \times 10^{8}$  & STCF     \\
$C^{\rm A}_{cu}$   & $1.4 \times 10^{7}$  & BESIII    & $1.2 \times 10^{8}$  & STCF     \\
$C^{\rm V}_{bs}$   & $4.8 \times 10^{8}$  & Belle    & $1.0 \times 10^{9}$  & Belle II \\
$C^{\rm A}_{bs}$   & $1.4 \times 10^{8}$  &BaBar & $4.6 \times 10^{8}$  & Belle II \\
$C^{\rm V}_{bd}$   & $1.2 \times 10^{8}$  & BaBar   & $4.0 \times 10^{8}$  & Belle II \\
$C^{\rm A}_{bd}$   & $4.7 \times 10^{7}$  & LEP & $5.4 \times 10^{8}$  & Belle II \\
$C^{\rm iso}_{\mu e}$ & $5.5 \times 10^{9}$ & TRIUMF      & $3.3 \times 10^{10}$ & Mu3e     \\
$C^{\rm L}_{\mu e}$    & $1.2 \times 10^{9}$ & TWIST     & $3.3 \times 10^{10}$ & Mu3e     \\
$C_{\tau e}$     & $1.0 \times 10^{7}$ & Belle II    & $4.0 \times 10^{7}$  & Belle II \\
$C_{\tau \mu}$   & $1.3 \times 10^{7}$ & Belle II     & $3.5 \times 10^{7}$  & Belle II \\
\hline
\end{tabular}
\caption{Present and prospective laboratory  limits on inverse  couplings $\F_{\rm eff} \equiv 2 \Lambda/C_{ij}$ in GeV for the respective flavor transition $i \to j$ and massless axions. Values are derived from  Tables~\ref{tab:exp} and \ref{tab:expfut}, see text for details. \label{tab:lablimits}}
\end{table}
\subsection{Limits from astrophysics}
\label{sec:astro}

Light dark particles can be copiously produced in stellar plasmas and  carry away energy, potentially  messing up stellar dynamics~\cite{Raffelt:1996wa}.  This argument has been widely used to constrain light dark sectors that couple to ordinary stellar matter, such as nucleons, electrons and photons. In particular, stringent  limits have been derived for the interactions of new light bosons such as the QCD axion, see Refs.~\cite{Carenza:2024ehj, Caputo:2024oqc} for recent reviews. Similar limits apply to all particles that  i) are sufficiently long-lived to escape from the interior  and ii) light enough to be produced in the plasma, which roughly  requires a mass  below the core temperature. Ordinary stars such as red giants and white dwarfs can probe interactions of new particles with masses in the keV range, while heavier particles with masses up to 100 MeV can be produced in the extreme environments of proto-neutron stars (PNS) formed during core-collapse or Type-II supernovae (SN). 

Such supernova explosions  accompany the terminal phase of massive stars heavier than about eight solar masses, which collapse under their own gravity after the nuclear fuel has exhausted.  When the inner core reaches nuclear densities,  the collapse is stalled and converted into a shock wave propagating outwards (the ``bounce"), which eventually ejects the  outer mantle in the SN explosion.  In the course of this process, most of the gravitational binding energy is emitted in the form of neutrinos on a timescale of few seconds~\cite{Janka:2017vlw}. In this late cooling phase, the PNS is essentially a  black-body that  radiates neutrinos from a surface with a temperature of about 5 MeV (the ``neutrinosphere")~\cite{Burrows:1988ba}.

 The observation of the late neutrino pulse from  the supernova SN1987A was in good agreement with theoretical expectations~\cite{Raffelt:1996wa} (see Ref.~\cite{Fiorillo:2023frv} for a recent discussion). Since then it has been used to constrain any additional energy-loss besides neutrino emission in the form of light dark particles. Numerical simulations indicate that a simple approximate criterion can be derived from the requirement that the total dark luminosity should not exceed the neutrino luminosity,  limiting the energy loss rate per unit mass $\eps$ (also called emissivity) to be below   $\epsilon_{\rm max} = 10^{19} \, {\rm erg \, s^{-1} \, g^{-1}}$~\cite{Raffelt:1996wa} at 1 second after the bounce. Crucially this criterion excludes only a compact range of couplings for given mass of the dark particle. This is because with growing  strength of interactions, the mean free path of the dark particle decreases, and eventually becomes so small that the  particle is effectively trapped within the PNS, reaching thermal equilibrium. In this ``trapping regime" it is still emitted from the surface of this ``dark sphere" as black-body radiation, violating the energy-loss criterion until the coupling is so large that the  dark sphere becomes as large as the neutrinosphere (for larger radii the temperature drops quickly so that the Stefan-Boltzmann luminosity $\propto r^2 T^4$ is lower than the neutrino luminosity). While the emitted flux of dark particles in this regime does not violate the energy-loss criterion, it might well be excluded by other observables, for example excessive energy deposition~\cite{Caputo:2022mah, Fiorillo:2025yzf} or direct detection~\cite{Engel:1990zd}.

While traditionally the SN cooling argument has been applied mainly to couplings of light particles to nucleons~\cite{Raffelt:2006cw}, it became clear that the temperatures and densities during SN explosions are sufficiently high to sustain also sizable numbers of moderately heavy flavors, such as muons and strange matter~\cite{1960SvA,Oertel:2016bki,Bollig:2017lki,Fischer:2020vie}. 
This fact was subsequently exploited to constrain flavor-conserving muon couplings of dark particles that can be produced from muon-initiated scattering processes~\cite{Bollig:2020xdr,Croon:2020lrf, Caputo:2021rux,Manzari:2023gkt}. Interestingly, the energy-loss argument can also be applied to  \emph{decays} of heavy flavors to light dark bosons, which has been pioneered in Ref.~\cite{MartinCamalich:2020dfe} for the case of hyperons (subsequently studied in Refs.~\cite{Camalich:2020wac, Alonso-Alvarez:2021oaj,Cavan-Piton:2024ayu,Fischer:2020vie}) and applied to muons in Ref.~\cite{Calibbi:2020jvd}. In contrast to dark particle production from scattering processes, which only permits to limit couplings within specific models, axion production from decays  directly constrains  the (model-independent) decay rate, just like laboratory experiments.

In the following we  discuss the  SN 1987A constraints of flavor-violating couplings of light dark ALPs, before we briefly review the astrophysical limits on flavor-conserving couplings for comparison.

\subsubsection{Hyperon decays}
The simple kinematics of two-body decays allow for a straightforward estimate  of the energy loss rate~\cite{MartinCamalich:2020dfe}, in contrast to the more complicated $2\to 3$ processes in axion production from nucleon bremsstrahlung~\cite{Raffelt:1996wa}. Taking the two-body decay $\Lambda \to n X$ of the lightest hyperon $\Lambda$ to neutron $n$ and dark boson $X$ as an example,  the energy-loss rate  $Q$  per unit volume  can be estimated very roughly as the product of i) the hyperon decay rate, ii) the  energy released per decay, approximately given by the hyperon-neutron mass difference, and iii) the hyperon number density $n_\Lambda$. The latter may be estimated as  the neutron number density $n_n$ times a Boltzmann factor, giving in total
\begin{align}
Q_{\Lambda} \sim  n_n (m_\Lambda-m_n)\Gamma(\Lambda\to n  X) \exp \left(- \frac{m_\Lambda - m_n}{T}\right) \, , 
\end{align}
which neglects effects like neutron degeneracy that will reduce the numerical value. The  emissivity $\epsilon$ is obtained by dividing the above estimate by the (nuclear) density $\rho \sim m_n n_n$, 
\begin{equation}
\label{eq:SNestimate}
\epsilon_{\Lambda} \sim   \frac{m_\Lambda - m_n}{m_n} \frac{{\rm BR} (\Lambda\to n  X)}{\tau_\Lambda}  \exp \left(- \frac{m_\Lambda - m_n}{T}\right) \, ,
\end{equation}
where we also  rewrote the decay rate in terms of the branching fraction and the total hyperon lifetime $\tau_\Lambda$.
Plugging in the hyperon-neutron mass difference of $176 \MeV$ and the peak PNS temperature of about 30 MeV obtained in numerical simulations~\cite{Bollig:2020xdr}, one finds
\begin{equation}
\label{eq:SNhypapp}
\epsilon_\Lambda \sim \frac{10^{19}  {\rm erg}}{{\rm g \, s}}  \left(\frac{{\rm BR}(\Lambda \to n X)}{6 \times 10^{-9}} \right) \, .
\end{equation}
Thus SN cooling restricts the branching fraction to be less than about $10^{-8}$, which is many orders of magnitude more stringent than the limits obtained from laboratory searches, roughly of order $10^{-5}$, cf.~Section~\ref{sec:lab}.

The above back-of-the-envelope estimate in Eq.~\eqref{eq:SNhypapp} can be extended to an accurate calculation of the decay rate in the PNS frame, including relativistic corrections, Pauli-blocking and medium effects modifying the baryonic dispersion relations in vacuum~\cite{Camalich:2020wac}. The thermodynamical input for the full radial  profile is taken from the spherically symmetric simulations of core-collapse SNe  reported in~\cite{Bollig:2020xdr}. Because these simulations do not include hyperons as an ingredient in the nuclear equation of state (EOS), the relevant thermodynamic quantities need to be recalculated using hyperonic extensions of the relevant EOS used in those simulations. This procedure has been validated in Ref.~\cite{Fischer:2024ivh}, which performed the first simulations of core-collapse SN explosions for a hyperonic equation of state  including the possible energy loss due to invisible $\Lambda$ decays.  The model-independent upper limit on the branching fraction obtained in this way reads
\begin{equation}
\label{eq:SNhypFin}
{\rm BR}(\Lambda\to n X)\lesssim 8.0 \times10^{-9} \, , 
\end{equation}
which is remarkably close to the rough estimate  in Eq.~\eqref{eq:SNhypapp}. When applied to the ALP couplings in Eq.~\eqref{scalar} and \eqref{vector}, the limits read  $\FV_{sd} \gtrsim 7.1 \times10^9$ GeV and $\FA_{sd} \gtrsim 5.2 \times10^9$ GeV~\cite{Camalich:2020wac}, the latter exceeding  present  laboratory limits by two orders of magnitude. 

Note that the cooling criterion would be violated even in  the deep-trapping regime, since  the decay rate is limited by the hyperon lifetime (i.e. ${\rm BR} <1 $) and the dark sphere, where $X$ is in thermal equilibrium with the plasma, has to lie within the region where hyperons remain in  thermal equilibrium, which is enclosed by the neutrinosphere. 

\subsubsection{Muon decays}

A similar analysis can be performed for the LFV muon decays $\mu\to e  X$. Using the same approximations as for hyperons in Eq.~\eqref{eq:SNestimate} one obtains the rough estimate for the dark boson emissivity from muon decays 
\begin{equation}
\epsilon_{\mu} \sim  Y_e \frac{m_\mu - m_e}{m_n} \frac{{\rm BR} (\mu \to e  X)}{\tau_\mu}  \exp \left(- \frac{m_\mu - m_e}{T}\right) \, ,
\end{equation}
where $Y_e = n_e/n_n \approx 0.12$ is the electron fraction. For a  PNS temperature of  30 MeV this gives
\begin{equation}
\label{eq:SNmuapp}
\epsilon_\mu \sim \frac{10^{19}  {\rm erg}}{{\rm g \, s}}  \left(\frac{{\rm BR}(\mu \to e X)}{6 \times 10^{-5}} \right) \, ,
\end{equation}
which, contrary to the case of hyperon decays, yields a bound of the same order as  laboratory experiments, cf.~Section~\ref{sec:lab}, and thus is of minor  importance.  A  related analysis constraining the same flavor-violating $\mu-e$ coupling, but considering dark boson production from $\mu-p$ bremsstrahlung has obtained  similar limits, giving ${\rm BR}(\mu \to e  X) < 9 \times 10^{-6}$~\cite{Zhang:2023vva}.

\subsubsection{Flavor-diagonal couplings}
Finally we  briefly discuss   astrophysical  constraints on flavor-diagonal ALP couplings, which can be compared to the limits from flavor-violation. In analogy to axion couplings to electrons in Eq.~\eqref{eaxion} we define axion couplings to  nucleons  as 
\eq{
{\cal L} = \frac{\partial_\mu a}{2 \Lambda} C_N \overline{N} \gamma^\mu \gamma_5  N  \, , 
}
for $N = n,p$. In the following we assume equal couplings to protons and neutrons for simplicity and constrain the inverse effective couplings defined as $\F_{N} \equiv 2 \Lambda/C_N$.

Star cooling limits from Red Giants (RGs) constrain ALP couplings to electrons at a level of   $\F_{e}\gtrsim 6.4 \times 10^{9} \GeV$~\cite{Capozzi:2020cbu}, while the SN1987A cooling argument constraints couplings to nucleons  as $\F_{N}\gtrsim 1.9 \times 10^{9} \GeV$~\cite{Fiorillo:2025gnd} (see also Refs.~\cite{Raffelt:2006cw, Carenza:2019pxu}),  and to muons as $\F_{\mu} \gtrsim 2.3 \times 10^{7} \GeV$~\cite{Caputo:2021rux}.  For ALPs with masses above the di-electron threshold, strong limits on electron couplings can be obtained from constraints on gamma-rays resulting from decaying ALPs that were  produced in SN1987A, which amounts to $\F_{e}\gtrsim (1 -3) \times 10^{9} \GeV$, depending on the ALP mass~\cite{Fiorillo:2025sln}. ALP couplings to photons are limited from stellar evolution in Globular Clusters to $\Lambda_\gamma \equiv 2 \Lambda/|C^{\rm eff}_{\gamma}| \gtrsim 9.9 \times 10^7 \GeV$~\cite{Dolan:2022kul}, which is slightly stronger than constraints from the CAST experiment~\cite{CAST:2017uph}. 

Bounds from star cooling are not valid for ALP masses above the typical inner core temperature, roughly  about 10 keV  (RGs), 1 keV (WDs) and 100 MeV (SNe), while the laboratory limits discussed in the Section~\ref{sec:lab}  extend to much larger masses up to the kinematical threshold. Another important difference is the required minimal decay length of ALP in order to efficiently extract energy, which is  of the size of the stellar core, very roughly about $10^4$ km for WDs and RGs and 10 km for the PNS formed during core-collapse SNe.
 
 
\subsection{Limits from cosmology}
\label{sec:cosmolimits}

Similar to  stellar interiors, light axions can  be thermally produced in the hot and dense plasma in the early universe, and are subject to stringent bounds from precision cosmology. While these limits have primarily been studied  for flavor-conserving couplings to SM particles,  also flavor-violating interactions are constrained by cosmology, as light ALPs can be efficiently produced from decays of heavy SM flavors in the thermal bath. In contrast to the bounds from laboratory experiments and astrophysics discussed previously,  the cosmological limits on flavor-violating ALP couplings depend strongly on the ALP mass. 

Very light ($\lesssim 0.1 \eV$)  axions such as the QCD axion are  ultra-relativistic at recombination, and thus contribute to dark radiation. This is  the energy density stored in relativistic degrees of freedom, which is conveniently parameterized in terms of the effective number of additional neutrino species $\Delta N_{\rm eff}$, defined as 
\eq{
\Delta N_{\rm eff} & = \left. \frac{8}{7} \left( \frac{11}{4} \right)^{4/3} \frac{\rho_a}{\rho_\gamma}\, \right|_{T = T_{\rm CMB}} \, , 
}
where $\rho_a$ ($\rho_\gamma$) is the axion (photon) energy density evaluated at  the temperature at recombination $T_{\rm CMB} \approx 0.3 \eV$. This observable is constrained by CMB observations and baryon acoustic oscillations (BAO), and the most recent combined analysis by the Planck collaboration sets the upper bound $\Delta N_{\rm eff} \le 0.3$ at 95\% CL~\cite{Planck:2018vyg}. This limit is expected to be improved to 0.1 at the Simons Observatory~\cite{SimonsObservatory:2018koc}.

Dark ALPs with masses larger than about 0.1 eV are no longer ultra-relativistic during recombination, and in general require a dedicated  analysis of their imprint on cosmological observables. As the axion 
mass only adds up to its energy density,  the resulting constraints are generically stronger than those derived from $\Delta N_{\rm eff}$ for the same couplings. For  masses roughly above 10 eV, the ALP starts to behave as Warm DM (WDM), and the limits are dominated by constraints on  structure formation from observations of the Lyman-$\alpha$ forest, which trace the matter power spectrum at small scales~\cite{Viel:2004bf,Boyarsky:2008xj}. These measurements limit the WDM fraction to about 10\%  of the total DM energy density for masses roughly below 1 keV~\cite{Baur:2017stq, Badziak:2025mkt}. Above masses of   about 10 keV the ALP starts to behave as cold dark matter, and its abundance is only limited by overclosure, i.e. $\Omega_{a} h^2 \le 0.12$.

Many studies in the literature  have derived limits on flavor-violating couplings of a very light (QCD) axion, which is produced as dark radiation and is constrained only through $\dNeff$, see e.g.~Refs.~\cite{Baumann:2016wac, DEramo:2018vss, Arias-Aragon:2020shv, DEramo:2021usm, Badziak:2024szg, Badziak:2024qjg, Badziak:2025mkt}. The majority of these works employed various approximations to calculate $\dNeff$, such as  instantaneous decoupling  or by solving Boltzmann equations for the axion number density assuming thermal equilibrium. While the latter approximation is appropriate for large densities as e.g. obtained from thermal freeze-out, it ceases to be accurate for smaller abundances induced by e.g. freeze-in. In this case, which is particularly relevant for obtaining accurate predictions for planned  CMB probes such as the Simons Observatory~\cite{SimonsObservatory:2018koc}, one needs to solve the momentum-dependent Boltzmann equations for the axion phase-space distribution function.   This  approach was used to improve  model-independent constraints on axion couplings to SM leptons (including LFV) in Ref.~\cite{Badziak:2024qjg} and  flavor-conserving couplings in Ref.~\cite{DEramo:2024jhn}. State-of-the-art results for flavor-violating quark couplings are still missing,  because of subtleties posed by IR divergencies in flavor-violating $2 \to 2$ scattering processes~\cite{Aghaie:2024jkj}, which are expected to give sizable contributions to the production rate~\cite{DEramo:2021usm} in the case of quarks where $\alpha_s$-corrections are sizable. 

Limits on axions with  masses larger than about $0.1 \eV$  are  stronger than those derived from $\dNeff$ alone, however they require a dedicated cosmological analysis.  For common QCD axion benchmark models such studies have been carried out in Refs.~\cite{Ferreira:2020bpb,DEramo:2022nvb,Notari:2022ffe,Bianchini:2023ubu}, while model-independent limits for massive axions have been derived only for axion couplings to leptons (including LFV) in Ref.~\cite{Badziak:2025mkt}, and photons and gluon couplings in Ref.~\cite{Caloni:2022uya}. Limits on quark couplings of massive axions, both flavor-conserving and flavor-violating have not been studied to date.

It should be   emphasized that cosmological limits on axion couplings disappear when i) the ALP is sufficiently heavy to avoid constraints from dark radiation and structure formation, which is the case for roughly $m_a \gtrsim {\rm few}  \keV$; ii) the axion is not stable on cosmological scales, i.e. $\tau_a \ll 10^{17} \sec$; iii)  sizable couplings of the ALP to other SM particles substantially alter its thermal history.

In the following, we first sketch how approximate limits for light ALPs below 0.1 eV can be derived from constraints on dark radiation, mainly following  the estimates in Ref.~\cite{Badziak:2024szg} based on evolving axion number densities. Then we collect the existing state-of-the-art limits on flavor-violating  transitions in the lepton and quark sector, in particular Ref.~\cite{Badziak:2025mkt} that provides results also for massive axions (but only covers lepton couplings).  Finally we  briefly summarize cosmological limits on flavor-conserving couplings, which have been obtained in Ref.~\cite{DEramo:2024jhn} (only considering small axion masses) and Ref.~\cite{Caloni:2022uya} (only evolving axion number densities).

\subsubsection{Limits on dark radiation}

In order to derive  limits on light ($m_a \lesssim 0.1 \eV$) axions from $\dNeff$, one needs to determine the axion energy density at recombination. While for accurate results one should use the momentum-dependent Boltzmann equations  for the phase-space distribution and find  the energy density as a momentum integral, a common approximation is to work instead with the integrated Boltzmann equation, which describes the evolution of the axion number density as a function of the collision terms  controlled by the relevant axion coupling. One can then approximately relate number to energy density using the same relation as in thermal equilibrium. Below we  collect the main results from these approximations for  light axion  with LFV couplings, while in Appendix~\ref{app:cosmo} we provide more details for axions with generic couplings.

 For sufficiently large  couplings, SM fermion decays and their inverse processes bring the axion into thermal equilibrium with the SM plasma. When the temperature drops, the rate of inverse processes fades, and the axions maintain their freeze-out (FO) abundance when they decouple from the thermal bath. Alternatively, when the couplings are so small that thermal equilibrium is never achieved, an axion abundance slowly builds up from freeze-in (FI) of SM fermion decays, until the temperature drops below the mass of the mother particle, so it becomes non-relativistic and its abundance is exponentially suppressed~\cite{Hall:2009bx}. The transition region for LFV decays $\ell \to \ell^\prime a$ can be estimated  very roughly as (cf.~Eq.~\eqref{fCeqdecay})
\eq{
\label{Tcrit}
(\F_{\ell \ell^\prime})^{\rm eq}_{\rm decay} & \approx 2 \times 10^8 \GeV \sqrt{\frac{m_\ell}{\GeV}}\, ,
}
setting $m_\ell^\prime = m_a = 0$ for simplicity. Couplings smaller than this critical value define the freeze-in regime,  in which one can estimate the present yield from decays as (cf.~\eqref{YFI})
\begin{align}
	Y_a^{\rm FI} &  \simeq   3 \times 10^{-2} \frac{ \, m_\ell M_{\rm Pl}}{ g_{*s} (m_\ell) \sqrt{g_*(m_\ell)} \F_{\ell \ell^\prime}^2 } 
	  \, ,
\end{align}
and 
\eq{
\label{DNeffdecay}
\dNeff & \simeq    0.6  \left( \frac{ \, m_\ell M_{\rm Pl}}{  g_{*s} (m_\ell) \sqrt{g_*(m_\ell)} } \right)^{4/3}  \F_{\ell \ell^\prime}^{-8/3} \, .
}
If instead couplings are sufficiently large to keep the boson in thermal equilibrium at early times, the relic abundance is generated by thermal freeze-out, and given by the equilibrium yield at the decoupling temperature $T_d$  (cf.~Eq.~\eqref{YFO2})
\begin{align}
Y_a^{\rm FO} &  \simeq \frac{45 \zeta(3) }{2 \pi^4 g_{*s} (T_d)} \, , 
\end{align}
which is determined by the transcendental equation~\eqref{Tdeqs}. The contribution to dark radiation is obtained by plugging this   yield into Eq.~\eqref{Neffapp}
\eq{
\label{eq:Neffmain}
\Delta N_{\rm eff} & \approx 74.8  \left( Y_a^{\rm FO}  \right)^{4/3}  \, .
}
One can  use these approximate relations to derive the following limits on axion couplings from the present Planck limit $\dNeff \le 0.3$ (valid for axion masses below $0.1 \eV$) 
\eq{
\label{cosmoestimate}
\F_{\tau \ell} & \gtrsim 8 \times 10^6 \GeV \, , & \F_{\mu e}  & \gtrsim 2 \times 10^8 \GeV \, , 
}  
where $\ell = e, \mu$ with negligible differences, and we assumed that $\tau$-decays are in the FO and $\mu$-decays in the FI regime. The fact that these limits are close to the critical value in Eq.~\eqref{Tcrit} already indicate that this approximation is quite rough, and a numerical solution of the Boltzmann equation is needed for accurate results.

Note that here we have implicitly assumed that the reheating temperature $T_R$ is small enough such that axion production is IR dominated (and thus independent of $T_R$). If the thermal universe instead begins with large temperatures, one can derive stringent limits on axion couplings under the assumption that  future CMB experiments will be able to set the limit  $\Delta N_{\rm eff} \le 0.027$ (obtained from evaluating Eq.~\eqref{eq:Neffmain} for $ g_{* s} (T_d)= 106.75$), thus establishing that the axion  was never in thermal contact with the SM~\cite{Baumann:2016wac}. This implies that the decoupling temperature of UV sensitive scattering processes such as $\ell \overline{\ell} \to h a$ is above $T_R$, i.e. $\Gamma_{\rm scat}^{\rm UV} (T_R) \le H(T_R)$, which e.g.  implies the limit (see Eq.~\eqref{TdUV})
\begin{align}
\Lambda_{\tau \ell} \gtrsim 9 \times 10^{8} \GeV \left( \frac{T_R}{10^{6} \GeV }\right)^{1/2} \, . 
\end{align}
This result can be compared to the $T_R$-independent limits in Eq.~\eqref{DNeffdecay}, which give for $\Delta N_{\rm eff} \le 0.027$ 
\begin{align}
\Lambda_{\tau \ell} \gtrsim 6 \times 10^{8} \GeV  \, .
\end{align}
In the following we will restrict to $T_R$-independent limits, which can be derived from present ($\Delta N_{\rm eff} \le 0.3$) and near-future constraints ($\Delta N_{\rm eff} \le 0.1$)  on  dark radiation.

\subsubsection{Lepton flavor violation}

We summarize the cosmological limits on leptonic ALP couplings in Table~\ref{cosmolimitsL}, which are taken from Ref.~\cite{Badziak:2025mkt}. Shown are the present Planck limits on axions with  $m_a \ll 0.1 \eV$ from $\dNeff$, the projected limits for the Simons observatory, and the present limits on heavier axions. 

Comparing the present LFV limits for massless axions with the estimates in Eq.~\eqref{cosmoestimate}, one can observe that the estimate for $\mu \to e$ transitions is remarkably accurate, while the limits on $\tau \to \ell$ couplings are weaker by an order of magnitude. This difference is mainly due to the fact that  for the present Planck limits $\tau$-decays are in the FO  regime, while $\mu$-decays are dominantly freeze-in. The approximations underlying the estimates in Eq.~\eqref{cosmoestimate} are twofold: evolving dark boson number densities (instead of phase-space distributions) and calculating energy density  from number densities (instead of integrating the distributions). As analyzed in Ref.~\cite{Badziak:2024qjg}, this approximations have different quality depending on whether the dominant production proceeds through FO or FI. Freeze-out from (inverse) decays is very sensitive to the actual momentum distribution, as low-momentum modes decouple earlier from the plasma, so that the integrated approach  overestimates the dark boson yield, up to O(10\%) as compared to the phase space approach. On the other hand in the  FI regime inverse decays are negligible, so the production rate is basically independent of the dark boson momentum, and thus the two approaches gives essentially the same result. The second approximation instead has the opposite behavior: since it works well for almost thermal distributions, it is accurate for FO, while for FI it underestimates the actual energy density, although the effect is rather mild for $\tau$-decays as shown in Ref.~\cite{Badziak:2024qjg}. To summarize, while our estimate for $\tau$-decays in Eq.~\eqref{cosmoestimate} in agreement with earlier calculations~\cite{DEramo:2021usm, Badziak:2024szg}, it is  too large as a result of treating only  integrated phase-space distributions, which is not a good approximation for freeze-out from decay processes. Note however that this should not be an issue for future  limits on $\dNeff$, which requires  the couplings to be   small enough such that FI dominates. Indeed the FI estimate we obtain imposing $\dNeff \le 0.1$ is $\F_{\tau \ell}  \simeq 4 \times 10^8 \GeV$, in good agreement with the actual bound obtained in Ref.~\cite{Badziak:2025mkt} that is smaller by  factor two. 

While our estimates were  limited to axions with masses smaller than $0.1 \eV$  that are constrained by dark radiation, Table~\ref{cosmolimitsL} makes clear that  constraints from structure formation (Lyman-$\alpha$)  for slightly larger masses can be much stronger, e.g. by a factor 4 for LFV muon couplings and axion masses of 10 eV. This is in sharp contrast to limits from colliders or SN1987A, which cannot resolve such tiny mass differences.

\begin{table}[h!]
\renewcommand{\arraystretch}{1.6}
  \setlength{\arrayrulewidth}{.35mm}
\centering

\setlength{\tabcolsep}{3.9 mm}

\centering
\begin{tabular}{|c||c|c|c|c|}
\hline
& $\dNeff \le 0.3$ & $\dNeff \le 0.1$  & $m_a = 0.3\eV$   & $m_a = 10 \eV$ \\
\hline
$\F_{\tau \ell}$ & $2.7 \times 10^{5}$   & $2.0 \times 10^{8}$           & $8.7 \times 10^{6}$    & $5.0 \times 10^{8}$  \\
$\F_{\mu e}$ & $1.6 \times 10^{8}$   & $3.3 \times 10^{8}$           & $3.1 \times 10^{8}$    & $6.3 \times 10^{8}$  \\
\hline
$\F_{e}$ & $2.4 \times 10^{6}$   & $4.4 \times 10^{6}$          & $6.6 \times 10^{6}$    & $1.1 \times 10^{7}$  \\
$\F_{\mu }$ & $1.1 \times 10^{7}$   & $2.6 \times 10^{7}$           & $2.9 \times 10^{7}$    & $5.3 \times 10^{7}$  \\
$\F_{\tau }$ & no bound   & $3.1 \times 10^{5}$           &  $2.1 \times 10^{5}$   & $6.6 \times 10^{6}$  \\
\hline
\end{tabular}
\caption{Cosmological 95\% CL lower limits on inverse  leptonic axion couplings $\F_{ij} \equiv 2 \Lambda/\sqrt{|C_{ij}^{\rm V}|^2 + |C_{ij}^{\rm A}|^2}$ and $\F_{i} \equiv 2 \Lambda/C_i$ in units of GeV, taken from Ref.~\cite{Badziak:2025mkt}. The  first two columns  show the present Planck limit from $\dNeff \le 0.3$ and the projected limits for the Simons observatory $\dNeff \le 0.1$ valid for light axions with  $m_a \ll 0.1 \eV$, while the last two columns show the present limits on massive axions.  \label{cosmolimitsL} }
\end{table}

\subsubsection{Quark flavor violation}
In the quark sector there is  presently no analysis that matches the rigor of Ref.~\cite{Badziak:2025mkt} for  LFV couplings, i.e. using the full set of cosmological observables and evolving phase space distributions. For this reason, we use the estimate for the freeze-in yield from flavor-violating quark decays in Eq.~\eqref{YFI} and the approximate relation to $\dNeff$ in Eq.~\eqref{Neffapp} to estimate present ($\dNeff \le 0.3$) and prospective ($\dNeff \le 0.1$) limits from dark radiation on  flavor-violating quark couplings. These estimates  agree well with the limits presented previously in Ref.~\cite{Baumann:2016wac} and are  summarized in Table~\ref{cosmolimitsQ}. 

\begin{table}[h!]
\renewcommand{\arraystretch}{1.6}
  \setlength{\arrayrulewidth}{.35mm}
\centering

\setlength{\tabcolsep}{13.3 mm}

\centering
\begin{tabular}{|c||c|c|}
\hline
& $\dNeff \le 0.3$ & $\dNeff \le 0.1$   \\
\hline
$\F_{b q}$ & $3  \times 10^{8}$   & $5 \times 10^{8}$       \\
$\F_{t q}$ & $2 \times 10^{9}$   & $3 \times 10^{9}$        \\
$\F_{c u}$ & $2\times 10^{8}$   & $3 \times 10^{8}$        \\
\hline
\end{tabular}
\caption{Estimated  95\% CL lower limits on inverse  flavor-violating quark couplings $\F_{ij} \equiv 2 \Lambda/\sqrt{|C_{ij}^{\rm V}|^2 + |C_{ij}^{\rm A}|^2}$ in units of GeV for light axions with  $m_a \ll 0.1 \eV$. The first column shows  the present Planck limit, the second column the projected limits for the Simons observatory. \label{cosmolimitsQ} }
\end{table}

We do not include $s \to d$ transitions, because  the NA62 limit exceeds by far even future cosmology projections~\cite{DEramo:2021usm}, apart from complications in the calculation of the abundance at temperatures below the QCD phase transition. On the other hand, there are cosmological limits from top decays $t \to q  X$ with $q = c,u$, which reach values of order $10^9 \GeV$ because of the large top mass.  These couplings are constrained by laboratory searches only indirectly by $K \to \pi  a$ trough RG running, cf.~Eq.~\eqref{eq:DeltacVsd}, which gives limits of roughly the same size. Constraints on $c \to u$ and $b \to q$  transitions have also been studied in   Ref.~\cite{DEramo:2021usm},  including also flavor-violating scattering processes besides decays. However, the calculation of the corresponding production rates is quite subtle, due to complications from IR divergencies that can potentially give large unphysical enhancement factors  if not properly cancelled~\cite{Czarnecki:2011mr, Aghaie:2024jkj}.

\subsubsection{Flavor-diagonal couplings}

Limits on flavor-diagonal  couplings to leptons have been discussed in Ref.~\cite{Badziak:2025mkt} and we summarize their results in the lower part of Table~\ref{cosmolimitsL} (they essentially agree with the results obtained in  Refs.~\cite{DEramo:2024jhn, Barbieri:2026ewj}). Again it is instructive to compare these numbers to the estimates provided in Appendix~\ref{app:cosmo}. Eq.~\eqref{fCeqscat} gives an  approximate expression for the critical temperature from diagonal scattering processes as 
\begin{align}
\label{Tcritestl}
(\F_{\ell })^{\rm eq}_{\rm scat} & \approx4  \times 10^7 \GeV \sqrt{\frac{m_\ell}{\GeV}} \, , 
\end{align}
which suggests that for present and near-future limits axion production from $e$- and $\mu$-couplings is in the freeze-in regime, while $\tau$-couplings can be large enough to keep the ALP in thermal equilibrium.  Using the estimates for the respective contribution to dark radiation in Eqs.~\eqref{YFO2},~\eqref{YFI} and~\eqref{Neffapp} one finds excellent agreement (within a factor two) with the results in Ref.~\cite{Badziak:2025mkt, DEramo:2024jhn}.

For quarks the estimate for the critical coupling is larger than for leptons by a factor $\sqrt{4 \alpha_s/\alpha}$, which implies that for present and near future bounds dark boson production is in the FO regime, and using the estimates in Eqs.~\eqref{YFO2} and~\eqref{Neffapp}, it is clear that even requiring $\dNeff \le 0.1$ there are no limits on diagonal quark couplings, as confirmed by the analysis in Ref.~\cite{DEramo:2024jhn}. 

Finally we note that  cosmology also provides  constraints on couplings to gluons and photons, which for massless axions restrict the inverse effective  couplings to  $\F_G \equiv 2 \Lambda/C_{GG} \gtrsim 4 \times 10^7 \GeV$~\cite{Caloni:2022uya} and $\F_\gamma \equiv 2 \Lambda/C_\gamma \gtrsim 1 \times 10^5 \GeV$~\cite{Barbieri:2026ewj}, respectively.  These  limits are at least  two orders of magnitude weaker than the constraints from SN1987A (nucleons) and Globular Clusters (photons), cf.~Section~\ref{sec:astro}.

\subsection{Summary}
\label{sec:summarylimits}

\begin{figure}[h]
\centering
\includegraphics[width=1.0\textwidth]{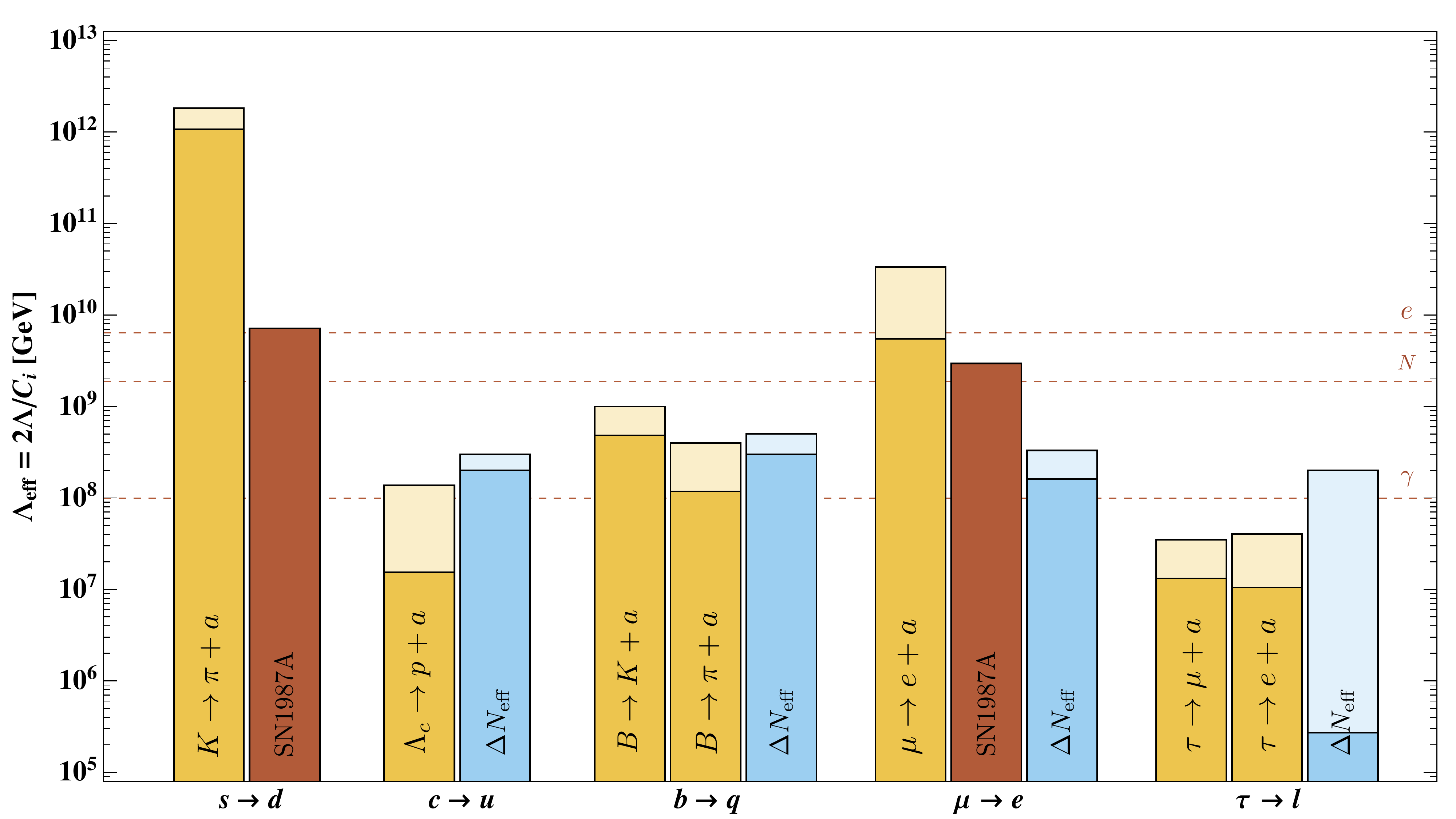}
\caption{Present and prospective limits on flavor-violating couplings of a light axion from laboratory searches (yellow), astrophysics (red) and cosmology (blue). For a given quark flavor transition $i \to j$ and chirality structure the bound is set on $\Lambda_{\rm eff} =  2 \Lambda /|\CVA_{ij}|$ in the notation of Eq.~\eqref{axionFV}. Dashed red horizontal lines indicate the astrophysical limits on  flavor-diagonal couplings to electrons, nucleons and photons in the same convention.}
\label{fig:limits}
\end{figure}

We summarize the most relevant present and future limits on light ALPs from laboratory experiments (yellow), astrophysics (red) and cosmology (blue) in Fig.~\ref{fig:limits}, where for comparison we also display the astrophysical limits on diagonal axion couplings to electrons, nucleons and photons.  These limits are valid for $m_a \ll 0.1 \eV$ (cosmology), $m_a \ll 100 \MeV$ (SN1987A) and $m_a \lesssim \MeV$ (laboratory). Similarly, the axion lifetime has to be sufficiently large for the ALP to be stable on cosmological  ($\tau_a \gtrsim 10^{17} \s$), stellar  ($\tau_a \gtrsim 10^{-5} \s$) and collider  ($\tau_a \gtrsim 10^{-9} \s$) scales, ignoring boost factors for the latter cases.

For quark transitions the shown laboratory limits constrain the vectorial couplings $\F_{ij}^V$, $\tau$-transitions restrict $\F_{ij}$, and  for $\mu \to e$ transitions we display limits on the isotropic couplings $\F_{ij}^{\rm iso}$ (limits on $\FA_{ij}$ or $\F^{\rm L,R}_{ij}$  are very similar except for $s \to d$ transitions, see Table~\ref{tab:lablimits}). For simplicity we only show the bounds for heavy flavor transitions $b \to q$ with $q = s,d$ and $\tau \to \ell$ with $\ell = \mu,e$ because cosmological limits are essentially identical and  laboratory limits differ only marginally. The show cosmological limits on $\Delta N_{\rm eff}$ are taken from Tables~\ref{cosmolimitsL} and~\ref{cosmolimitsQ}, which represent only rough estimates for quark transitions, see the discussion in Section~\ref{sec:cosmolimits}. Note that for $s \to d$ transitions no bound is shown, as it is subject to large uncertainties and anyway weaker than laboratory and SN1987A limits. Future limits refer to $\Delta N_{\rm eff} \le 0.1$, as projected for the Simons Observatory.  The limits from SN9187A are taken from Section~\ref{sec:astro}, and denote the limits on $\FV_{sd}$ from hyperon decays (the limit on axial couplings is very similar), while for $\mu \to e$ transitions we used the limit on the branching ratio from Ref.~\cite{Zhang:2023vva}, which is valid for isotropic couplings. 

In this comparison one notices the superiority of laboratory limits in the case of $s \to d$ and $\mu \to e$ transitions, which even outclass the stringent astrophysical limits on flavor-diagonal axion couplings (for couplings of similar size). In the other sectors  limits from laboratory experiments are in remarkably  close competition with cosmology, but a detailed comparison is complicated by the fact that the latter limits strongly depend on the axion mass and a complete analysis of these limits is not available. For light axions contributing to dark radiation it is nevertheless clear that laboratory searches have more sensitivity to axion couplings, since in the long run cosmological axion production is dominated by IR freeze-in (at least if $T_R$ is not too large), where $\Delta N_{\rm eff} \propto \F_{ij}^{-8/3}$ (see Eq.~\eqref{DNeffdecay}), while ${\rm BR} \propto \F_{ij}^{-2}$. On the experimental side ongoing searches at Belle~II will thus play a crucial role in probing $b \to q$ and $\tau \to \ell$ flavor transitions.

\section{Light dark matter}

The possibility to explain the observed  dark matter abundance  is one of the driving motivations of light dark axions. In this section we take a  a model-independent approach to axion dark matter, and discuss its stability (Section~\ref{sec:stability})  and  production  in the early universe (Section~\ref{sec:production}). We summarize the  parameter space in Section~\ref{sec:summaryDM} (Figs.~\ref{fig:DME1} and  \ref{fig:DME0}), including constraints from flavor, axion haloscopes, star cooling and searches for decaying DM.

\subsection{Dark matter stability}
\label{sec:stability}
We begin with the most important property of DM, namely that it is still around today. This means that DM particles have to be stable on cosmological scales, with lifetimes exceeding the age of the universe, roughly $1/H_0 \sim 10^{17} \sec$. In contrast to heavy WIMPs, which usually need to be stabilized by means of symmetries, a distinguished property of light axions is that their stability can  be a direct consequence of their tiny mass. Restricting for simplicity to dark axions with masses below the di-muon threshold, the only open decay channels are to electrons, neutrinos and photons. In the following we briefly discuss the associated decay rates controlled by the relevant flavor-conserving  ALP couplings in Eq.~\eqref{axionFC}, and the  experimental limits assuming that the ALP  fully accounts for the observed DM  abundance.

\subsubsection{Decays to electrons}

The inverse partial width  for ALP decays to an electron-positron pair is given by Eq.~\eqref{ratescalartoee}
\begin{align}
\frac{1}{ \Gamma_{a \to e^+  e^-}} \sim  10^{17}  \, {\rm s} \left( \frac{1 \MeV}{m_a} \right)  \left( \frac{\Lambda/C_e}{10^{15} \GeV} \right)^2 \, .
\end{align}
in terms of the couplings in Eq.~\eqref{eaxion} and we have neglected the phase space suppression. This implies that axions heavier than an MeV can have lifetimes larger than the age of the universe for UV scales exceeding  $10^{15} \GeV$. However, much more stringent limits on DM decays to electrons arise from  CMB observations~\cite{Slatyer:2016qyl},  soft $\gamma$-ray~\cite{Essig:2013goa} and $X$-ray data~\cite{Cirelli:2023tnx}, which require lifetimes larger than $10^{24} \sec$. These translate into effective suppression scales exceeding $10^{19} \GeV$, which suggests that DM axions either have extremely suppressed electron couplings or a mass below the di-electron threshold. 

\subsubsection{Decays to neutrinos}
Axion decays to neutrinos are suppressed by the tiny neutrino mass. Restricting for simplicity to one generation and taking the axion coupling to neutrinos analogous to  Eq.~\eqref{eaxion}, 
\begin{equation}
\label{nuaxion}
{\cal L}_{{\rm axion},\nu}  =  C_\nu  \frac{\partial_\mu a}{2 \Lambda} \, \overline{\nu}_L \gamma^\mu \gamma_5 \nu_L  \, ,
\end{equation}
the inverse partial width is given by 
\begin{align}
\frac{1}{\Gamma_{a \to \nu \overline \nu}} \sim  10^{20}  \, {\rm s} \left( \frac{1 \MeV}{m_a} \right)  \left( \frac{m_\nu}{0.09 \eV} \right)^2  \left( \frac{\Lambda/C_\nu}{10^{10} \GeV} \right)^2 \, .
\end{align}
Here we normalize to the upper bound on the sum of neutrinos masses from cosmology~\cite{diValentino:2022njd} and a value for the effective UV scale in the range of future LFV searches, cf.~Section~\ref{sec:lab}.

These decays will produce  monoenergetic neutrinos with energy smaller than $1\, \MeV$~\cite{Garcia-Cely:2017oco, Lin:2022xbu}. This is well below the kinematic threshold of $E_{\bar{\nu}_e} \ge 1.8\ \MeV$ for inverse $\beta$-decay  $\bar{\nu}_e+ p\to n + e^+$, which has been proposed to search for the cosmic neutrino background~\cite{Weinberg:1962zza}. Therefore elastic scattering on electrons is the only active process for such low-energy neutrinos, which makes it very challenging for current detectors such as Borexino~\cite{Borexino:2010zht} to provide  constraints on  DM decaying to neutrinos beyond the cosmological bound of $10^{17} {\rm s}$. 

\subsubsection{Decays to photons} 
Axion decays to photons are unavoidable, and depend on all ALP couplings to  charged particles. For very light axion masses, the decay rate is suppressed by  charged fermion masses, apart from  contributions due to the electromagnetic (EM) and color anomaly. In the limit  $m_a \ll m_e$ the decay rate reads, taking into account only the electron contribution (see Eq.~\eqref{eq:ALPCgamma}) 
\begin{align}
\Gamma_{a \to\gamma \gamma} & = \frac{\alpha_{\rm em}^2 }{64 \pi^3} \frac{m_a^3}{\Lambda^2} C_\gamma^2 \, ,  & 
C_\gamma & \approx C_{\gamma \gamma} - 1.96 C_{GG} - C_e \frac{m_a^2}{12 m_e^2} \, , 
\end{align}
in terms of the couplings defined in Eq.~\eqref{axionFC}.

In most QCD axion models the anomaly coefficients $ C_{\gamma \gamma} \equiv E$ and $C_{GG} = N$ are ${\cal O}(1)$ numbers (cf.~Appendix~\ref{app:PQ}), for example in the DFSZ-II
model $E=8$ and $N=3$~\cite{Zhitnitsky:1980tq,Dine:1981rt}. In these cases the anomalies give the dominant contribution to the decay rate, which scales with $\propto m_a^3$. If instead  the UV model is free of anomalies (see e.g. the explicit scenarios discussed in Sections~\ref{sec:majoron}
 and~\ref{sec:lfvfreezein}), the rate scales as $\propto m_a^7$. It is  useful to explicitly distinguish these two cases, with decay rates 
\begin{align}
 \frac{1}{\Gamma_{a \to\gamma \gamma}} & \sim  10^{18}  \, {\rm s} \left( \frac{10 \keV}{m_a} \right)^3  \left( \frac{\Lambda/C_\gamma}{10^{10} \GeV} \right)^2 \, , \\
\label{eq:gammaE0} 
 \frac{1}{\Gamma_{a \to\gamma \gamma}} & \sim 10^{27}  \, {\rm s} \left( \frac{10 \keV}{m_a} \right)^7  \left( \frac{\Lambda/C_e}{10^{10} \GeV} \right)^2 \, .
\end{align}
where in the second case $C_{\gamma \gamma} = C_{GG} = 0$.
These expressions imply  that light axions in  the range of future LFV searches can easily have  lifetimes that exceed the age of the universe by many orders of magnitude. Depending on the axion mass, there are however much more stringent limits on the photon decay rate of DM axions from a variety of observations.  In the following we assume for simplicity that this is the dominant decay channel, such that the inverse decay rate can be identified with the axion lifetime $\tau_a$. 

For axions in the keV-MeV range  there are  precision measurements of CMB temperature and polarization anisotropies, which limit the lifetime to roughly $\tau_{\gamma\gamma}\gtrsim3\times 10^{24}\text{ sec}$~\cite{Slatyer:2016qyl, Bolliet:2020ofj}. Even stronger constraints arise from searches for X-ray and low-energy gamma lines, which at present give limits roughly three orders of magnitude stronger than CMB observations, with details depending on the  axion mass. Here we use the results collected in Ref.~\cite{Panci:2022wlc} (see also the website~\cite{AxionLimits}), which summarizes  searches that have been conducted  with Chandra~\cite{Watson:2011dw, Horiuchi:2013noa}, Newton-XMM~\cite{Foster:2022ajl}, NuStar~\cite{Perez:2016tcq,Roach:2019ctw,Ng:2019gch,Roach:2022lgo}, and INTEGRAL~\cite{Laha:2020ivk}.  These constraints are expected to  strengthen with future X-ray telescopes, and we use the optimistic projections collected in Ref.~\cite{Panci:2022wlc} for GECCO~\cite{Coogan:2021rez}, THESEUS~\cite{Thorpe-Morgan:2020rwc} and Athena~\cite{Neronov:2015kca,Dekker:2021bos,Ando:2021fhj}, which could  probe lifetimes of order $10^{30} {\rm \, sec}$ for masses in the relevant mass range. For smaller masses there are limits from observations of the MUSE~\cite{Todarello:2023hdk} (few eV),  the Hubble Space Telescope (HST)~\cite{Todarello:2024qci} (few tens of eV) of the dwarf galaxy Leo T~~\cite{Wadekar:2021qae} (few hundreds of eV). Relevant future limits  are expected from exotic energy injection in the 21-cm power spectrum~\cite{Sun:2023acy} for masses between few tens of eV and few keV. We use the data provided on the website~\cite{AxionLimits} to collect the relevant constraints from these experiments, and combine them with the aforementioned bounds from X-ray telescopes. In this  way we can derive limits on the UV scale as a function of the axion mass for given couplings. We show these results for $C_{\gamma \gamma} -1.96 \, C_{GG} = 1, C_e =1$ in Fig.~\ref{fig:DME1},  and for $C_{\gamma \gamma} = C_{GG} = 0, C_e =1$ in Fig.~\ref{fig:DME0}. In these plots we also display the regions  where the axion can fully account for the observed DM abundance, which is  the requirement for the above constraints on decaying DM. In the following we discuss the associated DM production mechanisms in more detail.

\subsection{Dark matter production}
\label{sec:production}
The preferred production mechanism depends  on the DM mass:  axions with masses much below the keV scale have to be non-thermally produced in order to avoid constraints on warm DM (WDM), while  axions with masses larger than few keV can be thermal relics.

A classic non-thermal production mechanism  is misalignment, which has originally been proposed for the QCD axion~\cite{Preskill:1982cy, Abbott:1982af, Dine:1982ah}, and then generalized for  light ALPs in Ref.~\cite{Arias:2012az,Blinov:2019rhb}. While many variants of this mechanism exist in the literature (see e.g. Ref.~\cite{OHare:2024nmr} for a recent review),  here we restrict  for simplicity to pre-inflationary models, where PQ breaking occurs before inflation. In this case the axion abundance only depends on the original misalignment angle, the mass and the UV scale (or decay constant), for a given  temperature dependence of the ALP mass. This determines the UV scale as a function of the ALP mass up to an ${\cal O}(1)$ parameter in each scenario, which can be used to assess the sensitivity of precision flavor experiments to DM axions. We will discuss as examples the common case of a constant ALP mass, as well as  dynamical mass generation from  non-perturbative QCD effects. Another interesting ALP scenario is ``ALP-cogenesis"~\cite{Co:2020xlh}, which simultaneously explains the observed baryon and dark matter densities.  Instead for the QCD axion the decay constant is fixed in terms of the mass, so that the abundance only depends on two parameters. While standard misalignment is viable only for axion masses smaller than about 50 $\mu{\rm eV}$, in the scenario of ``kinetic misalignment"~\cite{Co:2019jts} the DM abundance is dominated by the initial velocity rather than the initial displacement, which allows for QCD axion DM also for much lower masses.

Heavy axions with masses above a few keV can also be produced  via ALP couplings to SM particles in the thermal bath. If these interactions are sufficiently strong to bring the axion into thermal equilibrium in the early universe, its relic abundance is determined by the freeze-out temperature according to Eq.~\eqref{OmegaFO}
\begin{align}
\Omega_{\rm FO}  h^2  \approx 0.12 \left( \frac{m_a}{170 \eV} \right) \left( \frac{106.75}{g_{*s} (T_d)} \right) \, .
\end{align}
It is therefore not possible to have light thermal relics in the few keV regime without extra dilution mechanisms. However,  the couplings to SM particles can  be sufficiently  small  to  avoid thermal contact in the early universe. In this case a  population of axions can be generated from  thermal scattering or decay processes of SM particles via freeze-in~\cite{Hall:2009bx}, which can easily match the observed DM abundance (see Appendix~\ref{app:cosmo}). This possibility  fits naturally to the large UV scales  required in the case of flavor-violating axions, allowing for a direct link between the size of flavor-violating decay rates and the DM relic abundance (see Section~\ref{sec:lfvfreezein}).

\subsubsection{Misalignment}

The equation of motion of a homogenous scalar field in an expanding universe reads
\begin{align}
\label{ALPevolution}
\ddot a + 3 H (T) \dot{a} + m_a^2 a = 0 \, , 
\end{align}
where we  restricted to a quadratic  potential with temperature-independent mass, and $H$ denotes the 
Hubble parameter. If inflation took place below the scale of PQ  breaking,  the initial ALP value $a_0$ is frozen by the inflationary dynamics as an initial condition for the thermal universe (we assume here that the initial velocity $\dot a_0$ is negligible). This value can be parametrized in terms of an angular variable, $a_0=\Lambda\theta_0$, where $\theta_0\in [0,\pi)$ is the initial misalignment angle, which in this scenario  is uniform across the whole Hubble patch. As long as the Hubble  rate is much larger than the ALP mass, $H (T) \gg m_a $, the evolution described by Eq.~\eqref{ALPevolution} an overdamped oscillator with Hubble friction $H(T)$, so the ALP is frozen at its initial value $a_0$.  As the temperature drops, the Hubble rate gets smaller and around $H \sim m_a$ the axion starts to oscillate. To determine the onset of oscillation at the temperature $T_{\rm osc}$, the condition $m_a = 1.6 H (T_{\rm osc})$ provides a good fit to the results of a more precise numerical treatment~\cite{Blinov:2019rhb}. The energy stored in these oscillations behaves just as cold DM and expands adiabatically, so that today's axion abundance can be obtained from rescaling the energy density at the onset of oscillations $\rho_a = 1/2 m_a^2 \Lambda^2 \theta_0^2$ by the ratio of entropy density today $s_0$ and at the start of oscillations $s(T_{\rm osc})$. The  final axion abundance depends through $T_{\rm osc}$ on the cosmological epoch at the time of oscillations. For sufficiently high reheating temperatures oscillations start during radiation domination, and with $H = 1.66 \sqrt{g_*} \, T^2/M_{\rm Pl} $ one obtains 
 \begin{align}
     T_{\rm osc} & = 2.1 \times 10^4 \GeV \left( \frac{m_a}{\eV} \right)^{\frac{1}{2}} \left( \frac{106.75} {g_*(T_{\rm osc)}} \right)^{\frac{1}{4}} \, ,      
\end{align}
and the final ALP abundance from misalignment reads
\begin{align}
\label{eq:mis}
  \Omega_{\rm mis} h^2 & = 0.12  \left(\frac{\Lambda \theta_0}{6.2 \times 10^{11}\GeV}\right)^2  \left(\frac{m_a}{ \eV}\right)^{\frac{1}{2}}  \left( \frac{106.75} {g_*(T_{\rm osc)}} \right)^{\frac{1}{4}}  \, .
  \end{align}
  Note that ALPs produced by misalignment have extremely small momentum, since they are associated with  field configurations that are homogenous up to the Hubble scale at the time where oscillations start~\cite{Sikivie:2006ni}. This gives a velocity dispersion today that is of the order 
  \begin{align}
\beta_0 \sim \frac{H(T_{\rm osc})}{m_a} \frac{T_0}{T_{\rm osc}} \sim  \frac{T_0}{T_{\rm osc}} \sim 10^{-17} \sqrt{\eV/m_a} \, ,
  \end{align}
  so that ALPs are extremely cold despite their small mass.
  
An  alternative mechanism that leads to enhanced misalignment production is obtained from a temperature-dependent ALP mass, which arises in scenarios where e.g. non-perturbative  effects dynamically generate the axion potential through chiral symmetry breaking, as for the QCD axion. In this case at $T=0$ the ALP mass is given by $m_a=\Lambda_\chi^2/\Lambda$, where $\Lambda_\chi$ is the scale of chiral symmetry breaking. At high temperatures the chiral symmetry is partially restored, so that the ALP mass is suppressed as $m_a(T)=m_a(\Lambda_\chi/T)^b$, where $b\approx 4$ in QCD (see Ref.~\cite{Gross:1980br} for the case of general gauge theories). The relic abundance from misalignment for this case has been studied in Refs.~\cite{Arias:2012az, Blinov:2019rhb}, and is given by  
\begin{equation}
\label{eq:misT}
\Omega_{\rm mis}^{\rm T} h^2=  0.12 \,  \theta_0^2 \left(\frac{\Lambda }{1.9 \times 10^{10}\GeV}\right)^{\frac{5}{3}}  \left(\frac{m_a}{ \eV}\right)^{\frac{1}{2}}  \left( \frac{106.75} {g_*(T^T_{\rm osc})} \right)^{\frac{5}{12}}  \, ,
\end{equation}
where now the onset of oscillations takes place at the temperature 
\begin{align}
     T^{\rm T}_{\rm osc} & = 2.3 \times 10^{-2} \GeV \left( \frac{m_a}{\eV} \right)^{\frac{1}{2}}  \left( \frac{\Lambda}{\GeV} \right)^{\frac{1}{3}} \left( \frac{106.75} {g_*(T^T_{\rm osc})} \right)^{\frac{1}{12}} \, .       
\end{align}
Therefore  the same abundance can be obtained for much smaller UV scales as compared to constant ALP masses, see Eq.~\eqref{eq:mis}. This analysis covers the QCD axion where $\Lambda_\chi = \Lambda_{\rm QCD}$,  giving (see e.g.~Ref.~\cite{DiLuzio:2020wdo})
\begin{equation}
\label{eq:misQCD}
\Omega_{\rm mis}^{\rm QCD} h^2=  0.12 \,  \theta_0^2 \left(\frac {7.5 \times 10^{-6} \eV}{m_a}\right)^{\frac{7}{6}}  \, ,
\end{equation}
which yields the ``natural" QCD axion mass window for $\theta_0 \in [0.1, \pi-0.1]$ as $m_a \in [0.14 , 51] \, {\mu \rm eV}$ or $f_a \in [1.1 \times 10^{11} , 4.1 \times 10^{13}] \GeV$. While the observed abundance can be achieved also for smaller masses by lowering $\theta_0$, for larger axion masses (smaller UV scales) one needs to rely on other production mechanisms, for example post-inflationary scenarios. Another possibility is ``kinetic misalignment"~\cite{Co:2019jts}, where the axion has large initial velocity $\dot{\theta}$, with a kinetic energy that exceeds the depth of the periodic potential. Thus the axion overcomes the potential barrier, and only starts to oscillate when its kinetic energy is sufficiently redshifted to become trapped in a single potential well. Therefore the  onset of oscillations  is delayed with respect to the standard case with negligible velocity, and given by
\begin{align}
\label{eq:kinQCD}
\Omega_{\rm kin}^{\rm QCD} h^2=  0.12 \,   \left(\frac {m_a}{5.7 \times 10^{-3} \eV}\right) \left( \frac{Y_\theta}{40} \right) \, ,
\end{align}
where $Y_\theta$ is the yield $n_\theta/s$ associated with the conserved Peccei-Quinn charge density $n_\theta = \dot{\theta} f_a^2$. This mechanism works as long as the initial velocity is large enough to overcome the potential barrier, corresponding to roughly $m_a \gtrsim 40 \, \mu {\rm eV}$~\cite{Co:2019jts}, while for smaller masses one has conventional misalignment. This means that the QCD axion can fully account for DM via pre-inflationary misalignment for essentially all masses not excluded by observations. 

Finally one can consider an ALP variant of kinetic misalignment, where the charge density is tied to the observed baryon asymmetry, dubbed ``cogenesis". Indeed the velocity $\dot \theta$ also controls the PQ charge asymmetry in the axion condensate, which can be converted  to a particle-antiparticle asymmetry of SM particles in the thermal bath via the ALP-SM couplings. This  asymmetry in turn is converted to a baryon asymmetry by the electroweak anomaly (i.e. sphaleron processes), which apart from the initial velocity only has a mild dependence on the ALP couplings to the SM.  Fixing  the baryon asymmetry to its observed value, one can predict the axion relic abundance in terms of the ALP mass as before, giving 
\begin{align}
\label{eq:cogen}
\Omega_{\rm cogen} h^2 =  0.12  \left(\frac{\Lambda }{2.1 \times 10^{6}\GeV}\right)^2 \left(\frac{m_a}{ \eV}\right)  \left(\frac{0.1}{ c_B}\right)   \, ,
\end{align}
where it is assumed that the electroweak phase transition is SM-like and the UV scale is temperature-independent.  Here the constant $c_B$ depends on the ALP coupling to SM particles, but is typically ${\cal O}(0.1)$. Note that cogenesis does not work for the QCD axion, since it requires scales as low as $f_a \approx 8 \times 10^5 \GeV$, which is robustly excluded by SN cooling~\cite{Springmann:2024ret, DiLuzio:2026sxp}.

\subsubsection{Freeze-in}
As argued in Section~\ref{sec:stability}, the stringent limits on thermal DM with masses above a few keV require both the absence of color and electromagnetic anomalies and UV scales larger than $\sim 10^{10} \GeV$ (cf.~Fig.~\ref{fig:DME0}), which strongly suppresses the axion couplings to  SM fermions. This means that DM  was never in thermal equilibrium with the SM in the early universe, but  a thermal DM population can still be generated by the freeze-in mechanism~\cite{Hall:2009bx}. Here axions are produced from  flavor-violating decays  and  $2\to 2$ scattering processes from  particles in the thermal bath, until the temperature drops below the mass of the initial SM states, so that their abundance become exponentially suppressed and  axion production in this channel shuts off. 
As discussed in Appendix~\ref{app:cosmo}, one can derive a simple analytic expression for the resulting freeze-in yield assuming a standard cosmology and taking the effective number of relativistic degrees of freedom in the SM approximately constant, cf.~Eq.~\eqref{YFI} . Restricting for simplicity to leptonic processes, the resulting abundance from scattering processes $\ell^+ \ell^- \to \gamma a$ and $\ell^\pm \gamma \to   \ell^\pm a$ is given by 
\begin{align}
\label{eq:OmFIscat}
\Omega_{\rm scat} h^2 \approx 0.12 \left(\frac{ 5.6 \times  10^8 \GeV}{\Lambda/C_{\ell \ell}}\right)^2 \left(\frac{m_a}{50\text{ keV}}\right) \left(\frac{m_{\ell
}}{m_\mu}\right)\left(\frac{18}{g_*(m_{\ell})}\right)^{\frac{3}{2}} \, .
\end{align}
Axion production rates from LFV decays are directly proportional to the LFV decay rate  $\Gamma (\ell \to \ell^\prime a) \approx C_{\ell^\prime \ell}^2m_{\ell}^3/64\pi \Lambda^2$, giving
\begin{align}
\Omega_{\rm decay} h^2 &  \approx  \frac{4.5 \times 10^{27}}{g_*(m_\ell)^{3/2}} \frac{m_a}{m_\ell^2} \Gamma_{\ell \to \ell^\prime a} \nonumber \\
& = 0.12   \left( \frac{  3.6 \times10^9 \GeV}{\Lambda/C_{\ell^\prime \ell}} \right)^2 \left( \frac{m_a}{50 \, \keV} \right)\left(\frac{m_{\ell}}{m_\mu}\right) \left(\frac{18}{g_*(m_{\ell})}\right)^{\frac{3}{2}}  \, , 
\label{eq:OmFIdecay}
\end{align}
so that flavor-violating decays dominate over  flavor-diagonal scattering processes for $C_{\ell \ell}\sim C_{\ell^\prime \ell}$, due to the absent suppression by $\alpha_{\rm em}$. Remarkably, the UV scales needed to account for the observed relic abundance  are in the range $\Lambda \sim 10^9-10^{10}\, \GeV$,  which will be probed by future laboratory experiments looking for two-body invisible muon decays. Requiring LFV decays to generate the observed relic abundance thus gives a target for experimental searches looking for the very same decay, which only depends on the ALP mass~\cite{Panci:2022wlc}. We will discuss this possibility  in more detail in  Section~\ref{sec:lfvfreezein}. 

\subsubsection{UV freeze-in}
 
Freeze-in scenarios are potentially sensitive to DM production processes that are dominated by high temperatures~\cite{Hall:2009bx}, also dubbed ``UV-freeze-in"~\cite{Bernal:2019mhf}. In the axion case, such processes arise from $2\to2$ scattering via the dimension-five operator  $a h \overline{\ell^\prime} \ell/\Lambda$, cf.~Appendix~\ref{app:cosmo}.   These interactions contributes to freeze-in production of axions via $\ell h \to \ell^\prime a$ and similar processes in a UV-sensitive manner, i.e. the rates  depend on the reheating temperature $T_R$. The resulting axion abundance reads (cf.~Eq.~\eqref{CtermsFI})   
\begin{align}
\label{eq:UVfreezein}
\Omega_{\rm UV} & \simeq  \frac{2 m_{\ell} T_R} {3 \pi^3 v^2}\times \Omega_{\rm decay}  \, ,
\end{align}
and is controlled by the reheating temperature that is only constrained to be above $\sim 4 \MeV$ in order to be consistent with Big-Bang-Nucleosynthesis (BBN)~\cite{Kawasaki:2000en, Hannestad:2004px}. This UV contribution can be neglected against the IR freeze-in contribution $ \Omega_{\rm decay} $ if $T_R$ is sufficiently low,  which is the case for 
$T_R \lesssim   10^7 \text{ GeV} (m_\mu/m_{\ell})$. Note that  small reheating temperatures may  affect the misalignment contribution to the relic abundance if axion oscillations take place before reheating, $T_{\rm osc} \gtrsim T_R$, with details depending on the assumed cosmology above $T_R$~\cite{Blinov:2019rhb, Visinelli:2009kt, Arias:2021rer}.

\subsubsection{Limits on warm dark matter}

While misalignment creates  axions with essentially vanishing momentum, freeze-in DM particles are produced from thermal processes and   have  large initial velocities. This leads to a suppression of  primordial fluctuations imprinted in the matter power spectrum at small scales, which is constrained by measurements of absorption features in the spectra of high-redshift quasars,
caused by neutral hydrogen in the intergalactic medium, commonly referred to as the Lyman-$\alpha$ forest~\cite{Viel:2004bf, Boyarsky:2008xj}. These analyses yield a stringent lower bound on the warm DM mass $m_{\rm WDM}^{\rm min} \approx 5.3 \, \keV$, which can be relaxed to  $m_{\rm WDM}^{\rm min} \approx 3.5 \, \keV$ under more conservative assumptions~\cite{Viel:2013fqw,Baur:2015jsy,Irsic:2017ixq}. These   limits on generic WDM have been recasted for different freeze-in processes by computing the exact DM velocity distribution in Refs.~\cite{DEramo:2020gpr, Ballesteros:2020adh, Decant:2021mhj}. Using the results of Ref.~\cite{Ballesteros:2020adh}, one obtains a lower limit  on the mass of DM particles produced from decays as 
\begin{align}
\label{WDMbound}
    m_a \gtrsim 7 \keV \left(\frac{m_{\rm WDM}^{\rm min}}{3 \,\text{keV}}\right)^{4/3}\left(\frac{106.75}{g_*(m)}\right)^{1/3} \, , 
\end{align}
where $m^{\rm min} _{\rm WDM}\approx 3.5 \keV$ or $5.3 \keV$ for the conservative and stringent WDM limits, respectively, and $m$ denotes either the mass  of the decaying particle in the thermal bath. These results are essentially consistent (within a factor of about two) with the  recent  analysis in Ref.~\cite{DEramo:2025jsb}, which provided  model-independent bounds on generic DM produced via two-body decays, using not only the Lyman-$\alpha$ forest, but also other constraints on WDM such as the number of Milky Way satellites and strong gravitational lensing with the James Webb Space Telescope (JWST). As all such small-scale probes of WDM are subject to significant observational and modelling uncertainties, we conservatively stick to the WDM limit in Eq.~\eqref{WDMbound} with $m^{\rm min} _{\rm WDM}\approx 3.5 \keV$. Finally we note the that WDM limits are relaxed if the total DM abundance is only partially produced from freeze-in, and mainly arises from non-thermal mechanisms like misalignment.

\subsection{Summary and discussion}
\label{sec:summaryDM}
\begin{figure}
\centering
\includegraphics[width=0.99\textwidth]{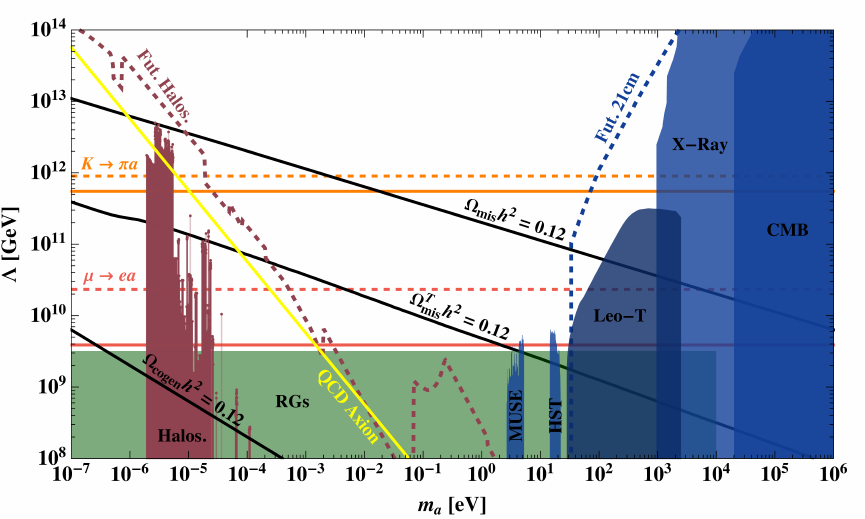}
\caption{Parameter space for axion dark matter for anarchic fermion couplings $\CV_{ij} = \CA_{ij} =1$ and $C_{\gamma \gamma}  - 1.96 \, C_{GG} =1$. Shown are exclusion limits on decaying DM (blue), star cooling constraints from RGs (light  green), haloscope searches (dark red) and precision flavor experiments looking for $K \to \pi a$ (orange) and $\mu \to e a$  (light red), where present limits are indicated as solid horizontal lines. Projected  limits from the 21-cm power spectrum,  flavor-violating decays   (NA62 and Mu3e) and future haloscopes  are indicated by dashed contours. Black solid lines denote the regions where the observed DM abundance can be obtained via misalignment for the indicated scenario. The yellow line denotes the QCD line, where the relic abundance can be reproduced with standard or kinetic misalignment. Not shown are the regions excluded from searches by XENON1T and XENONnT, since these are excluded also by X-ray telescopes (cf.~Fig.~\ref{fig:DME0}). }
\label{fig:DME1}
\end{figure}

\begin{figure}
\centering
\includegraphics[width=0.95\textwidth]{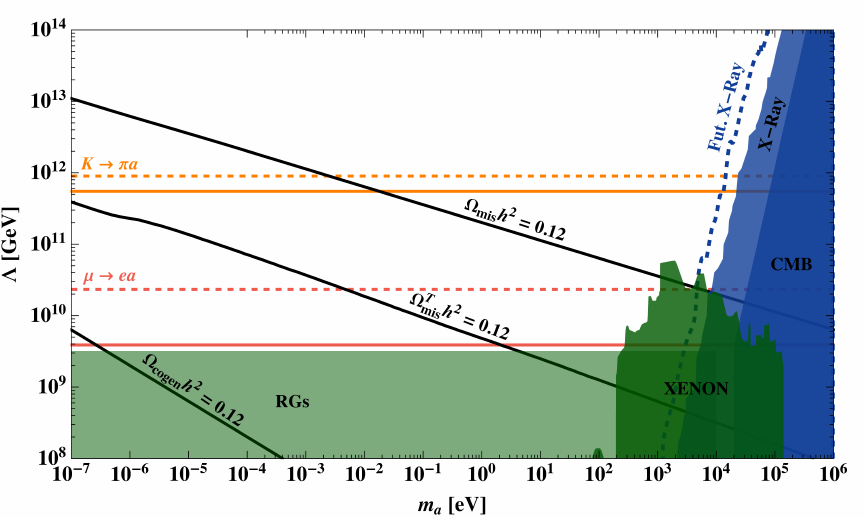}
\caption{Parameter space for axion dark matter for anarchic fermion couplings $\CV_{ij} = \CA_{ij} =1$ and $C_{\gamma \gamma} = C_{GG}  =0$. Shown are the same exclusion limits and DM lines as in Fig.~\ref{fig:DME0}, apart from limits by XENON1T and XENONnT (dark green), which probe axion  couplings to electrons.}
\label{fig:DME0}
\end{figure}

We summarize the generic parameter space for axion dark matter in Fig.~\ref{fig:DME1} and Fig~\ref{fig:DME0}, which shows the $m_a - \Lambda$ parameter space  for anarchic couplings to fermions ($\CV_{ij} = \CA_{ij} =1$), distinguishing between  anomalous ALPs ($C_{\gamma \gamma}  - 1.96 \, C_{GG} =1$) displayed in Fig.~\ref{fig:DME1} and  anomaly-free ALPs ($C_{\gamma \gamma} = C_{GG} =0$) shown in Fig.~\ref{fig:DME0}.  

We indicate as black solid lines the regions where the axion can fully account for the observed DM abundance via misalignment, taking as representative scenarios a constant ALP mass~\eqref{eq:mis}, a mass dynamically generated by QCD~\eqref{eq:misT} and the prediction of cogenesis~\eqref{eq:cogen} (for $c_B = 0.1$). The first two cases depend on the initial misalignment angle, which we fix close to its maximal value $\theta_0 = \pi - 0.1$, in order to show the lowest possible UV scale.  It is understood that also for larger values of the UV scale $\Lambda$ the ALP can saturate the observed abundance, with properly adjusted misalignment angles. Similar to the DM contours, also the QCD axion relation\footnote{This of course requires $C_{GG} \ne 0$.} in Eq.~\eqref{eq:QCDaxionmass} fixes the decay constant as a function of the axion mass, and we display this relation as a yellow contour. Everywhere on this line the DM relic abundance can be reproduced via standard~\eqref{eq:misQCD} or kinetic~\eqref{eq:kinQCD} misalignment.

If the ALP coupling  to photons is not suppressed, the standard haloscope searches for the QCD axion will cover significant parts of the parameter space, as shown in Fig.~\ref{fig:DME1}. The dark red region around few $\mu{\rm eV}$ is excluded by  current axion searches, in particular the ADMX experiment~\cite{ADMX:2009iij, ADMX:2018gho, ADMX:2019uok, ADMX:2021nhd, ADMX:2024xbv, ADMX:2025vom}, see \cite{AxionLimits} for more details. The future  haloscope campaigns will explore a much wider ALP mass region between $0.1-100\,\mu\text{eV}$, with  prospective limits taken from \cite{AxionLimits} and shown as a dark red dashed contour (the region between 0.1 and 1 eV denotes the reach of the LAMPPOST haloscopes \cite{Baryakhtar:2018doz}). Note that the sensitivities of large-scale helioscopes such as IAXO \cite{IAXO:2019mpb, IAXO:2025ltd}, and light-shining-through-wall experiments such as ALPS-II \cite{Bahre:2013ywa, ALPSII:2025eri} lie below the current stellar cooling bounds for our choice of parameters and therefore do not appear  in Fig.~\ref{fig:DME1}.

The haloscope searches are not sensitive to  axions with suppressed  couplings to photons, which is the case for anomaly-free ALPs shown in Fig.~\ref{fig:DME0}. In this case photon couplings have an additional suppression by the ALP mass, which in the usual haloscope mass range is strong enough to completely remove the sensitivity of even future haloscope searches.  However for larger ALP masses the suppression gets milder, and above few keV X-ray telescopes still exclude large parts of the parameter space in case electron couplings are not suppressed. The low mass range is in this case excluded mainly by star cooling, specifically the RG limit on electron couplings obtained in Ref.~\cite{Capozzi:2020cbu}. These limits are valid up to ALP masses of about 10 keV, but they are superseded by constraints from the XENON1T and XENONnT experiments~\cite{XENON:2019gfn, XENON:2020rca, XENON:2022ltv} that looked for electron recoils from  ALP absorption in the mass range from 0.1-10 keV.

In Figs.~\ref{fig:DME1} and~\ref{fig:DME0} we  show also present limits and prospects from precision flavor experiments, for flavor-anarchic ALP couplings $\CV_{ij} = \CA_{ij} =1$, see Table~\ref{tab:lablimits}. While limits from flavor physics are in principle relevant for several transitions in the indicated parameter range, we choose to show only the most sensitive limits from kaon (orange contour) and muon (light red contour) decays, with future prospects shown as dashed contours. Note that these limits are likely to be softened as a result of some degree of flavor suppression in the couplings, e.g. by small CKM matrix elements. 

In the absence of large hierarchies among the ALP couplings, Figs.~\ref{fig:DME1} and~\ref{fig:DME0} demonstrate that precision flavor experiments are highly complementary to other constraints and searches for axion DM, yielding stringent limits on the UV scale that are nearly independent of the ALP mass. For large masses above 100 eV, present and future  LFV limits are weaker than constraints from XENON and  decaying DM, while kaon decays probe scales that are about two orders of magnitude larger,  above the limits from Leo-T and XENON,  and only superseded by X-ray telescopes. The limits on decaying DM however quickly deteriorate for smaller ALP masses, and flavor searches are essentially the only probes of ALPs in the meV-eV range  and UV scales above the RG cooling limit (as observed  already in Ref.~\cite{Gelmini:1982zz}). For anomaly-free ALPs this holds for even smaller masses, due to the absence of haloscope limits. Instead for unsuppressed photon couplings future haloscopes probe scales comparable to LFV searches around meV masses, while kaon decays can compete with planned haloscopes down to few $\mu {\rm eV}$. This is precisely the range of QCD axion DM, so future searches for $K^+ \to \pi^+$ play a crucial role to explore this region of vanilla parameter space. Kaon decays are sensitive also to standard ALP misalignment DM, and in fact are the only probe for masses between 0.01 eV and 1 keV for anarchic couplings. Instead LFV searches will explore a region of parameter space where ALP production from  misalignment with constant mass  cannot saturate the observed DM abundance, even for maximal initial misalignment. They will however probe ALP DM from misalignment with dynamical mass generation in  a  window between few meV and 100 eV, and DM from cogenesis in almost the entire parameter range for coupling hierarchies allowed by star cooling and haloscopes.

\section{Explicit models}

 Finally we go beyond model-independent considerations and discuss three explicit models in which the various couplings in the general effective ALP Lagrangian are fixed in terms of few parameters.   
We begin with a scenario where the PQ symmetry is identified with a simple anomalous Froggatt-Nielsen flavor symmetry (Section~\ref{sec:axiflavon}), which links flavor-violating axion couplings of the QCD axion to the hierarchical pattern  
of SM Yukawa couplings. In Section~\ref{sec:majoron} we discuss the majoron,  where the PQ symmetry is identified with lepton number. Here the majoron  couplings are connected to observed neutrino parameters in a type-I seesaw scenario. Finally we discuss simple DM scenarios, where the size of flavor-violating axion couplings is phenomenologically linked to the observed relic abundance (Section~\ref{sec:lfvfreezein}), giving explicit targets for LFV searches with light axions.

\subsection{A QCD axion  from  flavor symmetries}
\label{sec:axiflavon}
As discussed in Section~\ref{sec:motivation}, the textbook example of a light dark particle is the QCD axion, which elegantly solves the Strong CP Problem and can fully account for the observed DM abundance. At low energies the QCD axion is described by the most general low-energy Lagrangian in Eq.~\eqref{eq:axionfull}, which contains flavor-violating couplings. While these terms can be directly probed by precision flavor experiments, it is not straightforward to write down explicit UV models in which these couplings are predicted. This is because in a complete setup with a spontaneously broken Peccei-Quinn symmetry, flavor-violating couplings arise by rotating the SM fermion PQ charges to the mass basis. This necessarily requires the knowledge of the Yukawa matrices in the basis where PQ charges are defined, and thus a theory of flavor. Nonetheless, particularly simple models of this kind can be constructed when the PQ symmetry is identified with a subgroup of a flavor symmetry that explains Yukawa hierarchies. In these kind of theories the SM fermions are assigned to representations of some flavor symmetry group, which is spontaneously broken by the vacuum expectation values (VEVs) of various scalar fields, the so-called \emph{flavons}. The SM Yukawa couplings  arise as higher-dimensional operators with suitable flavon insertions to be invariant under the flavor symmetry. Replacing  flavons by their VEVs determines Yukawas as powers of a small order parameter given by the ratio of VEV over the UV scale, up to the Wilson coefficients of the effective operators. In most cases the PQ subgroup is anomalous under QCD, so the associated Goldstone boson residing in a suitable linear combination of flavons is the QCD axion, which has also been dubbed ``familon"~\cite{Wilczek:1982rv}, ``axiflavon"~\cite{Calibbi:2016hwq} or ``flaxion"~\cite{Ema:2016ops} in this context.  
 
 This framework ties  PQ charges of SM fermions to the observed masses and mixings, and at the same time determines the unitary rotations that are needed to calculate axion couplings in the fermion mass basis, cf.~\eqref{eq:CVAbasis}. 
However, in this relation there is a two-fold ambiguity. First, the $U(1)_F$ charge assignment depends on the structure the full flavor symmetry group and its breaking pattern, i.e. the choice of the flavor model. Second, only the CKM and PMNS combinations of the unitary rotations are observables in the SM, so that most of the unitary rotations  in the quark and lepton sector depend on undetermined Wilson coefficients, which are  ${\cal O}(1)$ by assumption. Both aspects of this freedom can be reduced for more restrictive flavor symmetries, which drastically reduce the number of effective operators. In this section we briefly review these kind of theories, and discuss the simplest example of a classic $U(1)_F$ Froggatt-Nielsen symmetry~\cite{Ema:2016ops, Calibbi:2016hwq} as well as a more restrictive $U(2)_F$ symmetry~\cite{Linster:2018avp}. For early references see Refs.~\cite{Davidson:1981zd, Wilczek:1982rv, Berezhiani:1989fp}.  

\subsubsection*{The $\boldsymbol{U(1)_F}$  Axiflavon}
We assign $U(1)_F$ charges to SM fermions $f_i$ that we denote by $X_{f_i}$ for right-handed fields, and $-X_{f_i}$ for left-handed fields (with a relative sign for the sake of notational convenience). The scalar sector consists of the SM Higgs field, which is not charged under  $U(1)_F$, and a flavon field $\chi$ with charge $-1$. The SM Yukawa sector is then given by the effective operators
\begin{align}
\label{eq:LyukFN}
{\cal L}_{\rm yuk} & = \lambda^u_{ij}  \left( \frac{\chi}{\Lambda_F} \right)^{X_{q_i} + X_{u_j}}  \overline{q}_{Li} u_{Rj} H + \hdots \, , 
\end{align}
with analogous expressions for the down-quark and  lepton sectors. Replacing the flavon by its VEV parametrized as $\langle \chi \rangle = \Lambda = \eps \Lambda_F$, the quark yukawa matrices  become
\begin{align}
y^u_{ij}  & = \lambda^u_{ij}  \eps^{X_{q_i} + X_{u_j}}   \, , & y^d_{ij}  & = \lambda^d_{ij}  \eps^{X_{q_i} + X_{d_j}}  \, , 
\end{align} 
and by assumption $\lambda^{u,d}_{i,j} \sim {\cal O}(1)$, so that Yukawa hierarchies solely arise from $U(1)_F$ breaking parametrized by powers of $\epsilon$. 

Taking for simplicity all charges positive and hierarchical, $X_{q_3} \le X_{q_2} \le X_{q_1}$ etc, one obtains simple parametric relations for quark masses and mixings, which fixes all charges up to ${\cal O}(1)$ factors
\begin{align}
V_{{\rm CKM},i j} & \sim \eps^{|X_{q_i} - X_{q_j}|} \,, &
m^u_i & \sim   \eps^{X_{q_i} + X_{u_i}} \, , &
m^d_i & \sim   \eps^{X_{q_i} + X_{d_i}} \, .
\end{align}
A natural choice is to fix the order parameter as the Cabibbo angle, $\eps \approx V_{us} \approx 0.23$, and take $X_{q}  = \{3,2,0\}$ to reproduce the hierarchies in the CKM matrix, and similar choices for the remaining charges to fit quark masses. This means that the order parameter is not particularly small, and other completely equivalent choices for the charges can me made that give an equally good fit for suitable ${\cal O}(1)$ coefficients. 

While this simple framework can accommodate all Yukawa hierarchies observed in the SM, a central question is how it can be tested experimentally. Direct probes of the heavy  flavor dynamics at the scale $\Lambda_F$ need to rely on UV scales being close to the TeV scale, and there is no compelling reason why this scale should be  low (since the model aims to explain dimensionless numbers there is no preferred scale). More relevant are indirect probes, assuming that the same flavor symmetry that explains Yukawa hierarchies governs the interactions of new dynamics at low scales. A prime example is low-energy supersymmetry, but an interesting alternative is to take the  global flavor symmetry  at face value, and study the dynamics of the associated Goldstone boson.  
In the simple $U(1)_F$ model discussed above, the Goldstone $a(x)$ is readily identified by expanding the flavon as
\begin{align}
\chi (x) = \frac{\Lambda }{\sqrt 2} e^{- i a(x) /\Lambda} \, , 
\end{align}
where $\Lambda/\sqrt 2 = \langle \chi \rangle = \eps \Lambda_F$ and we dropped the radial mode for simplicity.  The Goldstone couplings to the SM are completely determined in terms of the charges $X_{f_i}$, and in particular the color and electromagnetic anomaly coefficients are given by the traces (cf.~Eq.~\eqref{eq:EN})  
\begin{align}
2 N & =   {\rm tr} \left(  2 \, {\bf X}_{q} +  {\bf X}_{u} +  {\bf X}_{d} \right) \, , \\
E & =   {\rm tr} \left(  \frac{5}{3}   {\bf X}_{q} + \frac{4}{3}  {\bf X}_{u} +  \frac{1}{3}  {\bf X}_{d} +    {\bf X}_{\ell} +  {\bf X}_{e} \right) \, . 
\end{align}
Because similar traces enter the determinants of Yukawa matrices, e.g.
\begin{align}
{\rm det} \, {\bf y}_u & = {\rm det} \, \boldsymbol{\lambda}_u \eps^{ {\rm tr} \left( {\bf X}_{q} +  {\bf X}_{u} \right)} \, , & 
{\rm det} \, {\bf y}_u & = {\rm det} \, \boldsymbol{\lambda}_u \eps^{ {\rm tr} \left( {\bf X}_{q} + {\bf X}_{d} \right)} \, , 
\end{align}
the anomaly coefficients can be directly related to the determinants of quark Yukawa matrices~\cite{Ibanez:1994ig, Binetruy:1994ru, Binetruy:1996xk}
\begin{align}
\label{detmumd1}
{\rm det} \,  {\bf y}_u \, {\rm det} \,  {\bf y}_d & = \lambda_{ud} \, \eps^{2 N}   \, , \\
 {\rm det} \,  {\bf y}_d / {\rm det} \,  {\bf y}_e & =  \lambda_{de} \, \eps^{\frac{8}{3} N - E}  \, , 
 \label{detmumd}
\end{align}  
where the quantities $\lambda_{ud} = {\rm det} \,  \boldsymbol{\lambda}_u {\rm det} \, \boldsymbol{\lambda}_d $ and $\lambda_{de} = {\rm det} \,  \boldsymbol{\lambda}_d/{\rm det} \,  \boldsymbol{\lambda}_e$ contain the ${\cal O}(1)$ coefficients defined in Eq.~(\ref{eq:LyukFN}). Given the strongly hierarchical structure of SM Yukawas, it is clear that the $U(1)_F$ symmetry has to be anomalous under QCD, $N \ne 0$, so that  the Goldstone boson of the flavor symmetry \emph{is the QCD axion} that solves the Strong CP Problem (dubbed \emph{axiflavon} in Ref.~\cite{Calibbi:2016hwq} and \emph{flaxion} in Ref.~\cite{Ema:2016ops}) with an axion decay constant set by the FN breaking scale, $f_a = \Lambda/(2N)$.

Indeed taking the numerical values for running Yukawas  at $10^9 \, \GeV$ from Ref.~\cite{Antusch:2025fpm}, one finds ${\rm det} \,  {\bf y}_u {\rm det} \,  {\bf y}_d \approx 5.7 \cdot 10^{-20} $, and  ${\rm det} \,  {\bf y}_d/ {\rm det} \, {\bf y}_e \approx 0.72$, so that clearly $N \ne 0$, since otherwise all Yukawa hierarchies would be due to small coefficients which instead are ${\cal O}(1)$ by assumptions. Quantitatively one finds
\begin{align}
\label{eq:2N}
2 N =\frac{\log {\rm det} \,  {\bf y}_u {\rm det} \,  {\bf y}_d- \log \lambda_{ud}}{\log  \eps} \approx \frac{-44- \log \lambda_{ud}}{\log  \eps} \, , 
\end{align}
with $\log \lambda_{ud} \ll 1$. Taking Eq.~\eqref{eq:2N}  at face value with $\eps = 0.23$ and $\lambda_{ud} = 1$ gives $N = 15$, although this value may vary substantially among different fits. For example,  the Cabibbo could arises from an ${\cal O}(1)$ coefficient, so that the order parameter would be much smaller, e.g. with $\eps = V_{cb}$ one obtains the value  $N = 7$ (note that despite appearance $N$ is an integer).

Remarkably  one has also a prediction for the axion-photon coupling  close to $E = 8/3\, N$, up to small model-dependent corrections, since 
\begin{align}
\frac{E}{N} & = \frac{8}{3} - 2 \frac{\log \frac{{\rm det} \,  {\bf y}_d}{{\rm det} \,  {\bf y}_e} - \log \lambda_{de}}{\log {\rm det} \,  {\bf y}_u {\rm det} \,  {\bf y}_d- \log \lambda_{ud}} \nonumber \\
& \approx  \frac{8}{3} -  \frac{0.64 + 2 \log \lambda_{de}}{44 + \log \lambda_{ud}}\, ,
\label{ENprediction}
\end{align}
and by assumption $\log \lambda_{de} \sim \log \lambda_{ud} \ll 1$. To estimate the uncertainty from ${\cal O}(1)$ coefficients, one may take flatly distributed ${\cal O}(1)$  coefficients in the range $[1/3,3]$ with random signs, resulting in a 99.9\% range for all $U(1)_F$ axiflavons
\begin{align}
E/N & \in \left[ 2.4 , 3.0 \right] \, ,  &
\left| E/N -1.92 \right| & \in \left[ 0.5 , 1.1 \right] \, , 
\label{ENpred}
\end{align}
to be compared with the usual KSVZ axion window~\cite{DiLuzio:2016sbl} 
\begin{align}
E/N & \in \left[ 5/3 , 44/3 \right] \, ,  &
\left| E/N -1.92 \right| & \in \left[ 0.3 , 12.7 \right] \, ,
\label{ENpredKSVZ}
\end{align}
and the explicit model of Ref.~\cite{Ema:2016ops} where  $E/N \approx 2.9$. Note that the prediction of $E/N = 8/3$ coincides with the one of the DFSZ-II axion model~\cite{Zhitnitsky:1980tq,Dine:1981rt}, or rather with any model where PQ charges are compatible with a $SU(5)$ GUT symmetry, i.e. ${\bf X}_{q} = {\bf X}_{u}  = {\bf X}_{e} $ and ${\bf X}_{d}  = {\bf X}_{\ell}$. This is no accident, as $SU(5)$ precisely implies that down-type and charged-lepton Yukawa have the same determinant (since ${\bf y}_d = {\bf y}_e^T$), which is  responsible for $E/N = 8/3$ according to Eq.~\eqref{detmumd}.

It is quite remarkable that the narrow prediction for $E/N$ in Eq.~(\ref{ENpred}) is largely insensitive on the details of the underlying flavor model, at least for the simplifying assumptions we made at the beginning of this discussion.  These are i) positive fermion charges ii)  vanishing Higgs charge  iii) no other fermions  charged under $U(1)_F$.  The first assumption can be relaxed to the requirement that only $\phi$ enters in the effective Yukawa operators but not $\phi^*$, which is enforced in supersymmetric embeddings, since the superpotential has to be holomorphic. The second assumption is actually not necessary, since possible Higgs charges would drop out of the ratio of determinants in Eq.~\eqref{detmumd} which is responsible for the $E/N$ prediction. Finally the third assumption is can be enforced when all additional particles are either bosons or vector-like fermions under $U(1)_F$. This can indeed be a natural feature of UV completions of the effective FN framework, see e.g. Refs.~\cite{Leurer:1992wg, Calibbi:2012yj}.

\begin{figure}
\centering
\includegraphics[width=0.92\textwidth]{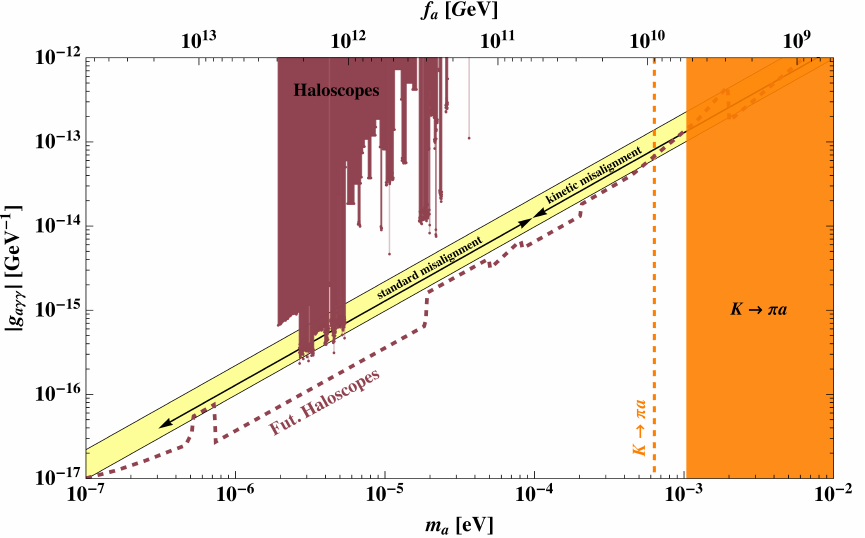}
\caption{Parameter space of the $U(1)_F$ axiflavon (within the yellow band).  Also shown is the region probed  by NA62 searches (orange),  present haloscope limits (dark red) and the mass range where the DM  abundance can be explained via standard or kinetic misalignment (black arrows).}
\label{fig:axiflavon}
\end{figure}

Besides the axion-photon coupling, the main  feature of the axiflavon is the prediction of flavor-violating axion couplings, which are given by rotating $U(1)_F$ charges to the fermion mass basis according to Eq.~\eqref{eq:CVAbasis}
\begin{align}
\CbVA_f& = - \frac{1}{2N} \left( {\bf U}_{fR}^\dagger {\bf X}_{f_R} {\bf U}_{fR}  \pm  {\bf U}_{fL}^\dagger {\bf X}_{f_L} {\bf U}_{fL} \right) \, .
\end{align}
As discussed in Section~\ref{sec:lab}, the by far strongest limits are set on $K \to \pi$ transitions controlled by the coupling $\CV_{sd}$. While the contribution from the right-handed quark sector is more model-dependent, in the left-handed sector the unitary rotations are CKM-like, i.e. $ {\bf U}_{dL}  \sim  {\bf U}_{uL}   \sim V_{\rm CKM}$, which implies for the relevant coupling
\begin{align}
\CV_{sd} \sim \eps \frac{X_{q_2} - X_{q_1}}{2N} \sim \frac{V_{us}}{2N} \, .
\end{align}
Since  the color anomaly is typically large, a reasonable estimate is $\CV_{sd} = 0.01 \kappa$, with a model-dependent ${\cal O}(1)$ parameter $\kappa$. Since the present limit on $f_a/|\CV_{sd}|$ is more than two orders of magnitude stronger than the constraints on $\mu e$-couplings or electron/nucleon-couplings from star cooling, it is suggestive that $K \to \pi a$ searches provide the strongest limits on the axiflavon (this is also the case in the explicit flaxion model of Ref.~\cite{Ema:2016ops}), giving a lower limit on $f_a$ as
\begin{align}
f_a \ge 6 \times 10^9 \GeV \times \kappa \, ,
\end{align}
or equivalently an upper limit on the axion mass
\begin{align}
m_a \le 1\times 10^{-3} \eV \times \frac{1}{\kappa} \, .
\end{align}
We summarize the parameter space of the $U(1)_F$ axiflavon in Fig.~\ref{fig:axiflavon}, embedded in the usual $m_a - g_{a \gamma \gamma}$ plane, but because of the limited range of $E/N$, the axiflavon lies within the yellow band. We also display the exclusion from $ K \to \pi  a$ searches at NA62 for $\kappa = 1$ (orange regions), as well as the future prospects (dashed orange line). Also indicated are present (dark red region) and future (dark red dashed) limits from haloscopes. Finally, we indicate  in the allowed region with black arrows the range where the axiflavon can fully account for the observed DM abundance,  either via standard misalignment (small masses) or kinetic misalignment (large masses). As can be seen from Fig.~\ref{fig:axiflavon}, the entire parameter space with axiflavon DM will be probed in a complementary way by future haloscopes and precision flavor experiments searching for $K \to \pi a$.

\subsubsection*{The $\boldsymbol{U(2)_F}$ Axiflavon}
While $U(1)_F$ flavor models come in many varieties, due to the freedom to make fermion charge assignments, the number of possibilities  is drastically reduced when  considering  additional non-abelian flavor symmetries. A class of very successful models based on $U(2)_F = SU(2)_F \times U(1)_F$ has been proposed in Ref.~\cite{Barbieri:1995uv}, and studied subsequently by many authors. While the original model has been strongly disfavored by precision measurements of CKM matrix elements at the B-factories~\cite{Roberts:2001zy}, a realistic model with slightly different charge assignments has been proposed in Ref.~\cite{Linster:2018avp}. In comparison with similar models in e.g.~\cite{Dudas:2013pja},  the original proposal for quarks and charged leptons is extended to the neutrino sector and does not rely on supersymmetry. In this framework, all hierarchies in masses and mixings originate from two small parameters associated with flavon VEVs. 

As discussed  in Ref.~\cite{Linster:2018avp}, the Goldstone boson of the spontaneously broken $U(1)_F \subset U(2)_F$ is a QCD axion, yielding a very predictive framework as all $U(1)_F$ charges are fixed (also different charge assignments are  viable, see e.g. Refs~\cite{Falkowski:2015zwa, Barbieri:2019zdz, Antusch:2023shi}, but to much less extent than in abelian models). We therefore briefly review the structure and predictions of the $U(2)$ axiflavon in the following, although the phenomenology closely resembles a standard DFSZ axion, due to the strong suppression of flavor violation thanks to the $SU(2)_F$ symmetry. 

The flavor quantum numbers of the fermions are compatible with a $SU(5)$ GUT symmetry, and the three generations transform as ${\bf 2} + {\bf 1}$ under $SU(2)_F$. The $U(1)_F$  charges are
\begin{align}
\label{eq:U2charges}
X_{{10}_3} & = 0 \, , & X_{{10}_a} & = X_{{\overline{5}}_a} = X_{{\overline{5}}_3} = 1 \, ,
\end{align}
where $a = 1,2$ is a generation index, and we use $SU(5)$ representations to collectively denote ${\bf X}_{10} \equiv - {\bf X}_q =  {\bf X}_u =  {\bf X}_e$ and  ${\bf X}_{\overline 5} \equiv - {\bf X}_\ell =  {\bf X}_d$.  The breaking of the flavor symmetry  is achieved by two scalar flavons $\phi$ and $\chi$, which transform under $U(2)_F$ as  $\phi$ = ${\bf 2}_{-1}$ and $\chi$= ${\bf 1}_{-1}$.  These fields acquire the VEVs
\begin{align}
\langle \phi \rangle & =  \begin{pmatrix} \eps_\phi \Lambda_F \\ 0 \end{pmatrix} \, , &
 \langle \chi \rangle & = \eps_\chi \Lambda_F \, ,
\end{align}
which are of  order $\eps_{\phi} \sim \eps_{\chi} \sim 0.01$,  with  precise values fixed by a numerical fit including ${\cal O}(1)$ Wilson coefficients.  To good approximation, the effective Yukawa matrices read
\begin{align}
{\bf y}_u  \approx
\begin{pmatrix}
0 & \lambda_{12}^u \eps_\chi^2  & 0  \\ 
- \lambda_{12}^u \eps_\chi^2 & \lambda_{22}^u \eps_\phi^2  & \lambda_{23}^u \eps_\phi \\
0 & \lambda_{32}^u \eps_\phi & \lambda_{33}^u \, , 
\end{pmatrix} 
\end{align}
and similar in the down and charge-lepton sector, with the same ``texture" zeros in the 11,13 and 31 entries. As a result of this zeros and the anti-symmetry in the 12 entry, the numbers of free parameters is greatly reduced with respect to pure $U(1)_F$ models, which implies that the unitary rotations are fixed to large extent.  Note that also the neutrino sector can be successfully reproduced by either adding right-handed neutrinos for Dirac neutrinos masses or just considering the effective Weinberg operator using a discrete $SU(2)_F$ subgroup, for other possibilities see e.g. Refs.~\cite{Antusch:2023shi, Giarnetti:2025idu}. 

 It is clear that the $U(1)_F$ symmetry is anomalous under QCD, so the associated Goldstone boson is a QCD axion. This follows immediately from the same argument as for the single $U(1)_F$ case, as the determinants of quark sector Yukawas still only depend on the $U(1)_F$ breaking flavon $\eps_\chi$, so ${\rm det} \,  {\bf y}_u \, {\rm det} \,  {\bf y}_d \sim \eps_\chi^9$, which implies $N \ne 0$, cf. Eq.~\eqref{detmumd1}. Moreover, since the charges are compatible with $SU(5)$, we obtain the prediction $E/N = 8/3$ exactly, just as in the standard DFSZ-II benchmark model. Since also $\phi$ carries $U(1)_F$ charge, the axion is embedded in the flavons as (ignoring the radial modes)
\begin{align}
\chi (x) & = \eps_\chi \Lambda_F e^{- \frac{ia(x)}{ \Lambda}}  \, , & \phi (x) & = \begin{pmatrix} \eps_\phi \Lambda_F \\ 0 \end{pmatrix} e^{- \frac{ia(x)}{ \Lambda}} \, ,
\end{align}
 with the $U(1)_F$ breaking scale $\Lambda \equiv  \sqrt2 \sqrt{\eps_\chi^2 + \eps_\phi^2} \, \Lambda_F$. From the charges in Eq.~\eqref{eq:U2charges} one can readily obtain the anomaly coefficients with Eq.~\eqref{eq:EN}
\begin{align}
2N & =  9\, , &  E &  =  12 \, , 
 \end{align}
 so that indeed $E/N = 8/3$. Note that the color anomaly coefficient is much smaller than in the pure $U(1)_F$ case, because the order parameters are  smaller and thus require smaller exponents to generate  large Yukawa hierarchies. One can identify the axion decay constant as $f_a \equiv  \Lambda/(2N)$ to match standard notation.

The axion couplings to fermions are obtained by rotating the $U(1)_F$ charges to the fermion mass basis using Eq.~\eqref{eq:CVAbasis}. Because of the ${\bf 2} + {\bf 1}$ flavor structure,  flavor violation is controlled entirely by rotations involving the third generation, giving
\begin{align}
{\bf C}^{\rm V}_{f} & =  {\bf C}^{\rm R}_{f}  + {\bf C}^{\rm L}_{f}  \, & {\bf C}^{\rm A}_{f} & =  {\bf C}^{\rm R}_{f}  - {\bf C}^{\rm L}_{f}   \, , 
\end{align}
with
\begin{align}
{\bf C}^{\rm P}_{f} & \equiv - \frac{{\bf U}_{f_P}^\dagger  {\bf X}_{f_P} {\bf U}_{f_P}}{2N}  =  - \frac{X_{f_P,a}}{2N} \mathbb{1} - \frac{X_{f_P,3} - X_{f_P,a}}{2N}  \boldsymbol{\eps}_{f}^{\rm P}   \, , 
\end{align}
for ${\rm P} = {\rm L,R}$ and the shorthand notation for unitary rotations
\begin{align}
 \eps_{f,ij}^{\rm P} & \equiv  (U_{f_P})_{3i} (U_{f_P})^*_{3j} \, .
\end{align}
As a consequence of unitarity the diagonal elements of these parameters satisfy
\begin{align}
0 & \le \eps^{\rm P}_{f,ii} \le 1 \, , & \sum_i  \eps^{\rm P}_{f,ii} & = 1 \, .
\end{align}
These expressions are valid for any axion model with PQ charges that are universal for two fermion generations (e.g. generalized DFSZ models~\cite{Saikawa:2019lng, Badziak:2021apn}). For the $U(2)_F$ axiflavon  one obtains
\begin{align}
\CV_{u_i u_j}  & =  - \frac{\eps^{\rm L}_{u,ij} - \eps^{\rm R}_{u,ij}}{9}  \, , &
\CA_{u_i u_j}  & = - \frac{2 \delta_{ij} -  \eps^{\rm L}_{u,ij}  -  \eps^{\rm R}_{u,ij} }{9}  \, , \nn \\
\CV_{d_i d_j}  & =  - \frac{ \eps^{\rm L}_{d,ij}  }{9}   \, , &\CA_{d_i d_j}  & =  - \frac{2 \delta_{ij} -  \eps^{\rm L}_{d,ij} }{9}   \, , \nn \\
\CV_{e_i e_j}  & =    \frac{ \eps^{\rm R}_{e,ij}}{9}    \, , &
\CA_{e_i e_j}  & =-  \frac{2 \delta_{ij} -  \eps^{\rm R}_{e,ij}}{9}   \, ,
\end{align}
with  the parametric structure
\begin{align}
\eps^{\rm L}_{u} \sim \eps^{\rm R}_{u} \sim \eps^{\rm L}_{d}  \sim \eps^{\rm R}_{e} \sim \begin{pmatrix}
\lambda^6  & \lambda^5   & \lambda^3  \\ 
 \lambda^5 & \lambda^4 & \lambda^2  \\
\lambda^3 & \lambda^2 &1
\end{pmatrix} \, .
\end{align}
Thus diagonal axial couplings are to very good approximation independent of the rotations, and one finds, denoting $C_{f_i} \equiv \CA_{f_i f_i}$, 
\begin{align}
C_{u,d,e,c,s,\mu} & = - \frac{2}{9} \, , &
 C_t  & = 0 \, ,  & C_{b,\tau}= - \frac{1}{9} \, .
\end{align} 
Instead flavor violation in all sectors is strongly suppressed by small CKM matrix elements
\begin{align}
C_{f_i \ne f_j} \sim (V_{\rm CKM})_{3i} (V_{\rm CKM})_{3j} \, ,  
\end{align}
which is  reminiscent for the suppression of FCNCs in supersymmetric extensions of the SM, for which the original $U(2)_F$ model was designed for~\cite{Barbieri:1995uv}. Indeed sfermion masses $ {\bf \tilde m}^2_{f,{\rm P}}$ have the same transformation property under the $U(3)^5$ flavor group as the axion couplings ${\bf C}^{\rm P}_{f} $ so they share the same CKM-like suppression. This suppression is particularly effective in $s \to d$ and $\mu \to e$ flavor transitions, so that the strongest limits on the $U(2)_F$ axiflavon do not come from precision flavor experiments but from astrophysics. Indeed the limit from RG cooling gives with $C_e = 2/9$ the limit
\begin{align}
f_a \ge 7.1 \times 10^8 \GeV \, , 
\end{align}
or equivalently 
\begin{align}
m_a \le 8 \, {\rm meV} \, , 
\end{align}
while the NA62 bound excludes only $f_a \gtrsim {\rm few} \times 10^7 \GeV$. Thus the $U(2)_F$ axiflavon has a phenomenology very similar  to the standard DFSZ-II axion, which is mainly constrained through its electron coupling (with fixed value $C_e = - 2/9$) and discovery prospects arise mostly  from haloscope searches through the photon coupling (determined by $E/N = 8/3$). 

\subsection{The Majoron from type-I seesaw models}
\label{sec:majoron}
Another  predictive scenario for flavor-violating couplings of a light dark ALP is the majoron~\cite{Chikashige:1980ui, Schechter:1981cv}, which arises as the pseudo-Goldstone boson from the  spontaneous breaking of lepton number in the context of type-I seesaw models. Its couplings to the SM fields are mostly fixed in order to reproduce neutrino masses and mixings, and since suppressed by the seesaw scale, the majoron can be sufficiently stable to play the role of DM~\cite{Frigerio:2011in}. For moderate values of the seesaw scale, these models are in the reach of  experiments looking for  muon decays to majorons, since LFV couplings are required in order to reproduce neutrino mixing. While there are many realizations of the majoron, in the following we review the simplest case with just two heavy sterile (right-handed) neutrinos discussed in Ref.~\cite{Calibbi:2020jvd}. 

\subsubsection*{Setup}
We add two right-handed neutrinos $N_{Ri}$  to the SM, which couple to leptons doublets with the $3 \times 2$ Yukawa  matrix ${\bf y}_N$ and have Majorana masses described by the $2 \times 2$ matrix ${\bf M}_N$
\begin{align}
\mathcal{L}_N =  i \overline{N_R}\slashed{\partial}N_R -\left( \overline{\ell_L} {\bf y}_N N_R H+
\frac{1}{2}\overline{N_R^c} {\bf M}_N  N_R + {\rm h.c.}\right).
\label{eq:Lseesaw}
\end{align}
In the seesaw limit,  where Majorana masses  are much larger than  Dirac masses, ${\bf M}_N \gg {\bf m}_D \equiv {\bf y}_N v/\sqrt{2} $, the sterile neutrinos are heavy and can be integrated out. The light neutrinos are predominantly part of the SM doublets, $\ell_L$, with the Majorana mass matrix obtained from the effective Weinberg operator according to the seesaw formula
\begin{equation}
\label{eq:seesaw}
{\bf m}_\nu = - {\bf m}_D  {\bf M}_N^{-1} {\bf m}_D^T \,,
\end{equation}
which in this case has rank 2, so one light neutrino is massless. 

The majoron $J$ arises when the Majorana mass matrix ${\bf M}_N$ is generated dynamically by the VEV of a new SM singlet scalar field $\sigma$. Replacing ${\bf M}_N\to {\bf g}_N \sigma$ in Eq.~\eqref{eq:Lseesaw}, with the singlet parametrized as  
\begin{equation}
\label{eq:maj}
\sigma (x) = \frac{f_N +  \hat{\sigma} (x)}{\sqrt{2}} e^{i J(x)/f_N} \, ,
\end{equation}
 the sterile neutrino mass matrix is given by
\begin{equation}
{\bf M}_N = \frac{{\bf g}_N f_N}{\sqrt{2}}\,. 
\end{equation}
The radial mode $\hat \sigma$ gets a mass of order $f_N$ and can be integrated out, while the majoron $J$ is a pNGB and thus expected to be light, with a mass $m_J$ induced by terms that break lepton number explicitly. Here we take the mass to be a free parameter, although it can be related to the electroweak scale and  other small couplings in complete models~\cite{Frigerio:2011in}. 

At tree-level, the majoron couples to sterile neutrinos proportional to ${\bf g}_N$, and thus to light neutrinos suppressed by the mixing $\propto v/f_N$. These  interactions  induce majoron couplings to charged leptons and quarks at one-loop~\cite{Chikashige:1980ui}; here we are only interested  in the seesaw limit of the general expressions~\cite{Pilaftsis:1993af}, which we match onto the effective axion Lagrangian~\eqref{eq:aferm}  upon identifying $a$ with $J$ and $f_a$ with $f_N$. Using the results presented in Refs.~\cite{Garcia-Cely:2017oco,Heeck:2019guh}, we find for the majoron couplings to quarks and leptons (see also Ref.~\cite{Herrero-Brocal:2023czw} for general results for one-loop couplings to charged leptons)
\begin{align}
\label{eq:Cqq:majoron}
{\bf C}^{\rm V}_{q} & = 0 \, , & {\bf C}^{\rm A}_{q} & =  - \frac{T^q_3}{16 \pi^2}  {\rm tr} \left( {\bf y}_N {\bf y}_N^\dagger \right) \mathbb{1}  \, , \\
{\bf C}^{\rm V}_{e} & =  \frac{{\bf y}_N {\bf y}_N^\dagger }{16 \pi^2} \, , & {\bf C}^{\rm A}_{e} & =\frac{1}{32 \pi^2}  \, {\rm tr} \left( {\bf y}_N {\bf y}_N^\dagger \right) \mathbb{1}-  {\bf C}^{\rm V}_e \, ,
\label{Ceemajoron}
\end{align}
where $T^{u,d}_3 = \pm 1/2$. Note that $\CV_{\mu e} = - \CA_{\mu e}$, so that the LFV couplings of the type-I seesaw majoron have the  V-A form, on which the best limits come from the TWIST experiment, cf.~Section \ref{sec:lab}. Also important are the ALP couplings to light neutrinos, which in the seesaw limit are diagonal in the neutrino mass basis and simply given by  $\CbVA_\nu  =\pm \mathbb{1}/\sqrt 2$. 

In order to make the majoron a viable DM candidate, one needs to ensure its stability on cosmological scales. Typically this requires to kinematically close the decay channels to charged leptons, so that only decays to neutrinos and photons are allowed. Both are suppressed, the former by light neutrino masses and  the latter  because lepton number has no electromagnetic anomaly. The corresponding decay rates read~\cite{Heeck:2019guh}  
\begin{align}
\label{Gnunu}
\Gamma(J\to \nu_i \bar{\nu}_i) = \frac{m_J}{16\pi f_N^2} m^2_i \sqrt{1-\frac{4 m_i^2}{m_J^2}}\,,
\end{align}
where $\nu_i$ are light neutrino mass eigenstates with mass $m_i$, and 
\begin{align}
\label{Ggaga}
\Gamma(J\to\gamma \gamma) \approx \frac{\alpha_{\rm em}^2 }{3072^2 \pi^7} \frac{m_J^7}{f_N^2 m_e^4} \left[ {\rm tr} \left( {\bf y}_N {\bf y}_N^\dagger \right) - 2 \left( {\bf y}_N {\bf y}_N^\dagger \right)_{11}  \right]^2  \,,
\end{align}
we we restrict to the leading contribution from the electron loop, which is a good approximation to the full loop function when $m_J \ll m_e$, cf.~Appendix~\ref{app:diphoton}. 

The free parameters ${\bf y}_N, {\bf g}_N, f_N$ are fixed to values that reproduce the observed masses and mixing angles  of light neutrinos. Here one can broadly distinguish two scenarios that lead to very different values for $f_N$, which is  the relevant mass scale suppressing all majoron couplings. In the standard seesaw scenario the elements of the Yukawa matrix ${\bf y}_N$ are all of similar size $y$, without any special structure. The mass scale of light neutrino masses is then given by  $m_\nu\sim y^2 v^2/M_N$, so that majoron couplings are of order $C/f_N \sim y^2/(16 \pi^2 f_N) \sim  m_\nu g_N/(16 \pi^2 v^2)$. Since the sum of neutrino masses are bounded from above by about $0.09 \eV$ from cosmology~\cite{diValentino:2022njd} and $g_N$ is a Yukawa coupling limited by perturbativity, one finds an effective suppression scale of about $10^{16} \GeV$, which makes the majoron completely invisible.  The standard set-up therefore cannot be probed by  present nor planned experiments. 

More interesting for phenomenology is therefore a low-scale seesaw scenario, in which light neutrino masses are additionally suppressed, so that $M_N$ and thus $f_N$ can be  lowered accordingly. This possibility can be easily understood in terms of symmetries, since light neutrino masses $m_\nu \propto y_N M_N^{-1} y_N^T$ transform non-trivially under under  unitary  symmetries  (in contrast to $ y_N y_N^\dagger$), and thus can be parametrically suppressed by small symmetry breaking terms. For example, one can employ generalized lepton number,  which is preserved by majoron couplings but broken by light neutrino masses, so these are proportional to small parameters that break the symmetry explicitly.   Such scenarios have been extensively studied in the literature, see Ref.~\cite{Deppisch:2015qwa} for a review. 

As a simple example of a majoron model with a low seesaw scale and enhanced majoron couplings we use the results of Ref.~\cite{Ibarra:2011xn}. To this extent one chooses  Majorana masses to have a pseudo-Dirac structure
 \begin{align}
{\bf M}_N & = \begin{pmatrix}
0 & M \\
M & 0
\end{pmatrix} = \frac{g_N}{\sqrt{2}} \begin{pmatrix}
0 & f_N \\
f_N & 0
\end{pmatrix} \, ,
\end{align} 
where $g_N \equiv \sqrt{2} M/f_N$ is a real  parameter, while the Yukawa couplings in Eq.~\eqref{eq:Lseesaw} have the general form 
\begin{align}
 {\bf y}_N & = \begin{pmatrix}
y_{e1} & y_{e2}  \\
y_{\mu 1} & y_{\mu 2}  \\
y_{\tau 1} & y_{\tau 2}  
\end{pmatrix} \, .
\end{align}
 In the $y_{\ell 1} \to 0$ limit the model has a global $U(1)$ symmetry, ${\bf M}_N \to P \,  {\bf M}_N \, P$, ${\bf y}_N \to e^{i \alpha} {\bf y}_N P$ with $P = {\rm diag} \{e^{i \alpha}, e^{-i \alpha} \}$. The majoron couplings, which are proportional to ${\bf y}_N {\bf y}_N^\dagger$, are invariant under this symmetry, while  neutrino masses are not, as they transform as ${\bf m}_\nu \to e^{2 i \alpha} {\bf m}_\nu$. This means that  neutrino masses must be necessarily  proportional to the symmetry breaking parameters $y_{\ell 1}$, which can be chosen sufficiently small to  enhance  majoron couplings while keeping light neutrino masses fixed. 
 
 Working in a basis where  the charged lepton matrix is diagonal, one can adjust the input parameters, $M$ and $y_{\ell i}$, such that all neutrino observables (2 mass differences + 3 mixing angles) are at their central experimental values. This leaves two free parameters, which one can choose to be $M$, the mass scale of sterile neutrinos, and the largest eigenvalue of the Dirac Yukawa matrix, $y^2 = \max \left[ {\rm eig} ( {\bf y}_N {\bf y}_N^\dagger )\right]$, which is only bounded by perturbativity, $y \lesssim 4$. Using the results of Ref.~\cite{Ibarra:2011xn}, one obtains for  Normal Ordering (NO) in the  limit $m_\nu  \ll y^2 v^2/M$
 \begin{align}
 \label{eq:yNyNdagger}
\left(  {\bf y}_N {\bf y}_N^\dagger \right)_{ij} & = y^2 \frac{m_3}{m_2 + m_3} A_i^* A_j \, , 
 \end{align}
where $m_2 = \sqrt{\Delta m_{21}^2}$ and  $m_3 = \sqrt{\Delta m_{31}^2}$ are the light neutrino masses\footnote{The lightest neutrino mass vanishes, $m_1 = 0$, since there are only two sterile neutrinos.}. Furthermore  
 \begin{align}
A_i & = U_{{\rm PMNS},i3} + i U_{{\rm PMNS},i2} \sqrt{m_2/m_3} \, , 
 \end{align}
in terms of PMNS matrix elements. The result for Inverted Ordering 
 (IO) is obtained from \eqref{eq:yNyNdagger} by replacing $m_3 \to m_2 = \sqrt{- \Delta m_{32}^2}$, $m_2 \to m_1 = \sqrt{- \Delta m_{21}^2 - \Delta m_{32}^2}$ and $U_{i3} \to U_{i2}$, $U_{i2} \to U_{i1}$.

\subsection*{Results}
One can use the latest global neutrino oscillation fit results~\cite{Esteban:2024eli,nuFIT}  in Eq.~\eqref{eq:yNyNdagger}, setting  neutrino masses, mixing angles and the Dirac CP phase  to their central experimental values. This gives the  matrix ${\bf y}_N {\bf y}_N^\dagger$ with a residual dependence on $y$ and a single Majorana phase $\alpha_m$. 
 For simplicity, we set the latter to zero, $\alpha_m = 0$, as it has not much impact on the results. The effective suppression scales for the  majoron couplings are in the NO case given by (similar results are obtained for IO)
 \begin{align}
  \frac{2 f_N}{C_{e}} & = \frac{1.1 \times 10^{10} \GeV}{ y^2}  \left(\frac{M/g_N}{10^7 \GeV} \right) \, , \nn \\ 
  \frac{2 f_N}{C_{\mu e}} & = \frac{1.6 \times 10^{10} \GeV}{ y^2}   \left( \frac{M/g_N }{10^7 \GeV} \right) \, .
  \label{eq:majoroncouplings}
   \end{align}
The other flavor-conserving and LFV couplings of the same order, so most relevant for majoron phenomenology are the leptonic couplings shown above. 

The majoron also couples to nucleons via its couplings to quarks in Eq.~\eqref{eq:Cqq:majoron},  which induce a  coupling to protons that reads
 \begin{align}
 \frac{2 f_N}{|C_{p}|}  & \approx \frac{0.68 \times 10^{10} \GeV}{ y^2} \left(\frac{M/g_N }{10^7 \GeV} \right) \, ,
 \end{align}
 and is constrained by SN cooling, cf.~Section~\ref{sec:astro}.
 
 Before we discuss  majoron phenomenology, we  comment on constraints on low-energy seesaw models from LFV processes mediated by  heavy sterile neutrinos with mass of order $M$, such as  $\mu\to e\gamma, \mu \to 3 e,$ and $\mu N\to eN$. The most relevant process is $\mu\to e\gamma$, which reads in the seesaw limit $M \gg v$~\cite{Ibarra:2011xn}, 
\begin{align}
\label{eq:BRmuegamma}
{\rm BR} (\mu \to e \gamma) & = \frac{3 \alpha_{\rm em}}{32 \pi}
\frac{v^4}{M^4} |({\bf y}_N {\bf y}_N^\dagger)_{12} |^2 \, . 
\end{align}  
Since the dependence on the sterile neutrino Yukawas is the same as the majoron process $\mu \to e J$, it is instructive to relate the two branching ratios as
\begin{align}
{\rm BR} (\mu \to e \gamma) & = 5.9 \times 10^{-14} \left( \frac{  \TeV}{  g_N M} \right)^2 \left( \frac{ {\rm BR} (\mu \to e J)}{5.8 \times 10^{-5}} \right) \, ,
\end{align}
where we took the current TWIST bound on $\mu \to e J$ as reference value. 
This should be compared to the current 90\% CL  limit set by the MEG-II experiment, ${\rm BR} (\mu \to e \gamma) <3.1\times 10^{-13}$ \cite{MEGII:2023ltw}.
The LFV  decays to majorons therefore  provide stronger limits on the seesaw scale $M$ than the LFV  processes mediated by heavy sterile neutrinos unless the seesaw limit is violated ($M \ll \TeV$) or $g_N \ll 1$. This result is a consequence of different scalings with the sterile neutrino mass, since ${\rm BR} (\mu\to e J) \propto f_N^{-2} \propto (g_N/M)^{2}$ is induced by a dimension-five operator, while ${\rm BR} (\mu \to e \gamma) \propto M^{-4}$ from a dimension-six  operator and similar for the other LFV processes. 

\subsection*{Majoron Phenomenology}

 Turning to majoron phenomenology, we first discuss its stability on cosmological scales, restricting for simplicity to NO. Summing over the two neutrino channels using Eq.~\eqref{Gnunu}, and taking $m_J \gg m_3 \approx 0.05 \eV$, one obtains for the inverse decay rate into neutrinos
 \begin{align}
 \left( \Gamma_{J \to \nu \overline \nu} \right) ^{-1} & = 3 \times 10^{18} \sec  \left( \frac{ \keV} {m_J}\right) \left( \frac{M/g_N}{10^{7} \GeV} \right)^2 \, , 
 \end{align}
 which should exceed the age of the universe, roughly $10^{17} \sec$. The inverse decay rate into photons is given by Eq.~\eqref{Ggaga}
 \begin{align}
 \left( \Gamma_{J \to \gamma \gamma} \right) ^{-1} & = \frac{7 \times 10^{33}}{y^4} \sec  \left( \frac{ \keV} {m_J}\right)^7 \left( \frac{M/g_N}{10^{7} \GeV} \right)^2 \, , 
 \end{align} 
 to be compared with the limits from X-ray telescopes, which at present are of order $10^{28} \sec$, and will improve by about two orders of magnitude in the future. 

For majoron production, one can consider misalignment for constant and dynamical majoron masses, see Section~\ref{sec:production}. Another possibility is freeze-in production from sterile neutrinos annihilation,  $N \overline{N} \to J J$, which is controlled by the majoron coupling to neutrinos proportional to $g_N$. Using the analytic results of Appendix~\ref{app:cosmo} for this case, on obtains  $\Omega_{\rm FI} h^2 \propto g_N^3 m_J/f_N$, in good agreement with the numerical results in Ref.~\cite{Frigerio:2011in}, which gives the required decay constant to obtain the observed abundance for given $g_N$ and $m_J$ as 
\begin{align}
\label{eq:FImajoron}
f_N^{\rm FI} = 2.2 \times 10^{10} \GeV \left( \frac{m_J}{\MeV} \right)  \left( \frac{g_N}{10^{-3}} \right)^3 \, .
\end{align}
In this case one has to respect the WDM bound, which is $m_J \ge 7 \keV$ from Eq.~\eqref{WDMbound} (the relevant scale in this case is the sterile neutrino mass $M$, and we take $g_*(M) = 106.75$). This ensure that the majoron coupling to sterile neutrinos is sufficiently small, so that the freeze-in regime is valid. If freeze-in gives the dominant contribution to the observed DM abundance, Eq.~\eqref{eq:FImajoron} fixes the coupling $g_N$ as a function of majoron mass and decay constant, so that ${\rm BR}(\mu \to e \gamma)$ is predicted from Eq.~\eqref{eq:BRmuegamma}. 

\begin{figure}
\centering
\includegraphics[width=1.0\textwidth]{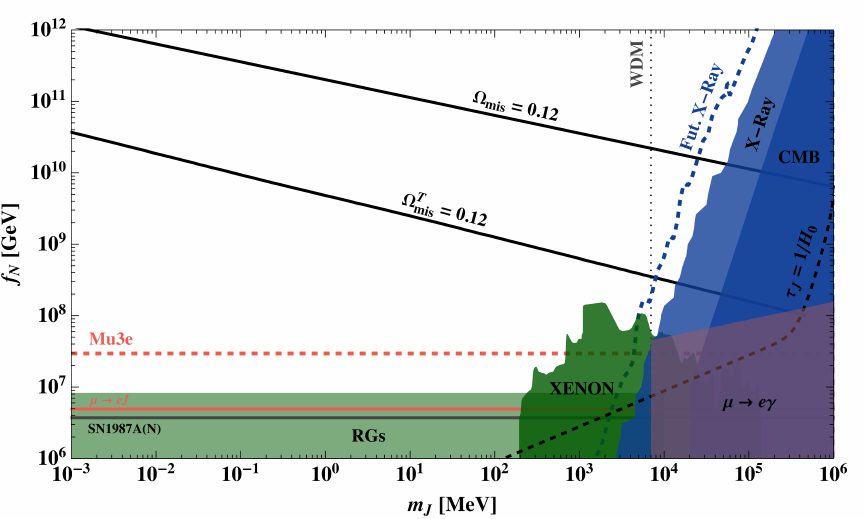}
\caption{Allowed parameter space for the low-energy seesaw majoron model. Shown are the  regions  excluded by  CMB (dark blue) and X-ray (blue) constraints on decaying DM, along with the reach of future X-rays searches (dashed blue). The black dashed contour denotes the region where the total lifetime exceeds $10^{17} \sec$, which includes neutrino decays (everywhere to the left of this line the majoron is a viable DM candidate).  Also shown is the present limit from  $\mu \to e J$ searches  (solid red) along with  future prospects at Mu3e (dashed red), and limits on the electron coupling by RGs (light green) and XENON (dark green). Black solid lines indicate the border of the regions where the majoron can fully account for the observed DM abundance via misalignment, which are the same as in Fig.~\ref{fig:DME1}. Conservative  limits on WDM are indicated by a  dotted  gray vertical line (on the right of this line the majoron can be cold DM). When the observed abundance is generated via the majoron coupling to heavy sterile neutrinos, $\mu \to e \gamma$ decays are predicted that are subject to MEG-II limits shown in  purple.}
\label{fig:majoron}
\end{figure}

In Fig.~\ref{fig:majoron}, we summarize the current status and future prospect of the constraints on the parameter space of the low-energy seesaw majoron model described above. Shown are the present and expected limits  from LFV experiments, star cooling constraints from RGs (on the electron coupling) and SN1987A  (on the nucleon coupling), limits from XENON and present limits from X-ray telescopes and CMB observations (on the photon coupling). We also indicate the border of the region where the total lifetime exceeds $10^{17} \sec$, which includes neutrino decays. As can be seen from Fig.~\ref{fig:majoron}, limits on decaying DM are dominated by XENON and X-ray line searches, while RG limits prevail in the low mass range over present LFV searches and SN cooling constraints. 
Black solid contours indicate the  regions where the majoron can fully account for the observed DM abundance. As in Fig.~\ref{fig:DME1} we have two representative lines that show the maximal misalignment contribution for $\theta_0 = \pi - 0.1$. On the right of the gray dotted line the majoron can be cold DM, produced e.g. via freeze-in from annihilation of sterile neutrinos, for a suitable choice of $g_N$ as determined by Eq.~\eqref{eq:FImajoron}. In this case one can predict $\mu \to e \gamma$ induced by heavy sterile neutrinos, but the parameter region excluded by MEG-II is (shown in dark red) is already ruled out by X-ray telescopes.  

 While the indicated misalignment regions will be partially probed by future X-ray line searches, the region of low majoron masses below 100 eV will only be tested by future LFV searches, which requires an alternative non-thermal mechanism (e.g. ALP kinetic misalignment) to produce majorons in sufficient quantities.

\subsection{ALP Dark Matter from LFV decays}
\label{sec:lfvfreezein}

In the previous scenarios flavor-violating ALP couplings were derived in by a top-down approach, by specifying a theoretical framework of spontaneous breaking of  a flavor symmetry or  lepton number. Here instead we follow a different approach that is more phenomenologically motivated, and ties flavor physics to cosmology. The main idea is to provide experimental targets for flavor-violating decays by connecting them directly to the observed DM relic abundance.  As discussed in Section~\ref{sec:production}, if a light axion has flavor-violating couplings that are too small to bring the axion into thermal equilibrium in the early universe,  decays of SM particles  produce an ALP abundance through thermal freeze-in proportional to the decay rate according to Eq.~\eqref{eq:OmFIdecay}
\begin{align}
\label{eq:freezeindecayapp}
\Omega_a h^2 & \sim 0.12 \left( \frac{{\rm BR} _{f \to f^\prime  a}}{8 \times 10^{-6}} \right) \left( \frac{m_a}{10 \keV} \right)  \left( \frac{m_\mu}{m_f} \right)^2  \left( \frac{\tau_\mu}{\tau_f} \right)  \, .
\end{align}   
Thus for a given flavor transition and ALP mass, the two-body decay rate is predicted in terms of the observed DM abundance. Remarkably, for muon decays the associated branching ratio  is in the reach of future experiments, cf.~Table~\ref{tab:expfut}. In turn, the present limits on the decay rate translate into limits on the couplings of order $10^9 \GeV$, which indeed is sufficiently small to avoid thermal contact of the ALP with the SM bath in the early universe, making this scenario self-consistent. 

Although the branching ratio is only predicted as a function of the ALP mass $m_a$, the associated range is  compact, bounded from above by the present limit on the decay rate and from below by the requirement that the decay is kinematically allowed (as lower rates require larger ALP masses to obtain the same abundance). This  window further shrinks when taking into account limits on warm and decaying DM.  Since DM is produced from decays of thermal particles, the Lyman-$\alpha$ limit in Eq.~\eqref{WDMbound} constrains $m_a$  to be larger than about 10 keV. For such heavy ALPs, limits on decaying DM from X-ray telescopes and CMB observations require that the ALP coupling to photons is suppressed (cf.~Fig.~\ref{fig:DME1}), which can be  realized naturally if the  PQ symmetry is free of color and EM anomalies. Even in this case, X-ray line searches give a stringent upper    upper bound on the effective electron coupling from Eq.~\eqref{eq:gammaE0} (see also  Fig.~\ref{fig:DME0})
\begin{align}
\tau_a \sim 10^{27}  \, {\rm s} \left( \frac{10 \keV}{m_a} \right)^7  \left( \frac{\Lambda/C_e}{10^{10} \GeV} \right)^2 \, .
\end{align}
This implies that a (slight) hierarchy between flavor-diagonal and flavor off-diagonal couplings is needed, in order to ensure sufficient DM stability. Note that in principle  diagonal couplings are not required in this scenario, however from a UV perspective extremely large hierarchies should be considered unnatural. 

Finally, one needs to ensure that the freeze-in contribution from decays indeed gives the dominant contribution to the DM abundance. The are two competing mechanisms, UV freeze-in production from $ f h \to f^\prime a$ scattering and misalignment. The UV freeze-in contribution is given in Eq.~\eqref{eq:UVfreezein}, and is negligible against  the decay contribution if the reheating temperature is sufficiently small,  roughly $T_R \lesssim 10^7 \GeV m_\mu/m_f$. The misalignment contribution instead is given in  Eq.~\eqref{eq:mis}
\begin{align}
  \Omega_{\rm mis} h^2 & \sim 0.12  \left(\frac{\Lambda \theta_0}{1.1 \times 10^{11}\GeV}\right)^2  \left(\frac{m_a}{ \keV}\right)^{\frac{1}{2}}  \, ,
  \end{align}
which  is under control for the relevant UV scales, even if the initial angle $\theta_0$ is not particularly small. Due to the upper limit on the reheating temperature, this contribution may be further diluted if  the onset of coherent oscillations happens prior to reheating, i.e. $T_{\rm osc} > T_R$, at least for the case of a period of early matter domination~\cite{Visinelli:2009kt, Blinov:2019rhb, Arias:2021rer}.

\subsubsection*{Setup}
While the basic mechanism sketched above can be applied to every SM flavor transition, in the following we focus on the lepton sector. Here the contributions to ALP production from flavor-diagonal scattering processes are suppressed with respect to decays by additional factors of $\alpha_{\rm em}$, cf.~Eq.~\eqref{eq:OmFIscat}, and thus are under perturbative control. Instead quark flavor-violating decays receive corrections from diagrams with additional gluons, which can give sizable corrections to ALP production rates in the early universe, in particular due to IR divergences which worsen the convergence of perturbative QCD calculations. For flavor-violating vertices parts of these corrections have been calculated in Ref.~\cite{Aghaie:2024jkj}, but it is questionable to what degree the gluon can be treated as a  weakly coupled degree of freedom when $g_s \gtrsim 1$, so perturbation theory can only be trusted at $T \gg \TeV$~\cite{Notari:2022ffe}. Nevertheless one may expect that production rates based solely on flavor-violating decays are a reasonably good approximation of the full calculation, so the result for the quarks sector in Ref.~\cite{Aghaie:2024jkj} should be valid. Still a complete analysis of next-to-leading-order (NLO) corrections in this case is desirable, in particular for cosmological limits, as discussed in Section~\ref{sec:cosmolimits}.

Following Ref.~\cite{Panci:2022wlc}, we  consider a ``leptophilic''  ALP $a$
that only couples to SM leptons according to the effective Lagrangian 
\begin{align}
{\cal L}_{\rm a} = \frac{\partial_\mu a}{2 \Lambda} \overline{f} \gamma^\mu \left( {\bf C}_f^{\rm V}  + {\bf C}_f^{\rm A} \gamma_5 \right) f - \frac{m_a^2}{2} a^2\, ,
\label{Lag} 
\end{align}
where ${\bf C}_f^{\rm A,V}$ are  hermitian matrices in lepton flavor space and $f = \ell, \nu$. These originate from rotating the charge matrices ${\bf X}_{L,R}$ of the underlying, spontaneously broken $U(1)_{\rm PQ}$ symmetry to the mass basis (cf.~Appendix~\ref{app:PQ})
\begin{align}
\CbVA_e & =  {\bf U}^\dagger_{R} {\bf X}_{R} {\bf U}_{R} \pm  {\bf U}^\dagger_{L} {\bf X}_{L} {\bf U}_{L} \, , & \CbVA_\nu& =  \pm  {\bf U}^\dagger_{\nu} {\bf X}_{L} {\bf U}_{\nu} \, , 
\end{align}
where ${\bf X}_{R}$ (${\bf X}_{L}$) are the  PQ charges of $SU(2)_L$ singlet (doublet) lepton fields, and ${\bf U}_{L,R,\nu}$ are unitary matrices are defined by ${\bf U}_{L}^\dagger {\bf m}_e {\bf U}_{R} = {\bf m}_e^{\rm diag}$, ${\bf U}_{\nu}^T {\bf m}_\nu {\bf U}_{\nu}  = {\bf m}_\nu^{\rm diag}$.  Left-handed charged lepton and neutrino rotations are related by the PMNS matrix ${\bf U}_{\rm PMNS}  = {\bf U}_{L}^\dagger {\bf U}_\nu $. As explained above, the ALP should have suppressed coupling to photons in order to satisfy the stringent constraints on decaying DM, which means that  leptonic PQ charges  satisfy ${\rm Tr} \, ({\bf X}_L - {\bf X}_R) = 0$, and thus are either vector-like or traceless. 

In the following we will discuss two simple scenarios: first we consider the case where only right-handed leptons of first and second generation are charged under $U(1)_{\rm PQ}$, so that without loss of generality ${\bf X}_R = {\rm diag} (1,-1,0)$ and ${\bf X}_L = 0$. We refer to this case as the ``two-flavor scenario''. Second we consider generic traceless charges for left-handed fields  ${\bf X}_L = {\rm diag} (1,{\rm X},-1-{\rm X})$ and ${\bf X}_R = 0$.
In the two-flavor case the rotation matrix is taken to be a general rotation in the 1-2 space parameterized by an angle $ 0 \le \beta \le \pi/2$, suitably defined such that in the mass basis 
\begin{align}
\label{eq:muedef}
\CbV_e & =\CbA_{e}  = \begin{pmatrix} s_{ \beta} &c_{ \beta}  & 0 \\ c_{ \beta}  &-s_{ \beta} &0  \\ 0 & 0 & 0 \end{pmatrix}   \, , &
\CbVA_{\nu} & =  0 \, .
\end{align}
We denote this case as the ``$\mu e$-scenario"; analogously one can consider the ``$\tau \mu$-scenario" and the ``$\tau e$-scenario". 
In the second case, dubbed the ``PMNS-scenario", we will take the rotation matrix in the LH sector to be the PMNS matrix, so that
${\bf U}_{\rm PMNS} \approx {\bf U}_L^\dagger$ and
${\bf U}_{\nu}$ is close to the identity. This gives in the mass basis
\begin{align}
\label{eq:PMNSdef}
\CbV_{e} & = - \CbA_{e}  = {\bf U}_{\rm PMNS} \,{\rm diag} (1,{\rm X},-1-{\rm X}) {\bf U}_{\rm PMNS}^\dagger\, , \nonumber \\
\CbV_{\nu} & = - \CbA_{\nu}  =  {\rm diag} (1,{\rm X},-1-{\rm X})  \, .
\end{align}
Both scenarios  depend on three free parameters: $\Lambda$, $m_a$ and $\beta$ (X) in the first (second) case, respectively. One combination of these parameters is fixed to reproduce the observed DM relic abundance, leaving a 2-dimensional parameter space in each scenario. 

\subsubsection*{Results}

\begin{figure}
\centering
\includegraphics[width=1.0\textwidth]{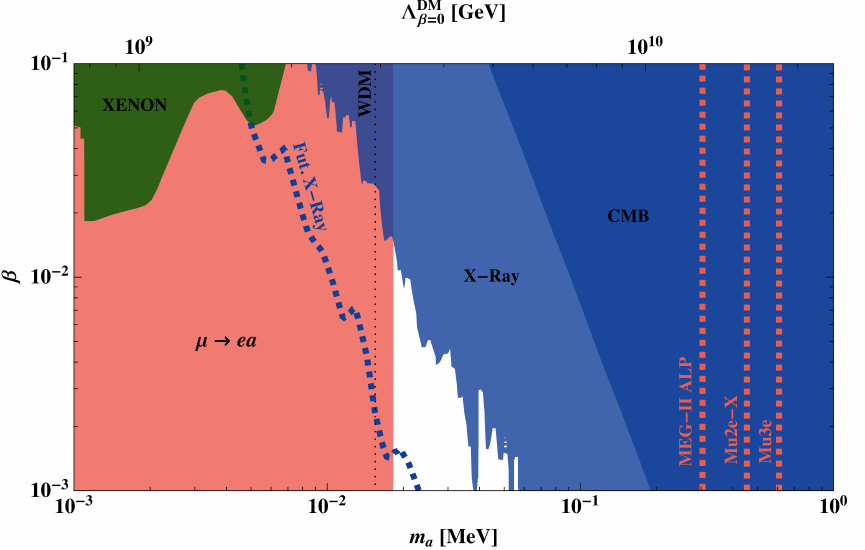}
\caption{Allowed parameter space for the $\mu e$-scenario defined in Eq.~\eqref{eq:muedef}, reproduced and updated from Ref.~\cite{Panci:2022wlc}. The UV scale  $\Lambda$ (top axis) is determined  requiring that the DM abundance  is produced via freeze-in, once  ALP mass $m_a$ (bottom axis) and the mixing angle  $\beta$ (left axis) are fixed  (we choose the reference value $\beta = 0$). The blue  regions are excluded by  CMB and X-ray constraints on decaying DM, while the reach of future X-rays searches is shown as a dashed blue line.  Conservative  constraints on WDM are indicated as a dotted  vertical line. Limits on the electron coupling set by XENON1T and XENONnT exclude the dark green region. Present limits from  $\mu \to e a$ searches exclude the red shaded region, with  prospects for  proposed searches at MEG~II, Mu2e-X and Mu3e   shown as dashed red lines. }
\label{fig:freezeinmue}
\end{figure}

\begin{figure}
\centering
\includegraphics[width=0.84\textwidth]{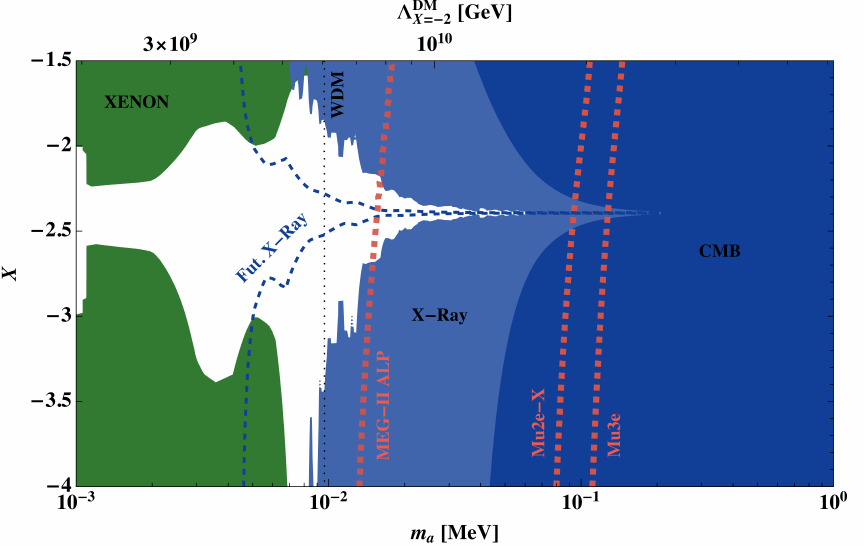}
\caption{Allowed parameter space for the PMNS-scenario defined in Eq.~\eqref{eq:PMNSdef}, reproduced and updated from Ref.~\cite{Panci:2022wlc}. The UV scale $\Lambda$ (top axis) is determined by requiring that the DM abundance  is produced via freeze-in, once  ALP mass $m_a$ (bottom axis) and the PQ charge $X$ (left axis) are fixed  (we choose the reference value $X = -2$). See Fig.~\ref{fig:freezeinmue}. }
\label{fig:freezeinPMNS}
\end{figure}

We summarize the results in Figs.~\ref{fig:freezeinmue} and \ref{fig:freezeinPMNS}, which show the allowed parameter space in the $\mu e$- and PMNS scenario, respectively. In both cases the UV scalet $\Lambda$ has been fixed to reproduce the observed relic density, solving the Boltzmann equation numerically (the results are in very good agreement with the analytical approximations in Eq.~\eqref{eq:freezeindecayapp}). For the $\mu e$-scenario (Fig.~\ref{fig:freezeinmue}) the required value of $\Lambda$ is essentially independent of the rotation angle $\beta$ and shown in the upper axis. In the PMNS scenario instead (Fig.~\ref{fig:freezeinPMNS})  the rotation angles are fixed to be the relatively large mixing angles in the neutrinos sector. This implies that limits from decaying DM are stringent, and a sufficient suppression of the decay rate into photons requires small ALP masses close above the WDM limit, unless there is an approximate cancellations in the axion-photon couplings for ${\rm X} \approx 2.6$. The allowed parameter space of these scenarios will be further tested by future X-ray telescopes, and in some cases almost entirely probed by precision measurements of rare muon decays. In particular,  the LFV muon decay width required for freeze-in sets a definitive target for proposed LFV experiments such as Mu3e~\cite{Calibbi:2020jvd,Perrevoort:2018ttp}, MEG~II~\cite{Jho:2022snj} and Mu2e-X~\cite{Hill:2023dym}. These experiments thus have the unique opportunity to search for the very same decay that could have produced axion DM in the early Universe. 

We do not show the results for the other two-flavor cases where the  axion is coupled solely to taus and muons or taus and electrons. In the $\tau \mu$-scenario the suppression of the photon coupling by the muon mass is sufficient in order to allow for diagonal couplings of the same order of the LFV ones, giving a largely unconstrained parameter space. However, the expected sensitivity to LFV tau branching fractions with final-state axions at Belle~II is  of  order $10^{-5}$  (cf.~Table~\ref{tab:expfut}),  while the target set by axion freeze-in is of order $10^{-9}$, due to the stringent bound from WDM. This leaves only future X-ray telescopes to probe this scenario. The $\tau e$-scenario is very similar to the $\mu e$-scenario, except that the WDM bound is slightly relaxed and again the expected experimental sensitivity on  LFV $\tau$-decays is too weak to probe the allowed parameter space. 

Finally we comment on scenarios with  flavor-violating axion couplings to quarks (as discussed above, there is the caveat that missing NLO corrections to ALP production from QCD corrections are potentially important).  The analogues for the effective two-flavor models and PMNS scenarios have been discussed in Ref.~\cite{Aghaie:2024jkj}, with important differences. In all scenarios  the axion decay rate into photons is additionally suppressed at least by a factor $m_a^4/m_\pi^4$, which enhances axion stability and  eases
constraints from X-ray line searches. The phenomenology of the two-flavor scenarios resemble the $\tau \ell$-scenarios, since flavor-violating searches are too weak to compete with the WDM constraint, which makes it very hard to probe these models in the near future,  also given the enhanced lifetime. Instead anomaly-free scenarios with free PQ charge and fixed CKM rotations are more interesting. Due to the smallness of CKM angles  flavor-diagonal couplings dominate ALP production, but $s\to d$ transitions are only suppressed by a single power of the Cabibbo-angle, making searches for $K \to \pi a$ very sensitive to this scenario.  Similar to the PMNS scenario, most of the viable parameter space will be probed complementary by future X-ray telescopes and flavor experiments.   

In all scenarios discussed so far we insisted on IR freeze-in, which puts an upper limit on the reheating temperature. By relaxing this bound one could also consider additional contributions to the DM abundance from UV freeze-in and misalignment, which however depends on additional parameters (the reheating temperature and the original misalignment angle). Such an analysis has been carried out in Ref.~\cite{Feruglio:2024dnc} for a light CP-violating scalar that  arises in scenarios where the Strong CP Problem is addressed by modular symmetries and flavor-violating couplings tare related to Yukawa hierarchies.

\section{Summary and Conclusions}

We  have discussed the flavor phenomenology of light dark particles, focusing on the interactions of pseudoscalars (``ALPs"), motivated as the remnants of spontaneously broken global symmetries. Below the electroweak scale their interactions with the SM is described  by an effective Lagrangian consisting of dimension-five operators (Eq.~\eqref{eq:axionfull}), which are suppressed by a UV scale that is set by the  symmetry breaking scale. In general these  operators contain also flavor-violating couplings, which arise in complete models from rotating Peccei-Quinn charges to the fermion mass basis, and are non-zero whenever these charges are not aligned with SM Yukawas in flavor space. 

The presence of flavor-violating interactions gives rise to two-body decays of SM particles with final state axions. For light ALPs below roughly 40 MeV, present constraints on photon and electron couplings imply that the associated decay length  exceeds ${\cal O} (100)$ meters, rendering the ALP invisible  on the scales of collider experiments, cf.~Figure.~\ref{fig:decaylength}. Existing (Table~\ref{tab:exp}) and planned (Table~\ref{tab:expfut}) searches for such two-body decays with missing energy put stringent limits on the associated decay rates, which can be translated to limits on the effective inverse ALP couplings. The resulting bounds are listed in Table~\ref{tab:lablimits} for each flavor transition, reaching $10^{12} \GeV$ in the case of $\s \to d$ transitions and  $10^{8} \GeV$  in $b \to q$ and $c \to u$ transitions. Although in the lepton sector the SM three-body decay with missing energy is  one of the main  channels, two-body decays probe UV scales up to $10^{7} \GeV$   in $\tau \to \ell$ and $10^{10} \GeV$ in $\mu \to e$ transitions. These limits can thus compete with the most stringent astrophysical limits on flavor-conserving couplings (electrons and nucleons) of order $10^{9 \div 10} \GeV$.

Flavor-violating couplings are also probed in the extreme environments of core-collapse supernovae, where nuclear densities and temperatures up to 40 MeV feature a sizable population of hyperons and muons. Decay of these particles to dark axions would contribute to efficient energy loss, which is constrained by the observation of the SN1987A neutrino pulse. This leads to model-independent constraints on the associated decay rates of muons and hyperons, which are of the same order as laboratory experiments for muons and  much more stringent  for hyperons.  In a similar way  the early universe acts as a cosmic laboratory probing flavor-violating decays, which would produce light axions  contributing to dark radiation, if sufficiently long-lived. Such contributions are constrained by precision CMB measurements, which probe  effective couplings in $c \to u$, $b \to q$ and $\tau \to \ell$ transitions of the same  order as  laboratory experiments (see Figure~\ref{fig:limits}).  

One of the main motivations of light dark ALPs is their ability to fully account for the observed dark matter relic abundance.  This primarily requires stability on cosmological scales, which can be naturally achieved for light masses and large UV scales. The axion also has to be produced in the early universe in sufficient quantities, and there are several mechanisms to achieve this. Most relevant is non-thermal production by misalignment, while thermal mechanisms have to ensure that the ALP is sufficiently cold in order to avoid constraints from structure formation, which roughly requires ALPs to be heavier than a few keV. Such heavy ALPs can  easily be in conflict with limits on decaying DM, in particular from X-ray telescopes, as illustrated in Figure~\ref{fig:DME1}, which shows a panorama of axion DM for sizable photon couplings. In this case thermal DM is strongly disfavored by searches for decaying DM, while the low mass region (below few keV) is constrained by astrophysical constraints and haloscopes. In this mass range precision flavor experiments play a key role to look for light axion DM, in particular  the QCD axion. They are even more important to probe ALP DM from anomaly-free symmetries, where  photon couplings are strongly suppressed for light axion masses, see Figure~\ref{fig:DME0}. 

In order to assess the relevance of the above discussions valid for a generic axion EFT, it is important to consider also  complete models where flavor-violating couplings are explicitly predicted. Here an important class of scenarios arises when the Peccei-Quinn symmetry that solves the Strong CP Problem is identified with a flavor symmetry that explains the hierarchical structure of  Yukawa couplings. Indeed global flavor symmetries are typically anomalous under QCD, which makes the associated Goldstone a QCD axion, also referred to as ``familon", ``axiflavon" or ``flaxion". Its couplings to photons and fermions are then largely predicted, in particular the flavor-violating couplings since by construction the transition between Peccei-Quinn and fermion mass basis is determined by the flavor symmetry. This gives rise to QCD axions that can be probed  by future searches for $K \to \pi a$ complementary to haloscopes, see Figure~\ref{fig:axiflavon}. Another scenario where lepton flavor-violating ALP couplings are predicted is the majoron, which arises from the spontaneous breaking of lepton number. In a type-I seesaw setup these couplings are directly connected to the pattern of neutrino masses and mixings, giving rise to a predictive scenario, where a light majoron can be searched for in LFV muon decays, although this requires a low decay constant (see Figure~\ref{fig:majoron}).  Finally, we have  discussed a third possibility to set targets for flavor-violating  two-body decays,  demanding that the very same decay has produced enough axions in the early universe  to fully account for the observed DM abundance via the freeze-in mechanism. For a given flavor transition this requirement gives rise to simple two-parameter models, which in the lepton sector can be almost entirely probed by future LFV searches and X-ray telescopes looking for photon lines from decaying DM, see Figure~\ref{fig:freezeinmue} and \ref{fig:freezeinPMNS}.

To conclude, flavor physics plays a key role in probing light new physics, which can be naturally related to the origin of dark matter, the Strong CP Problem or the spontaneous breaking of generic flavor non-universal symmetries. Flavor-violating decays then produce light dark particles in two-body decay of SM particles, which gives rise to missing energy signals that can be observed not only in high-precision flavor experiments, but also be probed in core-collapse supernovae or the early universe. These decays are controlled by dimension-five operators,  which makes dedicated laboratory searches  sensitive to very large UV scales and highly complementary to astrophysical and cosmological probes. Flavor-violating decays allow for efficient production of light DM, and can probe parameter regions that cannot be explored with common direct or indirect detection experiments, complementary to axion haloscopes. In the past, flavor physics was crucial to obtain indirect hints for heavy SM quarks before their  discovery, and it may well yield the first evidence for light  dark matter produced from  quark decays in the  future.  

\newpage
\section*{Acknowledgments}
 I am  grateful to all my collaborators, with whom I  enjoyed working on this topic over the past years, in particular to Marcin Badziak, Lorenzo Calibbi, Ferruccio Feruglio,  Jorge Martin Camalich, Paolo Panci,  Diego Redigolo,  Thomas Schwetz, Emmanuel Stamou, Mustafa Tabet and Jure Zupan. I also thank  Marcin Badziak,  Torben Ferber, Diego Redigolo and Jure Zupan for helpful discussions and Uli Nierste, Luca di Luzio, Jorge Martin Camalich and Jure Zupan for useful feedback on the manuscript. 
 
 This work has been partially supported by projects B3a and C3b of the DFG-funded Collaborative Research Center TRR257, ``Particle Physics Phenomenology after the Higgs Discovery" and has received support from the European Union's Horizon 2020 research and innovation programme under the Marie Sk{\l}odowska-Curie grant agreement No 860881-HIDDeN.

\begin{appendix}
\section{Axion couplings}
\label{app:axionbasis}
In this appendix, we consider the most general effective axion Lagrangian, and discuss how its Wilson coefficients transform under field redefinitions. Below the electroweak scale the general Lagrangian up to  local operators of dimension-five reads (omitting derivative axion-higgs couplings of the form $h^ \dagger \partial a \partial h$, which can be absorbed into axion-fermion couplings~\cite{Georgi:1986df})
\begin{align}
{\cal L } (f)& =  C_{GG} \frac{a}{\Lambda} \frac{\alpha_s}{4 \pi} G_{\mu \nu} \tilde{G}^{\mu \nu} + C_{\gamma \gamma} \frac{a}{\Lambda} \frac{\alpha_{\rm em}}{4 \pi} F_{\mu \nu} \tilde{F}^{\mu \nu} \nonumber \\
& + \frac{\partial_\mu a}{2 \Lambda} \overline{f} \gamma^\mu \left( {\bf C}_{\rm V} + {\bf C}_{\rm A}  \gamma_5  \right) f +  v \frac{a}{\Lambda} \overline{f} \left( {\bf C}_{\rm S}   + i {\bf C}_{\rm P}  \gamma_5 \right) f  \nonumber \\
& + i \overline{f} \gamma^\mu \partial_\mu f -  \left( \overline{f}_L  {\bf m}_f  f_R + {\rm h.c.}  \right) \, ,
\label{eq:genLagapp}
\end{align}
where $\Lambda$ denotes the UV scale, $v = 246 \GeV$ is the electroweak scale and $ \tilde{F}^{\mu \nu} = 1/2 \eps^{\mu  \nu \rho \sigma} F_{\rho \sigma}$ the dual field strength. We use a compact matrix notation where $f$ collects all charged fermions of the SM, $f=(u, d,e)^T$ with flavor indices
understood. Similarly we define the real diagonal fermion mass matrices   ${\bf m}_f = {\bf m}_u  \oplus  {\bf m}_d \oplus  {\bf m}_e $ and  the hermitian coupling matrices ${\bf C}_{\rm X} = {\bf C}_{{\rm X},u}  \oplus   {\bf C}_{{\rm X},d}  \oplus  {\bf C}_{{\rm X},e} $ for ${\rm X} = {\rm V,A, S,P}$.  Since $a$ is a pseudoscalar, this Lagrangian is CP-invariant if ${\bf C}_{\rm V}, {\bf C}_{\rm A},{\bf C}_{\rm P}$ are symmetric (and thus real)  matrices and ${\bf C}_{\rm S}$ is an antisymmetric (and thus purely imaginary) matrix.  Note that we have normalized ${\bf C}_{\rm S,P}$ appropriately, since above the electroweak scale the associated operators violate $SU(2)_L$, and thus must involve (at least) an additional power of the electroweak scale $v$. This also implies that the ALP necessarily couples to the Higgs boson, with couplings that can be recovered by replacing $v \to h(x)$ in Eq.~\eqref{eq:genLagapp}.   

 Furthermore we work with  the convention\footnote{We also use $g_{\mu \nu} = {\rm diag} (+1,-1,-1,-1)$, $\gamma_5 = i \gamma^0 \gamma^1 \gamma^2 \gamma^3$, $P_L = (1-\gamma_5)/2$. } $\eps^{0123} = +1 = - \eps_{0123}$, which agrees with the conventions in e.g. Refs.~\cite{Peskin:1995ev, Bauer:2017ris}. Note that other standard references such as Refs.~\cite{GrillidiCortona:2015jxo, DiLuzio:2020wdo} work with  $\eps^{0123} = -1$, which lead to relative signs with respect to our results. 

One can perform the following  axion-dependent chiral fermion field redefinitions with hermitian parameters in flavor space ${\bm \alpha}$ and ${\bm \beta}$ and the same matrix notation as above, 
\begin{align}
f & \to f^\prime \equiv e^{i a(x)/\Lambda \left( {\bm \alpha} + {\bm \beta}  \gamma_5 \right)} f \, .\label{PQtrafo}
\end{align}
To first order in $1/\Lambda$ this gives the same Lagrangian ${\cal L} (f^\prime)$ with  couplings to gluons and photons shifted as 
\begin{align}
C_{GG} & \to C_{GG} + \sum_{i = {\rm quarks}}  {\bm \beta}_{ii}  \, , \\
C_{\gamma \gamma} & \to C_{\gamma \gamma} +  2 \sum_{i = {\rm all}}  {\bm \beta}_{ii} N_c Q_i^2  \, ,
\end{align}
where  ${\bm \beta}_{ii}$  denotes the diagonal entries of the matrix ${\bm \beta}$, $N_c$ is a color factor (3 for for quarks, 0 for leptons), $Q_i$ is the electromagnetic charge of the fermion $f_i$, and $i$ runs over all fermions $(u,c,t,d,s,b,e,\mu,\tau)$, unless restricted to quarks. At the same time fermion couplings shift as
\begin{align}
{\bf C}_{\rm V} & \to  {\bf C}_{\rm V} - 2 {\bm \alpha} \, , & {\bf C}_{\rm A} & \to  {\bf C}_{\rm A} - 2 {\bm \beta} \, , \nonumber \\
{\bf C}_{\rm S} & \to  {\bf C}_{\rm S} - i [ {\bf m}_f/v, {\bm \alpha} ] \, , & {\bf C}_{\rm P} & \to  {\bf C}_{\rm P} - \{  {\bf m}_f/v, {\bm \beta} \} \, ,
\label{eq:Ctransform}
\end{align}
where the last line originates from 
\eq{{\bf m}_f \to e^{- i a(x)/\Lambda ( {\bm \alpha} -  {\bm \beta} \gamma_5)} {\bf m}_f  e^{ i a(x)/\Lambda ( {\bm \alpha} +  {\bm \beta} \gamma_5)} \, .} 
This implies that one can completely absorb derivative axion couplings with field redefinitions that satisfy ${\bm \alpha} =  {\bf C}_{\rm V}/2$ and ${\bm \beta} =  {\bf C}_{\rm A}/2$. Moreover, field redefinitions where only diagonal entries ${\bm \alpha}_{ii} \ne 0$ are non-vanishing affect only diagonal entries of ${\bf C}_{\rm V}$, since their commutator with diagonal fermion masses vanishes. This means that diagonal entries of ${\bf C}_{\rm V}$ are unphysical and have to drop out of physical quantities. 

Non-derivative couplings can only be absorbed when ${\bf C}_{\rm S}$ has vanishing diagonal entries, since the  field redefinitions  satisfy ${\bm \alpha}_{ij} =  i {\bf C}_{{\rm S},ij} v/(m_j - m_i)$ for $i \ne j$ and 0 (or rather arbitrary real numbers) otherwise and ${\bm \beta}_{ij} =   {\bf C}_{{\rm P},ij} v/(m_i + m_j)$, which  both are hermitian symmetric matrices. In particular all non-derivative flavor-violating couplings can be converted to derivative couplings.  Moreover, all non-derivative couplings can be absorbed in CP-conserving theories, where ${\bf C}_{\rm S}$ is an anti-symmetric matrix. 

\section{Peccei-Quinn symmetry}
\label{app:PQ}
In this appendix we discuss how the effective axion couplings in Appendix~\ref{app:axionbasis} arise from the point of view of a spontaneously broken Peccei-Quinn symmetry $U(1)_{\rm PQ}$. The axion $a$ is defined as the associated Goldstone boson, which transforms non-linearly under the PQ symmetry as 
\begin{align}
a (x) \to a(x) + \alpha f_{\rm PQ} \, , 
\end{align}
where $\alpha$ is the continuous parameter of the PQ transformation and $f_{\rm PQ}$ the PQ breaking scale. The associated Noether current $J_\mu^{\rm PQ}$ is given by 
\begin{align}
J_\mu^{\rm PQ} & = \frac{\partial {\cal L}}{\partial (\partial^\mu a ) } \frac{\delta a}{\delta \alpha} + \frac{\partial {\cal L}}{\partial (\partial^\mu \Psi ) } \frac{\delta \Psi}{\delta \alpha} \, , 
\end{align}
where $\delta a(x)$ denotes an infinitesimal field transformation and $\Psi(x)$ collectively denote fields that transform linearly under PQ. Restricting for simplicity to chiral fermions with   PQ charges $X_{\Psi}$, they transform under PQ as $\Psi(x) \to e^{i \alpha X_{\Psi}} \Psi(x) $, and the Noether current becomes
\begin{align}
\label{eq:noether}
J_\mu^{\rm PQ} & = \partial_\mu a  f_{\rm PQ} -  \sum_\psi X_{\psi}  \overline \psi  \gamma_\mu  \psi \, ,  
\end{align}
where the first term comes from the Goldstone kinetic term and the sum in the second term runs over all fermions charged under PQ. 

The first term implies in accordance  with the Goldstone theorem that the  Noether current  creates the Goldstone boson form the vacuum.   
\begin{align}
\langle 0 | J_\mu^{\rm PQ}| a(x) \rangle = - i f_{\rm PQ}  p_\mu e^{-i px}  \, , 
\end{align}
while the remaining terms determine the couplings of the Goldstone boson to fermions. 

In order to identify all Goldstone couplings, it is convenient to change field basis and work with new fermion fields $\psi_0 (x)$ defined as
\begin{align}
\label{psi0}
\psi_0(x) \equiv e^{-i X_\psi a(x)/f_{\rm PQ}} \psi (x) \, , 
\end{align}
which are invariant under PQ transformations. Since the Noether current (or rather the associated  charge density) is an observable, it cannot depend on the chosen field basis, which means  that in the new basis the second term in Eq.~\eqref{eq:noether} must arise from a Lagrangian coupling of the fermions to $\partial^\mu a$ 
\begin{align}
{\cal L}_{\rm ferm} & = - \frac{\partial^\mu a}{f_{\rm PQ}}   \sum_{\psi_0} X_{\psi}  \overline \psi_0  \gamma_\mu  \psi_0 =  - \frac{\partial^\mu a}{f_{\rm PQ}}   \sum_\psi X_{\psi}  \overline \psi  \gamma_\mu  \psi \, , 
\end{align}
which originates from plugging Eq.~\eqref{psi0} into the fermion kinetic terms. These  couplings of the Goldstone  respect the shift symmetry, and would be the only couplings if the symmetry would be exact. Indeed the Noether theorem states that in the field basis of Eq.~\eqref{psi0} 
\begin{align}
\label{eq:djmu}
\partial^\mu J_\mu^{\rm PQ} & = \frac{\delta {\cal L}}{\delta \alpha}  =  \frac{\partial {\cal L}}{\partial (\partial^\mu a)  } \frac{\delta (\partial^\mu a) }{\delta \alpha} + \frac{\partial {\cal L}}{ \partial a  } \frac{\delta a}{\delta \alpha} = \frac{\partial {\cal L}}{ \partial a  } f_{\rm PQ} \, .
 \end{align}
If the Lagrangian would be invariant under the PQ transformation, this would need to vanish, so the Goldstone only has derivative couplings. If however the PQ symmetry is anomalous, the Lagrangian is not invariant under PQ transformations, and the derivative of the Noether current does not vanish, but is given by (restricting to color and EM anomalies)
\begin{align}
\partial^\mu J_\mu^{\rm PQ} & = N  \frac{\alpha_s}{4 \pi} G_{\mu \nu} \tilde{G}^{\mu \nu} + E  \frac{\alpha_{\rm em}}{4 \pi} F_{\mu \nu} \tilde{F}^{\mu \nu} \, , 
\end{align} 
so that from Eq.~\eqref{eq:djmu} gives
\begin{align}
{\cal L}_{\rm anom} & = N \frac{a}{\vpq} \frac{\alpha_s}{4 \pi} G_{\mu \nu} \tilde{G}^{\mu \nu} + E \frac{a}{\vpq} \frac{\alpha_{\rm em}}{4 \pi} F_{\mu \nu} \tilde{F}^{\mu \nu} \, . 
\end{align} 
with anomaly coefficients given in terms of PQ charges (see Appendix~\ref{app:axionbasis} for our sign conventions)
\begin{align}
\label{eq:EN}
N & = -\sum_\psi T_c (\psi) \left( X_{\psi_L} -  X_{\psi_R}  \right) \, , \nonumber \\
 E & = - \sum_{\psi} Q_\psi^2 d_c (\psi)  \left( X_{\psi_L} -  X_{\psi_R}  \right) \, .
\end{align}
where $T_c(\psi)$ denotes the  Dynkin index of the $SU(3)_c$ representation of $\psi$ (with standard normalization $T_c = 1/2$ for triplets),  $d_c (\psi)$ its dimension ($d_c = 3$ for triplets) and $Q_{\psi}$ is the electric charge. Note that all fermions need to be vector-like under $SU(3)_c$ and $U(1)_{\rm EM}$.

One can now adopt the standard normalization of a QCD axion, $f_a  \equiv \vpq/(2N)$, and evaluate the fermion couplings for the SM in flavor space
\begin{align}
{\cal L}_{a} & =  \frac{a}{f_a} \frac{\alpha_s}{8 \pi} G  \tilde{G} + \frac{E}{N} \frac{a}{f_a} \frac{\alpha_{\rm em}}{8 \pi} F \tilde{F}  - \frac{\partial^\mu a}{2 N f_a}  \sum_{\psi_{\rm SM}} \overline \psi  \, {\bf X}_{\psi}  \gamma_\mu  \psi \, , 
\end{align} 
where we suppressed Lorentz and flavor indices for simplicity and  the last sum runs over ${\psi_{\rm SM} = q_L, u_R, d_R, \ell_L, e_R} $. Finally we go the fermion mass basis defined by unitary rotations $f_{L,R} \to {\bf U}_{f_{L,R}} f_{L,R}$ such that
\begin{align}
{\bf U}_{f_L}^\dagger {\bf m}_f {\bf U}_{f_R} =  {\bf m}_f^{\rm diag} \, , 
\end{align}
for each SM fermion sector $f=u,d,e, \nu$, in order to write the fermion couplings  as
\begin{align}
\label{eq:aferm}
{\cal L}_{a,{\rm ferm}} & =   \frac{\partial_\mu a}{2 f_a} \, \overline{f} \gamma^\mu \left(\CbV_f  + \CbA_f \gamma_5 \right)f \, , 
\end{align} 
with hermitian coupling matrices defined in terms of PQ charges as
\begin{align}
\label{eq:CVAbasis}
\CbV_f & = - \frac{1}{2N} \left( {\bf U}_{f_R}^\dagger {\bf X}_{f_R} {\bf U}_{f_R}  + {\bf U}_{f_L}^\dagger {\bf X}_{f_L} {\bf U}_{f_L} \right) \, , \nonumber \\
\CbA_f & = - \frac{1}{2N} \left( {\bf U}_{f_R}^\dagger {\bf X}_{f_R} {\bf U}_{f_R}  - {\bf U}_{f_L}^\dagger {\bf X}_{f_L} {\bf U}_{f_L} \right) \, .
\end{align}

\section{Two-body decay rates}
\label{app:2bodydecays}

In this appendix we collect  full expressions for the following two-body decay rates with axion final states:

\begin{enumerate}[label=C.\arabic*:]
  \item  Pseudoscalar meson transitions ${P\to P^{\prime}a}$
  \item Pseudoscalar to vector meson transitions, ${P\to \mathcal{V}a}$
  \item   Baryon transitions $ {B\to B^{\prime}a}$
  \item   Lepton transitions $ {\ell\to \ell^{\prime} a}$
  \end{enumerate}
  where we also consider polarized decays for baryon and lepton decays. For all decays we assign 4-momenta as
\begin{equation}
  \text{SM} (p)\to \text{SM}^\prime(p') + a(q) \, , 
\end{equation}
with $q=p-p^\prime$. In the case of hadronic decays the decay rates involve  hadronic matrix elements, which are parametrized in terms of various form factors that are evaluated at   
the ALP mass, $q^2=(p-p^\prime)^2=m_a^2$ (for the sake of light notation we  suppress this argument in the expressions below, apart from the dependence on the respective hadronic transition). We provide the definitions of the relevant form factors in Appendix~\ref{app:FFs}, and also collect their numerical values at $q^2 =  0$ relevant for very light axion masses. Apart from the form factors and the ALP couplings in Eq.~\eqref{axionFV} 
\begin{align}
{\cal L}_{a} & =   \frac{\partial_\mu a}{2 \Lambda} \, \overline{f} \gamma^\mu \left(\CV_{ij}  + \CA_{ij} \gamma_5 \right)f_j \, , 
\end{align}
the decay rates depend on  two mass ratios, for which we introduce the shorthand notation
\begin{align}
  \kappa_\alpha & \equiv m_\alpha^2 / M^2 \, , &
    \lambda_{\alpha \beta} & \equiv (1- \kappa_\alpha - \kappa_\beta)^2 - 4 \kappa_\alpha \kappa_\beta\,.
\end{align}
with $m_\alpha$ indicating the mass of the final-state particle $\alpha = \{ P^\prime, \mathcal{V}, B^\prime, \ell^\prime, a \}$ and $M$ the mass of the decaying particle.

\subsection[Meson decay rates ${P\to P' a}$]{Pseudoscalar meson decays $\boldsymbol{P\to P'a}$}
The  rate for the decay ${P\to P'a}$ with an underlying 
$q \to q^\prime$ flavor-changing transition is given by
\begin{equation}
  \label{eq:ratePPprime}
\Gamma (P \to P^\prime a) = \frac{m_P^3}{64 \pi \Lambda^2}
			\lambda^{1/2}_ {P^\prime  a } A_{P^\prime}  |\CV_{q^\prime q } |^2 \,,
\end{equation}
with the coefficient $A_{P^\prime}$ given by
\begin{align}
A_{ P^\prime}  &=  (1- \kappa_{P^\prime} )^2 |f_+^{P P^\prime}|^2  + \kappa_a^2 |f_-^{P P^\prime}|^2 \nonumber \\
& + 2 (1- \kappa_{P^\prime}) \kappa_a {\rm Re} \left[ f_+^{P P^\prime} f_-^{P P^\prime *} \right] \,,
\end{align}
and all form factors are evaluated at $q^2 = m_a^2$. Note that due to  parity conservation the rate does not depend  on the axial couplings $\CA_{ij}$ . In the limit of massless axions, the decay rate simplifies to
\begin{equation}
\Gamma(P \to P^\prime a) \xrightarrow[m_a = 0]{} 
\frac{ m_P^3}{64\pi \Lambda^2} \left(1- \frac{m_{P^\prime}^2}{m_P^2} \right)^3 |f_+^{P P^\prime} (0)|^2| \CV_{ q^\prime q } |^2\, .
\end{equation}
\subsection[Meson decay rates ${P\to {\mathcal V}a}$]{Pseudoscalar meson decays  $\boldsymbol{P\to {\mathcal V}a}$}
The  rate for the decay ${P\to {\mathcal V}a}$ with an underlying 
$q \to q^\prime$ flavor-changing transition is given by
\begin{equation}
\Gamma(P \to \mathcal{V}a) = \frac{m_P^3}{64 \pi \Lambda^2}\lambda^{3/2}_{\mathcal{V} a}  |A_0^{P  \mathcal{V} } (m_a^2)|^2   |\CA_{q^\prime q }|^2
 \,,
\end{equation}
In the limit of massless axions, the decay rate reduces to
\begin{equation}
 \Gamma (P \to \mathcal{V} a) \xrightarrow[m_a = 0]{}  \frac{m_P^3}{64\pi \Lambda^2}\left(1- \frac{m^2_\mathcal{V}}{m_P^2}\right)^3 |A_0^{P  \mathcal{V} } (0)|^2|\CA_{ q^\prime q }|^2  
         \,.
\end{equation}

\subsection[Baryon decay rates ${B\to B'a}$]{Polarized baryon decays $\boldsymbol{B\to B'a}$}
The  rate for baryon decays  ${B \to B^\prime a}$ with an underlying 
$q \to q^\prime$ flavor-changing transition is given by
\begin{align}
\Gamma (B \to B^\prime a)  &= \frac{m_B^3}{64 \pi \Lambda^2}\lambda^{1/2}_{B^\prime a}  \Big[ A_{B^\prime}^{V}  |\CV_{ q^\prime q}|^2  + A_{B^\prime}^{A} |\CA_{ q^\prime q}|^2 \Big] \, , \end{align}
with the shorthand notation
\begin{align}
A_{B^\prime}^{V} & = |f_1|^2 U_{B^\prime  a}^{-}+|f_3|^2 V_{B^\prime  a}^{-}+ {\rm Re}(f_1f_3^{*}) W_{B^\prime  a}^{-}\, ,  \nonumber \\
A_{B^\prime}^{A} & = |g_1|^2 U_{B^\prime  a}^{+}+|g_3|^2 V_{B^\prime  a}^{+}+{\rm Re}
(g_1g_3^{*}) W_{B^\prime  a}^{+} \, ,
\end{align}
with form factors evaluated at $q^2 = m_a^2$  and the kinematical coefficients 
\begin{align}
U_{B^\prime  a}^{\pm} & = \left( (1\mp \sqrt{\kappa_{B^\prime}})^2 - \kappa_a \right) \left( 1 \pm \sqrt{\kappa_{B^\prime}} \right)^2 \, , \nonumber \\
V_{B^\prime  a}^{\pm} & = \left( (1\mp \sqrt{\kappa_{B^\prime}})^2 - \kappa_a \right) \kappa_a^2  \, , \nonumber\\
W_{B^\prime  a}^{\pm} & = - 2 \kappa_{a} \left( (1\mp \sqrt{\kappa_{B^\prime}})^2 - \kappa_a \right) \left(  \sqrt{\kappa_{B^\prime}} \pm 1 \right)  \, .
\end{align}
In the limit of massless axions the decay rate simplifies to
\begin{align}
\Gamma (B \to B^\prime a)  & \xrightarrow[m_a = 0]{} 
\frac{ m_B^3}{64 \pi \Lambda^2}\left(1- \frac{m_{B^\prime}^2}{m_B^2} \right)^3 \nn \\
& \times \left[|f_1  (0)|^2|\CV_{ q^\prime q}|^2+|g_1 (0)|^2|\CA_{q^\prime q}|^2\right]\, .
\end{align}
For a fraction $P_B$ of polarized initial baryons $B$ the  decay rate depends on the cosine of the angle $\theta$ between the three-momentum of the final state baryon $B^\prime$  and the polarization vector of $B$. The differential decay rate is given by
\begin{align}
	\frac{d\Gamma (B \to B^\prime a)}{d\cos\theta}
  &= \frac{m_B^3}{32\pi\Lambda^2}\lambda^{1/2}_{B^\prime a}\left[  R^-_{B^\prime a} 
    |\CV_{ q^\prime q}|^2 + R^+_{B^\prime a} 
    |\CA_{ q^\prime q}|^2 \right] \nn \\
  & 
    - \frac{m_B^3}{16\pi\Lambda^2}\lambda_{B^\prime a} P_B \cos\theta  \, {\rm Re} (S_{B^\prime a}) {\rm Re}( \CV_{q'q} \CAs_{q'q} ) \nn \\
  &
    + \frac{m_B^3}{16\pi\Lambda^2}\lambda_{B^\prime a}  P_B \cos\theta  \,  {\rm Im} (S_{B^\prime a})  {\rm Im}( \CV_{q'q}  \CAs_{q'q} )\, , 
    \end{align}
with the shorthands 
\begin{align}
R^-_{B^\prime a} & =U_{B^\prime  a}^{-}  |f_1|^2  +V_{B^\prime  a}^{-} |f_3|^2  + W_{B^\prime  a}^{-}{\rm Re}(f_1f_3^{*}) \, , \nn \\
R^+_{B^\prime a} & =U_{B^\prime  a}^{+}  |g_1|^2  +V_{B^\prime  a}^{+} |g_3|^2  + W_{B^\prime  a}^{+}{\rm Re}(g_1g_3^{*}) \, , \nn \\
S_{B^\prime a}  & = \hat{U}_{B^\prime  a}^{} f_1g_1^{*} +\hat{G}_{B^\prime  a}^{+} f_3g_1^{*}
      +\hat{G}_{B^\prime  a}^{-} f_1g_3^{*}+\hat{V}_{B^\prime  a}^{} f_3g_3^{*} \, , 
\end{align}
and the kinematical coefficients  
\begin{align}
\hat{U}_{B^\prime  a}^{}&=\kaB -1 \,  ,   \nn \\
 \hat{V}_{B^\prime  a}^{}&=\kappa_a^2  \, ,  \nn \\
\hat{G}_{B^\prime  a}^{\pm}&=  - \kappa_a \left(\sqrt{\kaB} \pm 1 \right) \, .
\end{align}
In the limit of massless axions, the decay rate simplifies to
\begin{align}
\frac{d\Gamma (B \to B^\prime a)}{d\cos\theta}   \xrightarrow[m_a = 0]{} &	\frac{ m_B^3}{32\pi\Lambda^2}\left(1- \frac{m^2_{B^\prime}}{m_B^2} \right)^3  \nn \\
& \times \Big[ |f_1  |^2|\CV_{q'q}|^2+|g_1|^2|\CA_{q'q}|^2  \nn \\
& +2 P_B \cos\theta \, {\rm Re}\left( f_1   g_1^*    {\CV_{q'q} }\CAs_{q'q} \right) \Big]\, ,
\end{align}
with form factors evaluated at $q^2 = 0$.
\subsection[Lepton decay rates ${\ell\to \ell'a}$]{Polarized lepton decays $\boldsymbol{\ell\to \ell'a}$\label{polarizeddecay}}
The  rate for lepton decays  ${\ell \to \ell^\prime a}$ is given by 
\begin{equation}
\Gamma(\ell \to \ell^\prime a) = \frac{ m_{\ell}^3}{64 \pi \Lambda^2}\lambda^{1/2}_{\ell a}   \left(
A_{\ell^\prime a}^{-}  \left|\CV_{\ell^\prime\ell}\right|^2 
+ A_{\ell^\prime a}^{+}  \left|\CA_{\ell^\prime\ell}\right|^2 \right) \, ,
\end{equation}
with the kinematic coefficients
\begin{equation}
A_{\ell^\prime a}^{\pm}  =\left(1\pm \sqrt \kal\right)^2 \left( 1 \mp 2 \sqrt{\kal} + \kal -\kappa_a \right) \, .
\end{equation}
In the limit of massless axions the decay rate simplifies to 
\begin{equation}
 \Gamma (\ell \to \ell^\prime a) \xrightarrow[m_a = 0]{}  
\frac{m_{\ell}^3}{{64 \pi \Lambda^2 }}\left(1- \frac{m_{\ell^\prime}^2}{m_\ell^2} \right)^3\left(
\left|\CV_{\ell^\prime\ell} \right|^2 + \left|\CA_{\ell^\prime\ell}\right|^2
\right) \,.
\end{equation}
For a fraction $P_{\ell^-}$ of polarized initial leptons $\ell$ the  decay rate depends on the cosine of the angle $\theta$ between the three-momentum of the final state lepton $\ell^\prime$  and the polarization vector of $\ell$. The differential decay rate reads
\begin{align}
\label{totaldecayrate}
\frac{d\Gamma (\ell \to \ell^\prime a )}{d\cos\theta} &= \frac{m_{\ell}^3}{128 \pi   \Lambda^2} \lambda^{1/2}_{\ell^\prime a} 
\left[
 A_{\ell^\prime a}^- \left|\CV_{\ell^\prime\ell} \right|^2 
+ A_{\ell^\prime a}^+ \left|\CA_{\ell^\prime\ell}\right|^2  \right] \nn \\
& + \frac{m_{\ell}^3}{64 \pi   \Lambda^2} \lambda_{\ell^\prime a} 
   P_{\ell^-}   \cos\theta A_{\ell^\prime a}^\theta {\rm Re}( {\CV_{\ell^\prime\ell} }\CAs_{\ell^\prime\ell} ) \, ,
\end{align}
with the kinematic coefficient
\begin{align}
A_{\ell^\prime a}^\theta   &=  1 -\kal \, .
\end{align}
In the limit of massless axions, the polarized differential two-body rate simplifies to
\begin{align}
\frac{d\Gamma (\ell \to \ell^\prime a) }{d\cos\theta}  \xrightarrow[m_a = 0]{} &
\frac{ m_\ell^3}{128\pi \Lambda^2}\left(1- \frac{m_{\ell^\prime}^2}{m_\ell^2} \right)^3\left(\left|\CV_{\ell^\prime\ell} \right|^2  + \left|\CA_{\ell^\prime\ell}\right|^2 \right) \nn \\
& + \frac{ m_\ell^3}{64 \pi \Lambda^2}\left(1- \frac{m_{\ell^\prime}^2}{m_\ell^2} \right)^3  P_{\ell^-} \cos\theta \, {\rm Re}( \CV_{\ell^\prime\ell} \CAs_{\ell^\prime\ell}) \, . 
\end{align}
For polarized anti-leptons the same expressions hold upon the replacement $P_{\ell^-} \to -P_{\ell^+}$. 
\section{Hadronic matrix elements }
\label{app:FFs}
Below we give the definitions for the relevant hadronic matrix elements in terms of form factors, where the parametrization for vector and axial-vector currents are taken from the indicated references. The form factors are all functions of the momentum transfer $q = p - p^\prime$, where $p$ ($p^\prime$) is the momentum of the initial (final) particle and depend on the hadron transition; below we suppress this dependence. In Table~\ref{tab:FFtable} we collect the numerical values of all relevant form factors at $q^2 = 0$.

\subsection*{$\boldsymbol{P \to P^\prime }$}
For mesonic pseudoscalar ($P$) $\to$ pseudoscalar ($P^\prime$) decays the hadronic matrix elements for  
vector and axial-vector currents read~\cite{Gubernari:2018wyi}
\begin{align}
\langle P^\prime (p^\prime) | \overline{q}^\prime \gamma^\mu q | P(p) \rangle &= (p + p^\prime)^\mu f_+ (q^2) +  q^\mu f_- (q^2)\, ,  \\
\langle P^\prime (p^\prime) | \overline{q}^\prime \gamma^\mu \gamma_5 q | P(p) \rangle &= 0\, .  
\end{align}

\subsection*{$\boldsymbol{P \to \mathcal{V}}$}
For mesonic pseudoscalar ($P$) $\to$  vector ($\cal V$) decays the hadronic matrix element for  vector and 
axial-vector currents are parametrized as~\cite{Gubernari:2018wyi}
\begin{align}
\langle \mathcal{V} (p^\prime,\lambda)  | \overline{q}^\prime \gamma^\mu  q | P(p) \rangle & = 
P_1^{\mu} \mathcal{V}_1 \, ,  \\
\langle \mathcal{V} (p^\prime,\lambda)  | \overline{q}^\prime \gamma^\mu \gamma_5  q | P(p) \rangle & = 
- P_2^{\mu}\mathcal{V}_2 -  P_3^{\mu}\mathcal{V}_3- P_P^{\mu}\mathcal{V}_P\, , 
\end{align}
where $\lambda$ denotes the polarization of the vector particle $\mathcal V$ and we have suppressed the $q^2$ dependence of the form factors $\mathcal{V}_i(q^2)$. The kinematical functions read
\begin{align}
P_P^{\mu}&=i(\epsilon^{*}\!\cdot\!q) q^{\mu}\,,  \\
P_1^{\mu}&=- 2\epsilon^{\mu}_{~~\alpha\beta\gamma}\epsilon^{*\alpha}p^{\prime\beta}q^{\gamma} \, , \\
P_2^{\mu}&=i\left[\left(m_P^2-m_{{\mathcal{V}}}^2\right)\epsilon^{*\mu}-(\epsilon^{*}\!\cdot\!q)\left(p^{\prime}+p\right)^{\mu}\right]\,,  \\
P_3^{\mu}&=i(\epsilon^{*}\!\cdot\!q) \big[q^{\mu}-\frac{q^2}{m_P^2-m_{\mathcal{V}}^2}(p^{\prime}+p)^{\mu}\big]\,,
\end{align}
where $\epsilon^*_\mu=\epsilon_\mu^*(p',\lambda)$ denotes the polarization vector of the 
outgoing $\mathcal V$, and the  form factors are given by
\begin{align}
\mathcal{V}_P(q^2)&=\frac{-2m_{\mathcal{V}}}{q^2}A_0 \,,  \\ 
\mathcal{V}_1(q^2) & =\frac{-V}{m_P+m_{\mathcal{V}}}\,,  \\
\mathcal{V}_2(q^2) & =\frac{-A_1}{m_P-m_{\mathcal{V}}}\,, \\
\mathcal{V}_3(q^2)&=\frac{m_P+m_{\mathcal{V}}}{q^2}A_1 -\frac{m_P-m_{\mathcal V}}{q^2}A_2    \equiv\frac{2m_{\mathcal{V}}}{q^2}A_3 \,,
\end{align}
where we have suppressed the $q^2$ dependence of the form factors $V(q^2)$ and $A_i(q^2)$. They satisfy $A_3(0)=A_0(0)$, which ensures finite matrix elements at $q^2=0$. 
\subsection*{$\boldsymbol{B \to B^\prime}$}
For baryon  ($B$) $\to$  baryon ($B^\prime$) decays the hadronic matrix element for  vector and 
axial-vector currents 
are parametrized as~\cite{Detmold:2016pkz, Detmold:2015aaa, Meinel:2017ggx}
\begin{align}
\label{BFF}
\langle B^\prime (p^\prime) | \overline{q}^\prime \gamma^{\mu} q | B(p) \rangle &= 
\overline{u}_{B^\prime}(p^\prime) f^\mu (q^2) {u}_{B}(p)\, , \\
\langle B^\prime (p^\prime) | \overline{q}^\prime \gamma^{\mu}\gamma_5 q | B(p) \rangle &= 
\overline{u}_{B^\prime}(p^\prime) g^\mu (q^2) \gamma_5{u}_{B}(p)\,,  
\end{align}
where $u_B(p)$ and $u_{B'}(p')$ denote the Dirac spinors for $B$ and $B'$ respectively, and 
\begin{align}
f^\mu (q^2) & = f_1(q^2) \gamma^{\mu}-i\frac{f_2(q^2)}{m_B}\sigma^{\mu\nu}q_{\nu}+\frac{f_3(q^2)}{m_B}q^{\mu} \, , \\
g^\mu (q^2) & = g_1 (q^2)\gamma^{\mu}-i\frac{g_2(q^2)}{m_B}\sigma^{\mu\nu}q_{\nu}+\frac{g_3(q^2)}{m_B}q^{\mu}  \, . 
\end{align}
 
 \begin{table}
\centering
\renewcommand{\arraystretch}{1.6}
  \setlength{\arrayrulewidth}{.35mm}
\setlength{\tabcolsep}{0.8 mm}

\begin{tabular}{|c||c cc  | cccc |}
\hline
$q \to q^\prime$   & $f_+(0)$  & $P \to P^\prime$ & Ref. & $f_1(0)$   & $g_1(0)$  & $B \to B^\prime$  & Ref.  \\
\hline
$s \to d$ &  $0.9706 (27)$ &$K \to \pi$     &  \cite{FlavourLatticeAveragingGroupFLAG:2024oxs}     & $-1.22(6)$   & $-0.89(2)$     &    $\Lambda \to n$  &     \cite{Cabibbo:2003cu, Ledwig:2014rfa}\\

 &         &          &   &$-1.00(5)$   & $0.34(1)$  & $\Sigma^+ \to p$ &  \cite{Cabibbo:2003cu, Ledwig:2014rfa}\\

 &            &           &    & $1.00(5)$   & $1.26(5)$& $\Xi^- \to \Sigma^-$&  \cite{Cabibbo:2003cu, Ledwig:2014rfa}\\

 &             &             &     &$-0.71(4)$   & $-0.89(3)$&$\Xi^0 \to \Sigma^0$& \cite{Cabibbo:2003cu, Ledwig:2014rfa}\\

&             &            &     &$1.22(6)$   & $0.24(5)$& $\Xi^0 \to \Lambda$ & \cite{Cabibbo:2003cu, Ledwig:2014rfa}\\
\hline
$c \to u$ &  $0.612 (35)$ &$D \to \pi$   &  \cite{Lubicz:2017syv}   & $0.672(39)$   & $0.602(31)$             &  $\Lambda_c \to p$ &  \cite{Meinel:2017ggx} \\
\hline
$b \to d$ &  $0.21 (7)$ &$B \to \pi$     &   \cite{Gubernari:2018wyi}\        &   $0.23 (8)$           & $0.12 (13)$ & $\Lambda_b \to p$ &  \cite{Detmold:2015aaa} \\
\hline
$b \to s$ &  $0.335 (36)$ &$B \to K$     &    \cite{Bailey:2015dka}      &   $0.16 (4)$               & $0.11 (9)$ & $\Lambda_b \to \Lambda$ & \cite{Detmold:2016pkz} \\
\hline
\end{tabular}
\caption{\label{tab:FFtable} Numerical values of relevant form factors at $q^2 = 0$. In addition for $P \to {\cal V}$ transitions Ref.~\cite{Straub:2015ica}  gives for $B \to \rho$ the value $A_0(0) = 0.356 (42)$ and for $B \to K^*$ the value $A_0 (0) = 0.356 (46)$.}
\end{table}

\section{Axion decay rates into photons}
\label{app:diphoton}
In this appendix we collect the decay rates of a light axion into photons for  general CP-conserving ALP couplings. For the most general case including CP-violating couplings see Ref.~\cite{Feruglio:2025xvc}. 

The effective ALP Lagrangian at the electroweak scale is defined as in Appendix~\ref{app:axionbasis}
\begin{align}
\label{gammaEFT}
\mathcal{L}_{a} &= C_{GG}   \frac{a}{\Lambda} \frac{\alpha_s}{4\pi} G \tilde{G} +    C_{\gamma \gamma} \frac{a}{\Lambda} \frac{\alpha_{\rm em}}{4\pi} F \tilde{F}   + \overline f_i  \left( \frac{ia}{\Lambda} v y_i  + \frac{\partial_\mu a}{2 \Lambda}  C_i   \gamma^\mu\right) \gamma_5 f_i  \, , 
\end{align}
where we restricted to flavor-diagonal fermion couplings. 
We explicitly consider  both pseudoscalar and derivative couplings, although one can  restrict to one type using  the field redefinitions as discussed in Appendix~\ref{app:axionbasis}. 

The decay rate of a light ALP with mass $m_a \ll \GeV$ into two photons is given at one-loop    by
\begin{equation}
\label{decayrate}
\Gamma(a \to \gamma \gamma) = \frac{\alpha^2}{64\pi^3}\,\frac{m_a^3}{\Lambda^2}\, |\Cga^{\rm eff}(m_a)|^2 \,,
\end{equation}
with the (in general complex) decay amplitude 
\begin{align}
\Cga^{\rm eff}(m_a)&=\tilde{C}_{\gamma \gamma}
+ \frac{  \tilde{C}_{d} - \tilde{C}_{u}}{2}\frac{m_a^2}{m_{\pi^0}^2 - m_a^2}  + \sum_{i=e,\mu} \tilde{C}_i J_{1/2}^{PV}\left(\frac{m_a}{2 m_i} \right)  \nn \\
&+\tilde{C}_{GG}\left[- \frac{5}{3} -  \delta\,    \frac{m_{\pi^0}^2}{m_{\pi^0}^2 - m_a^2} \right] \nn \\
& +\tilde{C}_{GG}\left[ \frac{13 m_{\pi^0}^2}{12m_\eta^2}  + \delta\, ( 9 - 13 \delta - 9 \delta^2  ) \frac{ m_{\pi^0}^4}{12m_\eta^2 (m_{\pi^0}^2 - m_a^2 )}\right]   
\nn\\
& - \frac{m_a^2}{24m_\eta^2} \left[ 13 \tilde{C}_{u} + 13 \tilde{C}_{d} - 6 \tilde{C}_{s}  \right] \nn \\ 
& + \frac{m_a^2}{24m_\eta^2}  \left( 4 \tilde{C}_{d} + 6 \tilde{C}_{s} - 22 \tilde{C}_{u}  \right)  \delta \frac{m_{\pi^0}^2}{m_{\pi^0}^2 - m_a^2}  \nn \\ 
& - \frac{m_a^2}{24m_\eta^2} \tilde{C}_{GG}\left[  26   +  18\,\delta\, \frac{ m_{\pi^0}^2}{ m_{\pi^0}^2 - m_a^2} \right] \,.
\end{align}
with  the  shorthand notation
\begin{align}
\tilde{C}_i & \equiv C_i - \frac{y_i v}{m_i}  \, , \\
\tilde{C}_{GG} &\equiv  C_{GG} + \sum_{{i = u,d,s,c,b,t}}  \frac{y_i v}{2 m_i}  \,,   \\ 
\tilde{C}_{\gamma \gamma} &\equiv  C_{\gamma \gamma} + \sum_{{i = u,d,s,c,b,t,e,\mu,\tau}} N_c^i Q_i^2  \frac{y_i v}{m_i}   \, .  
\end{align}
We have also defined the loop function
\begin{align}
J_{1/2}^{PV}(z)&=1-\frac{1}{z^2}(\arcsin z)^2 \xrightarrow[z \ll 1]{} -  \frac{1}{3} z^2 + \hdots 
\end{align}
with $ | J_{1/2}^{PV} (z) |  \le 1.47$ and have neglected terms of order $m_a^2/M^2$, where $M$ is the mass of the $W$-boson or a heavy fermion above the GeV scale ($\tau, c,b,t$). Furthermore, $N_c^i$ denotes the color factor of the fermion $i$ (3 for quarks, 1 for leptons), $Q_i$ the fermion electric charge, and $\delta =(m_d - m_u)/(m_d+m_u) \approx 0.36$, where we used current-quark masses  $m_u=2.2 \MeV$, $m_d=4.7 \MeV$, $m_s=93.5 \MeV$~\cite{ParticleDataGroup:2024cfk}.

Since the decay rate is an observable, it cannot depend on the chosen field basis, and thus must be invariant under the field redefinitions in Eq.~\ref{PQtrafo} 
\begin{align}
f_i & \to f^\prime_i \equiv e^{i a(x)/\Lambda \left( \beta_i \gamma_5 \right)} f_i \, .
\end{align} 
This is indeed the case, as one can easily check that the effective couplings $\tilde{C}_i, \tilde{C}_{GG}, \tilde{C}_{\gamma \gamma}$ are all invariant under this transformation.

For ALP masses much below the pion mass, $m_a \ll m_\pi$, one can further simplify the expressions, dropping  terms of order $m_a^2/m_{\pi,\mu}^2$
\begin{align}
\Cga^{\rm eff}(m_a)& \approx
\tilde C_{\gamma \gamma} +\tilde{C}_{GG}\left[ - \frac{5}{3} -  \delta \right] +  \left[C_e -  \frac{y_e v}{m_e}    \right]  J_e   \nn \\
& +  \tilde{C}_{GG}\left[  \frac{13 m_\pi^2}{12m_\eta^2}  + \delta\, ( 9 - 13 \delta - 9 \delta^2  ) \frac{ m_\pi^2}{12m_\eta^2}\right]  \,, 
\end{align}
where we have defined 
\begin{align}
\label{eq:Jedef}
J_e \equiv  J_{1/2}^{PV}\left( \frac{m_a}{2 m_e} \right) = 1 - \frac{4 m_e^2}{m_a^2 } \arcsin^2{\frac{m_a}{2 m_e}}  \, ,
\end{align}
with $J_e \approx -  m_a^2/(12 m_e^2) $ for $m_a \ll m_e$. These expressions further simplify for  fermion couplings proportional to SM Yukawas $y_i = y \times m_i/v$, and are given by 
\begin{align}
\Cga^{\rm eff}(m_a)& \approx
C_{\gamma \gamma} - 1.96 \, C_{GG} + 2.16 y + \left( C_e - y   \right) J_e   \,.
\end{align}
It is instructive to study how large $\Cga^{\rm eff}$ can become in perturbative, renormalizable UV completions with universal fermion couplings. In this case $C_{\gamma \gamma} = C_{GG} = C_i = 0$, while  the pseudoscalar fermion couplings are restricted by perturbative unitarity, $y_i v/\Lambda < \sqrt{8 \pi/3} \approx 2.9$.~\cite{Cornella:2019uxs}. Thus $y$ is limited by the heaviest fermion the ALP couples to, for the case of the top quark $\Lambda/y > 56 \GeV$. This translates into an upper  limit on the effective photon coupling in renormalizable models with universal fermion couplings, given by $\Lambda_{\gamma} \equiv 2 \Lambda/|C_\gamma^{\rm eff}| \gtrsim 100 \GeV$ for $5 \MeV \le m_a \le 100 \MeV$ and $\Lambda_{\gamma}  \gtrsim 50 \GeV$ for $m_a \ll m_e$. 

Finally we take $m_a \ll m_e$ and consider pure derivative couplings, i.e. set $y=0$, which gives the effective photon coupling of a light (QCD) axion at low energies as
\begin{align}  
\label{eq:ALPCgamma}
\Cga &  \approx  C_{\gamma \gamma} - 1.96 \, C_{GG} - C_e \frac{m_a^2}{12 m_e^2} \, .
\end{align}
This coupling corresponds to a real Wilson coefficient in the low-energy effective Lagrangian at $Q \ll m_e$
\begin{align}
\mathcal{L}_{a} &=  C_{\gamma} \frac{a}{\Lambda} \frac{\alpha_{\rm em}}{4\pi} F ^{\mu \nu} \tilde{F}_{\mu \nu}  = \frac{1}{4} g_{a \gamma \gamma} a F ^{\mu \nu} \tilde{F}_{\mu \nu} \, , 
\end{align}
which defines the photon coupling commonly used in axion literature $g_{a \gamma \gamma} = \alpha_{\rm em} C_\gamma/(\pi \Lambda)$. Note that for the QCD axion $\Lambda = 2 C_{GG} f_a$, so that $g_{a \gamma \gamma} = \alpha_{\rm em} C_\gamma/(2 \pi C_{GG} f_a)$.

We  emphasize that the  result for $\Cga$ in Eq.~\eqref{eq:ALPCgamma} captures only partially the complete NLO corrections in the chiral expansion that have been derived in Ref.~\cite{GrillidiCortona:2015jxo} for two-flavor $\chi$PT with a matching between two-flavor and  three-flavor  $\chi$PT, and genuine three-flavor $\chi$PT in Ref.~\cite{Lu:2020rhp}. For $C_e = 0$ these computations determine the axion-photon coupling as $C_{\gamma \gamma} - 1.92(4) C_{GG}$ and $C_{\gamma \gamma} - 1.94(3) C_{GG}$, respectively. These results are in excellent agreement with recent determinations using lattice QCD, which give $C_{\gamma \gamma} - 1.77(8) C_{GG}$~\cite{Brandt:2026wir} and $C_{\gamma \gamma} - 1.965(35) C_{GG}$~\cite{Lin:2026hbw}. 

 \section{Axion cosmology  }
\label{app:cosmo}

In this appendix we briefly discuss the cosmology relevant for thermal  production of a light axion in the early universe, giving approximate expressions for  axion yields in freeze-out and freeze-in scenarios, partially taken from Refs.~\cite{Feruglio:2024dnc, Badziak:2024szg}.

\subsection*{Boltzmann equation}
The number density $n_a$ of  axions is determined by the integrated Boltzmann equation (see e.g. Ref.~\cite{Cadamuro:2010cz})
 \begin{align}
\frac{dn_a}{dt}+3Hn_a= \left( n_a^{\rm eq}-n_a \right)   \sum_i \Gamma_i \, ,
\end{align}
	where $n_a^{\rm eq} =   T^3 \zeta(3)/\pi^2 \approx 0.12 \, T^3$ is the axion  number density in equilibrium, $H$ is the Hubble parameter $H = T^2/M_{\rm Pl} 1.66 \sqrt{g_* (T)}$
with 	$g_* (T)$ denoting the total number of relativistic  degrees of freedom,
and $\Gamma_i$ are the specific axion production rates, which are related to the respective collision terms ${\cal C}_i$ by  $\Gamma_i = {\cal C}_i/n_a^{\rm eq}$.

Upon entropy conservation, $d(sa^3)/dt = 0$, the time derivative can be written in terms of a derivative with respect to temperature as $dT/dt = - H T$, which is valid when the effective number of relativistic entropy degrees of freedom is approximately constant, $dg_{*s} (T)/dT \approx 0$.  This relation can be used to rewrite the Boltzmann equation in terms of the axion yield $Y_a = n_a/s$, giving 
 \begin{align}
\frac{d Y_a}{dT} = - \left( 1 -\frac{Y_a}{Y_a^{\rm eq}} \right)   \sum_i \frac{{\cal C}_i (T)}{s T H} \, ,
\end{align}
with the entropy density $s = 0.439 \, T^3 g_{*s} (T)$ and the equilibrium yield $Y_a^{\rm eq}$ that is independent of temperature in this approximation. This equation can be integrated with the boundary condition that the yield at the onset of the thermal evolution at the reheating temperature $T_R$ is vanishing, $Y_a(T_R) = 0$. This gives for the final axion yield $Y^0_a$ today (at $T \approx 0$) 
\begin{align}
\label{BMsol}
Y^0_a & =  Y_a^{\rm eq} \left[ 1 - \exp \left( - \sum_i \int_0^{T_R} \frac{  {\cal C}_i (T)}{s T H Y_a^{\rm eq}} dT  \right) \right]  \nonumber \\
&  =  Y_a^{\rm eq} \left[ 1 - \exp \left( - \sum_i \int_0^{T_R} \frac{  \Gamma_i (T)}{ T H } dT  \right) \right]  \, ,
\end{align}
which can be further simplified depending whether the axion is in thermal equilibrium at early times or not, i.e. whether the ratio $ \Gamma_i/H$ is larger or smaller than unity. 

For sufficiently strong couplings to the thermal bath, $ \Gamma_i/H \gg 1$ at large $T$, the axion abundance is dominated by the first term in Eq.~\eqref{BMsol}, which will be true as long as the axion is in thermal equilibrium. However, as the temperature drops the production rates decrease, until $\Gamma_i (T_d) \approx H (T_d)$ at some temperature $T_d$ where  the axion decouples from the thermal bath,  with its equilibrium yield at the time of decoupling.  Thus in the regime of freeze-out one has
\begin{align}
\label{YFO}
Y^{\rm FO}_a =  Y_a^{\rm eq}(T_d) =  \frac{45 \, \zeta(3) }{2 \pi^4 g_{*s} (T_d)} \approx \frac{0.28}{g_{*s} (T_d)} \, ,
\end{align}
where the decoupling temperature is defined by the condition  $\Gamma_i (T_d) = H (T_d)$. Qualitatively $T_d$ decreases for smaller couplings, so $g_{*s}$ decreases and gives a larger yield. The final abundance in the freeze-out regime is obtained by multiplying the yield by  $m_a h^2 s_0/\rho_{\rm crit} \approx 0.27 (m_a/\eV)$,
\begin{align}
\label{OmegaFO}
\Omega^{\rm FO}_a  h^2  \approx 0.12 \left( \frac{m_a}{170 \eV} \right) \left( \frac{106.75}{g_{*s} (T_d)} \right) \, ,
\end{align}
so clearly the axion cannot be cold DM. Instead it will be constrained by limits on dark radiation if the axion is ultra-relativistic at recombination (roughly $m_a \lesssim 0.1 \eV$) and for heavier masses by structure formation (the Lyman-$\alpha$ forest). The bound from dark radiation is formulated as a limit on the effective numbers of additional neutrino species 
\eq{
\Delta N_{\rm eff} & = \left. \frac{8}{7} \left( \frac{11}{4} \right)^{4/3} \frac{\rho_a}{\rho_\gamma}\, \right|_{T = T_{\rm CMB}} \approx 4.40 \left.\frac{\rho_a}{\rho_\gamma}\, \right|_{T = T_{\rm CMB}}  \, , 
}
where $\rho_a$ is the axion energy density, $\rho_\gamma = 0.658 \, T^4$ is the energy density in photons, evaluated at  the temperature at recombination $T_{\rm CMB} \approx 0.3 \eV$. Assuming that approximately the same relation between number and energy density as in thermal equilibrium holds for $n_a$ and $\rho_a$, one can express $\rho_a$ through the axion yield at $T_{\rm CMB}$, 
\eq{
\Delta N_{\rm eff} & \approx 12.2  \left[  g_{*s} (T_{\rm CMB}) Y_a (T_{\rm CMB}) \right]^{4/3}  \, .
}
Finally neglecting the axion contribution to the entropy degrees of freedom, so that $g_{*s} (T_{\rm CMB})  \approx g_{*s}^{\rm SM} (T_{\rm CMB}) = 43/11$, and approximating the axion yield at recombination with the asymptotic yield $Y^0_a$ at $T \approx 0$ gives~\cite{DEramo:2021lgb} the simple estimate 
\eq{
\label{Neffapp}
\Delta N_{\rm eff} & \approx 74.8  \left( Y_a ^0 \right)^{4/3}  \, .
}
For sufficiently small couplings the axion never gets into thermal equilibrium, $ \Gamma_i/H \ll 1$ for all $T$, and the argument of the exponential in Eq.~\eqref{BMsol} is small. Expanding to first order, the  axion yield in the freeze-in regime is approximately given by  
\begin{align}
\label{YFI}
Y^{\rm FI}_a =   \sum_i \int_0^{T_R} \frac{  {\cal C}_i (T)}{s T H} dT    =  \sum_i  \frac{1.4}{g_{*s} \sqrt{g_*}} \int_0^{T_R} \frac{{\cal C}_i (T) M_{\rm Pl}}{T^6} dT 
\, .
\end{align}
In order to obtain the most accurate value (closest to the numerical solution), here the effective number of relativistic degrees of freedom should be  evaluated at the characteristic temperature of the most relevant production process.  For decays this is the mass of the decaying particle, while for scattering processes it is the threshold center-of-mass energy (unless the process is UV sensitive, in which case it is $T_{\rm max} = T_R$).  The final axion abundance in the freeze-in regime is thus given by the integral
\begin{align}
\Omega_a^{\rm FI} h^2 =  m_a  \sum_i  \frac{4.6 \times 10^{27}}{g_{*s} \sqrt{g_*}}  \int_0^{T_R} \frac{{\cal C}_i (T)}{T^6} dT  \, .
\label{abundance1}
\end{align}
Note that one may use again Eq.~\eqref{Neffapp} to estimate the contribution to $\dNeff$, however this expression is  expected to work well only for  freeze-out (where axion distributions  are close to thermal), while for  freeze-in Eq.~\eqref{Neffapp} underestimates the actual energy density~\cite{Badziak:2024szg}.  This is because Eq.~\eqref{Neffapp} is based on the assumption that  $\rho_a \approx \pi^2/30 (\pi^2/\zeta(3) n_a)^{4/3}$, for which the typical energy of axions  $\rho_a/n_a \sim 5 \, n_{a}^{1/3}$. But for freeze-in production $n_a^{1/3}$ is much smaller than the actual typical energy $(n_{a}^{\rm eq})^{1/3} \sim  T$ (since in freeze-in axions are  produced from particles in the thermal bath), so that the actual energy density of the axions is larger than obtained from Eq.~\eqref{Neffapp}. For this reason, in the context of freeze-in it is more appropriate to calculate the energy density from the evolution of the full phase space distribution, instead of using the Boltzmann equation for number densities~\cite{Badziak:2024qjg, DEramo:2024jhn}.

\subsection*{Collision terms} The form of the collision terms depend on the underlying axion production process. Here we consider all relevant processes off SM fermions for generic axion couplings in Appendix~\ref{app:axionbasis}. For flavor-violating couplings we consider flavor-violating (inverse) fermion decays $f \to f^\prime + a$,  while for flavor-diagonal couplings the leading processes are scatterings with photons (or gluons), $f + \gamma \to f + a$, fermion annihilations to axions and photons (or gluons) $f+  \overline{f} \to a + \gamma$, and scattering on Higgs bosons,  $f + h \to f + a$ and $f + \overline{f} \to a + h$, which is relevant only  in the high-energy limit. The respective collision terms read in terms of decay rates and cross-sections (using Maxwell-Boltzmann instead of Fermi-Dirac distributions)
\begin{align}
\label{Collterms}
{\cal C}_{f \to f^\prime a} & =  \frac{T m_f^2}{\pi^2} K_1 \left(\frac{m_f}{T} \right) \Gamma_{f \to f^\prime a} \, , \nn  \\
{\cal C}_{f \gamma \to f a} & =  \frac{T}{8 \pi^4}\int_{m_f^2}^\infty  \left(1- \frac{m_f^2}{s} \right)^2 s^{3/2}  \sigma_{f \gamma \to f a} (s) K_1 \left(\frac{\sqrt{s}}{T}\right) \, ds \, , \nn  \\
{\cal C}_{f \overline{f} \to \gamma a} & =\frac{T}{8 \pi^4 } \int_{4 m_f^2}^\infty  \left( 1 - \frac{4 m_f^2}{s} \right) s^{3/2} \sigma_{f \overline{f} \to \gamma a} (s) K_1 \left(\frac{\sqrt{s}}{T}\right) \, ds \, ,  \nn \\
{\cal C}_{f \overline{f} \to h a} & =  {\cal C}_{f h  \to f a}  = \frac{T}{8 \pi^4 } N_c^f \int_{0}^\infty   s^{3/2} \sigma^0_{f \overline{f} \to h a} K_1 \left(\frac{\sqrt{s}}{T}\right) \, ds \, ,
\end{align}
where $N_c^f$ is a color factor and $K_1(x)$ denotes the modified Bessel function of the second kind. Expressions for gluons instead of photons are analogous. Note that the expressions for colored particles are only valid for temperatures much above the QCD phase transition, so that quarks and gluons are appropriate weakly-coupled degrees of freedom. 

The decay rates and cross-section are defined of the generic Lagrangian couplings (see Appendix~\ref{app:axionbasis}), including also axion-fermion-fermion-Higgs couplings
\begin{align}
{\cal L } (f)& =   \frac{a}{\Lambda}  \left( 1 + \frac{h}{v} \right)  \overline{f} \left( {\bf C}_{\rm S}   + i {\bf C}_{\rm P}  \gamma_5 \right) f   \, ,
\end{align}
and read in the limit $m_f^\prime \ll m_f$ and $m_a=0$
\begin{align}
\Gamma_{f \to f^\prime\xi} & = \frac{m_f}{16 \pi \Lambda^2} \left(  \left| {\bf C}_{{\rm S}, f^\prime f}\right|^2   +  \left| {\bf C}_{{\rm P}, f^\prime f}\right|^2 \right) \, , \\
\sigma_{f \gamma \to f \xi}  & =   \frac{\alpha_{\rm em} Q_f^2}{8 s \Lambda^2}  \left[ {\bf C}_{{\rm S},  f f}^2 L_1 \left( m_f^2/s \right) +   {\bf C}_{{\rm P}, f f}^2  L_2 \left( m_f^2/s \right)   \right] \, , \\
\sigma_{q g \to q \xi}  & =   \frac{\alpha_s }{48 s \Lambda^2}  \left[ {\bf C}_{{\rm S}, q q}^2 L_1 \left( m_q^2/s \right) +   {\bf C}_{{\rm P}, q q}^2  L_2 \left( m_q^2/s \right)   \right] \, , \\
\sigma_{f \overline{f} \to  \gamma a}  & =   \frac{\alpha_{\rm em} Q_f^2}{s \Lambda^2}  \left[  {\bf C}_{{\rm S}, f f}^2  L_3 \left( m_f^2/s \right) +    {\bf C}_{{\rm P}, f f}^2 L_4\left(m_f^2/s \right)  \right] \, , \\
\sigma_{q \overline{q} \to  g a}  & =   \frac{4 \alpha_s}{9 s \Lambda^2}  \left[  {\bf C}_{{\rm S}, qq}^2  L_3 \left( m_q^2/s \right) +    {\bf C}_{{\rm P}, qq}^2 L_4\left(m_q^2/s \right)  \right] \, , 
\end{align}
where $Q_f$ denotes the electric charge of the fermion $f$ and the loop functions read
\begin{align}
 L_1 (x) & =  \frac{ -x^4 + 6 x^3 + 20 x^2 - 22x - 2 (3x+1)^2 \log x - 3 }{(1-x)^3} \, , \nn \\
 L_2(x)  & = \frac{  -x^2 + 4 x - 2 \log x - 3 }{1-x} \, , \nn \\
 L_3 (x) & =  \frac{4x}{\sqrt{1-4x}} + (1-4x) \tanh^{-1} (\sqrt{1-4x})  \, , \nn \\
 L_3 (x) & =  \frac{\tanh^{-1} (\sqrt{1-4x})}{1-4x} \, .
\end{align}
 Also relevant is scattering off Higgs bosons, which in the limit of $\sqrt{s} \gg m_H, m_f$ has the cross-section 
\begin{align}
 \sigma_{f h \to f a}^0 & = 2 \sigma^0_{f \overline{f} \to h a}  = \frac{  {\bf C}_{{\rm S}, ff}^2 +  {\bf C}_{{\rm P}, ff}^2}{32 \pi v^2 \Lambda^2} \, ,
\end{align}
which are not suppressed in the high-energy limit (as long as $\sqrt{s} \ll \Lambda$) in contrast to the ones involving gauge bosons above. Note that the factor of two between the two processes is compensated in the collision terms in Eq.~\eqref{Collterms} due to the different spin degrees of freedom.

\subsection*{Axion yields} 
We now use the results from the previous section in order to estimate the axion yields in the freeze-out  (Eq.~\eqref{YFO}) and freeze-in regime (Eq.~\eqref{YFI}) .

Starting with the latter case,  we need temperature integrals of the collision terms. For decays these integrals can be easily carried out, and are dominated by the region where $T \approx m_f$, giving $\int K_1 (m_f/T)/T^5 dT \propto m_f^{-4}$ for dimensional reasons, as long as $T_R \gg m_f$. Similarly one can do the temperature integrals for  scatterings involving vector bosons, as the remaining $s$-integral is convergent in the UV where  $\sigma_i \propto 1/s$ and can be done analytically. Instead the scattering on Higgs bosons is UV sensitive as the cross-section is constant, so one has first to perform the $s$-integral, giving $\int s^{3/2} K_1 (\sqrt{s}/T) \propto T^5$, and then perform the temperature integral which is linearly divergent and thus given by $T_{\max} = T_R$ (this justifies to work in the limit $\sqrt {s} \gg m_H, m_f$). The results for the integrated collision terms read 
\begin{align}
\label{CtermsFI}
 \int_0^{\infty} \frac{{\cal C}_{f \to f^\prime a} }{T^6} dT & = \frac{3}{2 \pi} \frac{\Gamma_{f \to f^\prime a}}{m_f^2} = \frac{3}{32 \pi^2 m_f \Lambda^2}  \left(  \left| {\bf C}_{{\rm S}, f^\prime f}\right|^2   +  \left| {\bf C}_{{\rm P}, f^\prime f}\right|^2  \right)\, , \nn \\
 \int_0^{\infty} \frac{{\cal C}_{f \gamma \to f a} }{T^6} dT & =  \frac{\alpha_{\rm em} Q_f^2}{168 \pi^3 m_f  \Lambda^2 }  \left[ (63 \pi^2 - 600)\,  {\bf C}_{{\rm S}, ff}^2 + 16\,  {\bf C}_{{\rm P}, ff}^2 \right] \, , \nn \\
 \int_0^{\infty} \frac{{\cal C}_{f \overline{f} \to \gamma a} }{T^6} dT & =   \frac{\alpha_{\rm em} Q_f^2}{160 \pi^2 m_f \Lambda^2 } \left( 13\,  {\bf C}_{{\rm S}, ff}^2 + 15\,  {\bf C}_{{\rm P}, ff}^2\right) \, , \nn \\
  \int_0^{T_R} \frac{{\cal C}_{f \overline{f} \to h a} }{T^6} dT & = \frac{4 T_R}{\pi^4} N^f_c \sigma^0_{f \overline{f} \to h a} = \frac{T_R }{16 \pi^5 v^2 \Lambda^2 } N_c^f  \left(  {\bf C}_{{\rm S}, ff}^2 + {\bf C}_{{\rm P}, ff}^2\right) \, .
 \end{align}
Expressions for gluons are obtained from the corresponding photon rates upon replacing $\alpha_{\rm em} Q_f^2 \to 4 \alpha_s$. However,  it is highly questionable whether the gluon can be treated as a  weakly coupled degree of freedom at temperatures of the order of the quark masses~\cite{Notari:2022ffe}, so that the integrated collision terms for these cases are likely to be subject of substantial non-perturbative corrections. 

Finally we adopt the standard convention for derivative axion couplings defined  in Eq.~\eqref{eq:aferm}
\begin{align}
{\cal L}_{{\rm axion}}  =  \frac{\partial_\mu a}{2 f_a} \, \overline{f}_i \gamma^\mu \left( \CV_{ij} + \CA_{ij} \gamma_5 \right)f_j \, ,
\end{align}
which can be matched to ${\bf C}_{{\rm S}}$ and ${\bf C}_{{\rm P}}$ according to Eq.~\eqref{eq:Ctransform}, giving in the limit $m_{f^\prime} \ll m_f$
\begin{align}
{\bf C}_{{\rm S}, f^\prime f} & = i m_f \CV_{f^\prime f} \frac{\Lambda}{2 f_a} \, , & {\bf C}_{{\rm P}, f^\prime f} & = - m_f \CA_{f^\prime f}  \frac{\Lambda}{2 f_a}  \, ,  \nonumber \\
{\bf C}_{{\rm S}, f f} & = 0 \, , & {\bf C}_{{\rm P}, f f} &  =  -2 m_f \CA_{f f}  \frac{\Lambda}{2 f_a}  \, .
\end{align}
Introducing the shorthand notation
\begin{align}
C_f & \equiv \CA_{ff} \, , & C_{f^\prime f} & \equiv \sqrt{|\CV_{f^\prime f}|^2 + |\CA_{f^\prime f}|^2 }\, ,  
\end{align}
the final freeze-in yields read
\begin{align}
\label{YFI}
Y_a^{\rm FI} &   =  10^{-3} \frac{ \, m_f M_{\rm Pl}}{ f_a^2 g_{*s} (m_f) \sqrt{g_*(m_f)} } 
\begin{cases}  
6.6 \, C_{f f^\prime}^2 	& {f}^\pm \rightarrow {f}^{\prime \pm} a \\
 8.6 \, \alpha_{\rm em} Q_f^2 \, C_{f }^2 &  {f}^\pm \gamma\rightarrow {f}^\pm a \\ 
13.3 \,  \alpha_{\rm em} Q_f^2 \, C_{f }^2 & {f}^+{f}^-\rightarrow \gamma a  \\   
\end{cases}  \, ,
\end{align}
including charge multiplicities factors of 2 for the first two processes. Similarly there is a factor 2 for the  UV sensitive process $f h \to f a$, which after summed with  $f \overline{f} \to h a$ gives the yield 
\begin{align}
\label{YFIhiggs}
Y_a^{\rm FI} &   =  0.86 \times 10^{-3} \frac{ \, m_f^2 T_R M_{\rm Pl}}{ f_a^2 v^2 g_{*s} (T_R) \sqrt{g_*(T_R)} } C_f^2 \, .
\end{align}
For leptons it is clear that decay processes dominate the freeze-in axion abundance for comparable couplings $C_f \sim C_{f^\prime f}$, because of the $\alpha_{\rm em}$ suppression. This is different for scattering processes with gluons, which formally can be obtained by replacing $\alpha_{\rm em} Q_f^2 \to 4 \alpha_s$. Since $\alpha_s (m_q)$ is sizable, it is however unclear whether quark decays still dominate over such  processes. Moreover, as discussed above, this also questions the validity of the gluonic scattering expressions, which are potentially subject to large non-perturbative corrections.

Turning to freeze-out, one needs to determine the decoupling temperature from the condition $ H (T_d) = \Gamma_i (T_d) = {\cal C}_i (T_d) /n_a (T_d) $, with collision terms given for all processes in Eq.~\eqref{Collterms}. To estimate the decoupling temperature, one only needs the production rates at high temperatures. Denoting the total production rate from scattering by $\Gamma_{\rm scat} (T)$, one finds 
\begin{align}
\label{GShigh}
\Gamma_{\rm scat} (T)  & =  \Gamma_{{\ell}^+{\ell}^-\rightarrow \gamma a} (T)  +  \Gamma_{{\ell}^+ \gamma\rightarrow {\ell}^+ a}  (T) + \Gamma_{{\ell}^- \gamma\rightarrow {\ell}^- a}  (T) \nonumber \\
& \xrightarrow{T \, \gg \, m_\ell} \frac{\alpha C_\ell^2   m_\ell^2 T}{ \pi^2 \zeta(3) f_a^2}  \log \frac{2  T}{m_\ell}  \, , 
\end{align} 
where for simplicity we restricted to leptons. Similarly one finds for lepton decays 
\begin{align}
\label{GDhigh}
\Gamma_{\rm decay} (T)  & = \frac{ {\cal C}_{{\ell}^+ \rightarrow \ell^{\prime + }a} + {\cal C}_{{\ell}^- \rightarrow \ell^{\prime - }a} }{n_a}  \xrightarrow{T \, \gg \, m_\ell}   \frac{C_{\ell \ell^\prime}^2 m_{\ell}^4}{32 \pi  \zeta(3) f_a^2 T} \, ,
\end{align}  	
so that production rates at high temperature are dominated  by scattering processes for similar values of couplings, $C_{\ell \ell^\prime} \sim C_\ell$. This is can be easily understood from the collision terms in the high-temperature  limit, which in general read for two-body decays and $2\to2$ scattering processes
\begin{align}
{\cal C}_{\rm decay} \sim T m^2 K_1 \left(\frac{m}{T}  \right)\Gamma \sim  \frac{T^2 m^4}{f_a^2} \, , 
\end{align}
and (taking $\sigma \sim m^2/f_a^2 \times 1/s$ in the high-energy limit)
\begin{align}
{\cal C}_{\rm scat} \sim T \int \sqrt{s} \frac{m^2}{f_a^2} K_1 \left(\frac{\sqrt{s}}{T} \right) ds  \sim  \frac{T^4 m^2}{f_a^2} \, . 
\end{align}
Since decoupling happens at moderately late times, i.e.  temperatures slightly below the characteristic mass scale of the process, $T \sim m/10$,  we just dress the high-temperature expressions above with  a Boltzmann suppression factor, also dropping the logarithm for simplicity, i.e. we take  
\begin{align}
\Gamma_{\rm scat} (T) 
& \xrightarrow{T \, \lesssim \, m_\ell} \frac{\alpha C_\ell^2  m_\ell^2  T}{ \pi^2 \zeta(3) f_a^2} e^{-\frac{m_\ell}{T}} \, , \nn \\
\Gamma_{\rm decay} (T)   & \xrightarrow{T \, \lesssim \, m_\ell}   \frac{C_{\ell \ell^\prime}^2 m_{\ell}^4}{32 \pi  \zeta(3) f_a^2 T} e^{-\frac{m_\ell}{T}}\, .
\end{align}  
From these expressions one can readily find the decoupling temperature for scattering and decays by  solving the transcendental equations
\begin{align}
\label{Tdeqs}
m_\ell  & \approx T_{d,{\rm scat}} \log \left[ \frac{\alpha C_\ell^2  m_\ell^2 M_{\rm Pl} }{19.7 \sqrt{g_{*s} (T_{d,{\rm scat}} )} f_a^2 T^d_S } \right] \, ,  \\
m_\ell & \approx T_{d,{\rm decay}}  \log \left[ \frac{C_{\ell \ell^\prime}^2 m_\ell^4 M_{\rm Pl} }{201 \sqrt{g_{*s} (T_{d,{\rm decay}} )} f_a^2 T_{d,{\rm decay}}^3 } \right] \, , 
\end{align}
and obtain the freeze-out yield from
\begin{align}
\label{YFO2}
Y^{\rm FO}_a =  Y_a^{\rm eq}(T_d)  \approx \frac{0.28}{g_{*s} (T_d)} \, ,
\end{align}
where $T_d = {\rm min} (T_{d,{\rm scat}}, T_{d,{\rm decay}})$. 

Similarly, for the UV sensitive scattering processes one finds
\begin{align}
\label{GUV}
\Gamma^{\rm UV}_{\rm scat} (T)  & =  \Gamma_{{\ell}^+{\ell}^-\rightarrow h a} (T)  +  \Gamma_{{\ell}^+ h\rightarrow {\ell}^+ a}  (T) + \Gamma_{{\ell}^- h\rightarrow {\ell}^- a}  (T)  \nonumber \\ 
& = 3 \times \frac{4 T^3}{\zeta(3) \pi^2} \sigma^0_{ \ell \overline{\ell} \to h a}  \, ,
\end{align} 
which decouple at the temperature
\begin{align}
\label{TdUV}
T_d^{\rm UV} & = 5 \times 10^8 \GeV \left( \frac{f_a/C_\ell}{10^{10} \GeV } \right)^2  \left( \frac{m_\tau}{m_\ell }\right)^2  \, ,
\end{align}
where we  took maximal entropy degrees of freedom in the SM, $g_{* s} (T^{\rm UV}_d) \approx g_{* s}^{\rm UV} = 106.75$.

Finally one can estimate the  temperature $T^{\rm eq}$ at which scatterings or decays bring axions into thermal equilibrium in the very early universe at $T \gg m_\ell$, which happens only for sufficiently large couplings $C/f_a$. For this task we evaluate the defining equations $H(T^{\rm eq}) =  \Gamma_i(T^{\rm eq})$ in the  high-temperature regime,  using Eqs.~\eqref{GShigh} and \eqref{GDhigh}. This gives the solutions
\begin{align}
T^{\rm eq}_{\rm scat} & \approx  \frac{\alpha C_{\ell }^2  m_\ell^2 M_{\rm Pl}}{ \pi^2 1.66  \zeta(3)  \sqrt{g_{* s}^{\rm UV}} f_a^2} \log \frac{2 \alpha C_{ \ell}^2  m_\ell M_{\rm Pl}}{ \pi^2 1.66  \zeta(3)  \sqrt{g_{* s}^{\rm UV}} f_a^2} \, , \\
T^{\rm eq}_{\rm decay} & \approx  \left( \frac{C_{\ell \ell^\prime}^2 m_\ell^4 M_{\rm Pl}}{32 \pi 1.66  \zeta(3)  \sqrt{g_{* s}^{\rm UV}} f_a^2} \right)^{1/3} \, ,
\end{align}
where we worked at logarithmic accuracy in $T^{\rm eq}_{\rm scat}$ and took in both cases the maximal entropy degrees of freedom in the SM, $g_{* s} (T^{\rm eq}) \approx g_{* s}^{\rm UV} = 106.75$, since $T^{\rm eq} \gg \TeV$. The point $T^{\rm eq} \approx m_\ell$ marks the transition regime between freeze-out and freeze-in, which we can estimate extrapolating the high-temperature expressions above. It  corresponds to the axion decay constant $(f_a/C_i)^{\rm eq}_i$ given by
\begin{align}
\label{fCeqscat}
& (f_a/C_{\ell })^{\rm eq}_{\rm scat}  \approx 0.058 \sqrt{ \alpha m_\ell M_{\rm Pl}} = 2 \times 10^7 \GeV \sqrt{\frac{m_\ell}{\GeV}} \, ,  \\
& (f_a/C_{\ell \ell^\prime})^{\rm eq}_{\rm decay}  \approx  0.022 \sqrt{m_\ell M_{\rm Pl}}= 8 \times 10^7 \GeV \sqrt{\frac{m_\ell}{\GeV}}\, .
\label{fCeqdecay}
\end{align}
For couplings much larger than these values, axion production is in the freeze-out regime with a yield approximately given by Eq.~\eqref{YFO2}, while for much smaller values  freeze-in production dominates with yield given in Eq.~\eqref{YFI}.

\end{appendix}

\newpage
\bibliographystyle{JHEP}

\bibliography{arxiv_v2}

\end{document}